\documentclass[twocolumn,twocolappendix]{aastex63}

\usepackage{amsmath,amsopn,amsxtra,txfonts}
\usepackage{comment}
\usepackage{threeparttablex}
\usepackage[caption=false]{subfig}

\usepackage{natbib} 
\usepackage{hyperref}
\usepackage{gensymb}
\usepackage{textcomp} 
\usepackage{float}
\usepackage{xcolor}

\newcommand{\hi}{H\textsc{~i}}
\newcommand{\hii}{H\textsc{~ii}}
\newcommand{\kms}{ \ifmmode{\rm km\thinspace s^{-1}}\else km\thinspace s$^{-1}$\fi}
\newcommand{\K}{\ensuremath{ \, \mathrm{K}}}

\newcommand{\kpc}{\ensuremath{\, \mathrm{kpc}}}
\newcommand{\cm}{\ensuremath{ \, \mathrm{cm}}}

\newcommand{\s}{\ensuremath{ \, \mathrm{s}}}

\newcommand{\R}{\ensuremath{ \, \mathrm{R}}}

\renewcommand{\thefootnote}{\roman{footnote}}

\newcommand{\vlsr}{\ifmmode{{v}_{\rm{LSR}}}\else ${v}_{\rm{LSR}}$\fi}

\newcommand{\dg}{\ifmmode{^{\circ}}\else $^{\circ}$\fi}

\begin{document}

\author[0009-0005-3076-1104]{April Horton}
\affiliation{Department of Physics \& Astronomy, Texas Christian University, Fort Worth, TX 76129, USA}

\author[0000-0003-0536-3081]{Suraj Poudel}
\affiliation{Department of Physics \& Astronomy, Texas Christian University, Fort Worth, TX 76129, USA}

\author[0000-0001-5817-0932]{Kathleen A. Barger}
\affiliation{Department of Physics \& Astronomy, Texas Christian University, Fort Worth, TX 76129, USA}

\author[0000-0001-9982-0241]{Scott Lucchini}
\affiliation{Center for Astrophysics $|$ Harvard \& Smithsonian, 60 Garden Street, Cambridge, MA 02138, USA}

\author{David L. Nidever}
\affiliation{Department of Physics, Montana State University, P.O. Box 173840, Bozeman, MT, 59717-3840, USA}

\author{Erica Chwalik}
\affiliation{Department of Physics, Montana State University, P.O. Box 173840, Bozeman, MT, 59717-3840, USA}

\author[0000-0003-4237-3553]{Frances H. Cashman}
\affiliation{Space Telescope Science Institute, 
3700 San Martin Drive,
Baltimore, MD 21218, USA}
\affil{Department of Physics, Presbyterian College, Clinton, SC 29325, USA}

\author[0000-0003-0724-4115]{Andrew J. Fox}
\affil{AURA for ESA, Space Telescope Science Institute, 3700 San Martin Drive, Baltimore, MD, 21218, USA}
\affil{Department of Physics \& Astronomy, Johns Hopkins University, 3400 N. Charles St., Baltimore, MD 21218, USA}

\author[0000-0001-9158-0829]{Nicolas Lehner}
\affiliation{Department of Physics and Astronomy, University of Notre Dame, Notre Dame, IN 46556, USA}

\author[0000-0002-7955-7359]{Dhanesh Krishnarao}
\affiliation{Physics Department, Colorado College, Colorado Springs, CO 80903, USA}

\author[0000-0003-2730-957X]{Naomi McClure-Griffiths}
\affiliation{Research School of Astronomy and Astrophysics, Australian National University, Canberra, ACT 2611, Australia}

\author{Elena D'Onghia}
\affiliation{Department of Astronomy, University of Wisconsin-Madison,Madison WI 53706, USA}

\author[0000-0002-7982-412X]{Jason Tumlinson}\affiliation{Space Telescope Science Institute, 
3700 San Martin Drive,
Baltimore, MD 21218, USA}
\affil{Department of Physics \& Astronomy, Johns Hopkins University, 3400 N. Charles St., Baltimore, MD 21218, USA}

\author[0000-0003-2308-8351]{Jo Vazquez}
\affiliation{Department of Physics \& Astronomy, Texas Christian University, Fort Worth, TX 76129, USA}

\author{Lauren Sdun}\affiliation{Department of Physics \& Astronomy, Texas Christian University, Fort Worth, TX 76129, USA}

\author{Stone Gebhart}\affiliation{Department of Physics \& Astronomy, Texas Christian University, Fort Worth, TX 76129, USA}

\author{Katherine Anthony}
\affiliation{Department of Computing, Mathematics and Physics,
Messiah University, 
Mechanicsburg, PA, 17055, USA}
\affiliation{Department of Physics \& Astronomy, Texas Christian University, Fort Worth, TX 76129, USA}

\author{Bryce Cole}\affiliation{Department of Physics \& Astronomy, Texas Christian University, Fort Worth, TX 76129, USA}

\author[0000-0002-1272-3017]{Jacco Th. van Loon}
\affiliation{Lennard-Jones Laboratories, Keele University, ST5 5BG, UK}

\author[0000-0002-6300-7459]{John M. Dickey}
\affiliation{School of Natural Sciences, Private Bag 37, University of Tasmania, Hobart, TAS 7001, Australia}

\author[0000-0001-6846-5347]{Callum Lynn}
\affiliation{Research School of Astronomy and Astrophysics, Australian National University, Canberra, ACT 2611, Australia}

\author[0000-0002-2712-4156]{Hiep Nguyen}
\affiliation{Research School of Astronomy and Astrophysics, Australian National University, Canberra, ACT 2611, Australia}

\author{Min-Young Lee}
\affiliation{Korea Astronomy and Space Science Institute
776 Daedeok-daero, Yuseong-gu, Daejeon 34055, Republic of Korea}

\title{New Interpretation for the Orientation of the LMC's Gaseous Arms B and E using ULLYSES\footnote{Based on observations made with the NASA/ESA \textit{Hubble Space Telescope}, obtained at the Space Telescope Science Institute, which is operated by the Association of Universities for Research in Astronomy, Inc. under NASA contract No. NAS5-26555.}}

\begin{abstract}

The Large Magellanic Cloud (LMC) has experienced disruption from tidal and ram-pressure forces as it travels through the halo of the Milky Way. In this project, we combine radio emission-line observations from the GASS and GASKAP surveys with UV absorption-line observations from the HST Ultraviolet Legacy Library of Young Stars as Essential Standards (ULLYSES) program to trace the material in front of the LMC. Along our 8~stellar sightlines near 30~Doradus, we observe gaseous structures likely associated with two arm-like features flowing in and around the LMC's disk. We detect the nearside gas in neutral, low, and medium ionization species. The lower-ionization species likely undergo both thermal and non-thermal broadening while the moderately-ionized phase is influenced by more non-thermal processes. The total integrated column density of Al\textsc{~iii} decreases with increasing angular offset from 30~Doradus, with sightlines within $\Delta\theta\lesssim0.25\arcdeg$ containing more moderately ionized gas. We demonstrate from a Gaussian decomposition technique on the \hi~emission that both arms likely trace an additional ${\sim}1\fdg0$ in Galactic longitude toward the 30~Doradus region than previously predicted. We constrain the orientation of the arms by suggesting that they likely converge around $(l,b) =(280\fdg5, -31\fdg2)$ and at least partially cross in front of the LMC. Our observations are consistent with two competing origins of the arms: 1) outflowing material is swept back by tidal and ram-pressure forces or 2) tidally stripped inflows fuel the ongoing stellar activity inside the LMC. Future studies are needed to distinguish between these scenarios. 

\end{abstract}

\section{Introduction}

\begin{figure*}[t]
  \centering
  \includegraphics[width=1\textwidth]{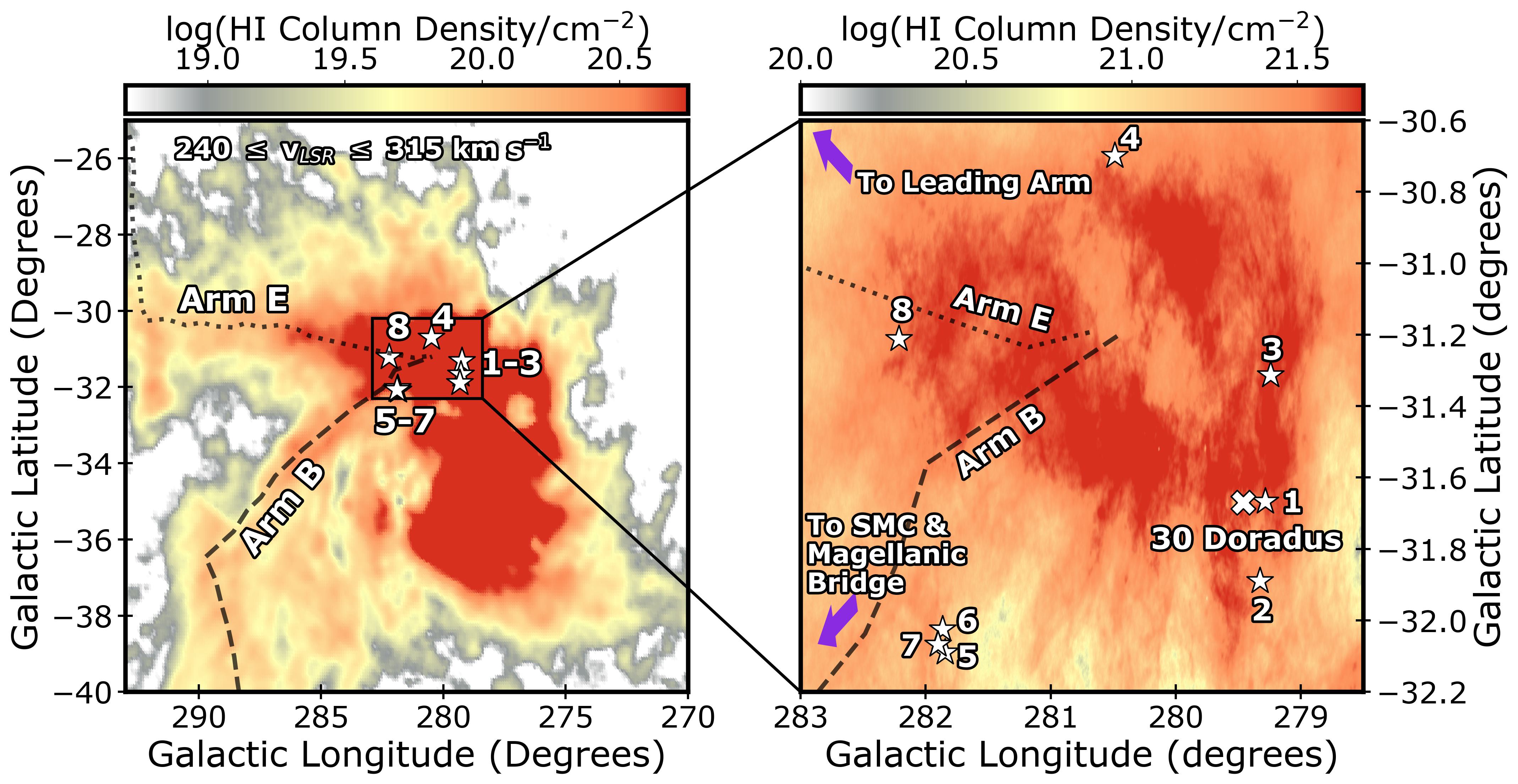}
  \caption{\textit{Left}: An integrated \hi \, emission map of the LMC from the GASS survey over the velocity range  $+240\le\  v_{\rm LSR}\le+315~\kms$. \textit{Right}: An integrated \hi \, emission map of the 30~Doradus region from the GASKAP survey over a velocity range of $+185\le\  v_{\rm LSR}\le+335~\kms$. We label the locations of the 8~sightlines in our project with a white star and the center of 30~Doradus with a white cross. The dashed and dotted black lines highlight the predicted locations of arm~B and arm~E, respectively, from the Gaussian decomposition of the LAB \hi~survey in \citet{2008ApJ...679..432N}. The numbered sightlines correspond to the IDs in Table\ref{tab:targets}.}
  \label{fig:sightlines}
\end{figure*}

One prominent nearby example of interacting galaxies is the Magellanic Clouds (MCs). The MCs are comprised of two galaxies, the Large Magellanic Cloud (LMC) and the Small Magellanic Cloud (SMC),
which have recently approached the Milky Way (MW) as a bound pair. Given that the LMC is located at a distance of $d_\odot\approx50\,\rm{kpc}$ \citep{1999ASSL..237..125W,2013IAUS..289..169P,2014AJ....147..122D}, it is a prime candidate to study in detail how galaxy interactions shape galactic evolution. Previous studies on the LMC have demonstrated that a combination of tidal and ram-pressure forces influence its gas stripping. Hydrodynamical simulations suggest that tidal interactions with both the MW and SMC significantly redistribute the gas in the LMC's disk \citep{2012MNRAS.421.2109B,2018ApJ...857..101P}, while observational studies of \hi~emission and UV absorption lines indicate that ram-pressure stripping plays a crucial role in removing material from the outskirts of the LMC as it travels through the Milky Way's halo \citep{2014ApJ...787..147F,2021ApJ...921L..36L,2022MNRAS.515..940W}. Building on these insights, this study analyzes the velocity structure and spatial distribution of two arm-like \hi~gaseous filaments flowing in and around the LMC's disk to better understand their origin.

These arm-like features are overdense \hi~structures referred to as arm~B and arm~E \citep{2003MNRAS.339...87S} (see Figure~\ref{fig:sightlines}). Arm~B and arm~E are possibly kinematically and spatially connected to the displaced gaseous Magellanic Stream (MS) and Leading Arm (LA), respectively, which extend outside of the LMC's disk \citep{1986PASA....6..471M,1998ApJ...503..674K,1998ApJ...507L..35G,2003MNRAS.339...87S,2008ApJ...679..432N}. Within the LMC, these arms can be traced to different supergiant \hi~shells \citep{2008ApJ...679..432N} near the 30~Doradus starburst region 
\text{ $(l,b)=(279\fdg47, -31\fdg67$}) that is being fueled by the NGC~2070 cluster \citep{1997ApJS..112..457W}. At the center of NGG~2070, resides the R\,136 subcluster containing a younger population of stars approximately 1--2\,Myr old \citep{2016MNRAS.458..624C,2020MNRAS.499.1918B}. NGC~2070 also houses an older generation of stars at $\sim$3--7\,Myr \citep{1996ApJ...466..254B,1997ApJS..112..457W,1999A&A...341...98S,2015ApJ...811...76C}. This hints that multiple star-formation episodes---and even ongoing star formation (e.g. \citealp{1992MNRAS.257..391H,1992A&A...261L..29R})---power the energetics of 30~Doradus. The strong stellar activity creates a turbulent environment surrounding 30~Doradus. Both inflowing and outflowing gas have been detected in this region \citep{2002ApJ...569..214H,2025arXiv250305968P} which complicates deciphering the dynamics of the arms.

The arms~B and~E trace gas close to 30~Doradus, which has fueled a debate on their origins. The first scenario is that the supergiant shells in the 30~Doradus region drive outflows (e.g. \citealp{2004A&A...423..895O}). Once the gas is dislodged from the LMC's disk and into the outskirts of the galaxy, it is more susceptible to the tidal and ram-pressure forces that are then able to funnel gas into the LA and MS \citep{2003MNRAS.339...87S,2008ApJ...679..432N}. This situation is plausible because the LMC contains high-velocity outflows, even in quiescent regions, ubiquitously across its face \citep{2007MNRAS.377..687L} and because the LMC is traveling through the MW's halo \citep{2021ApJ...921L..36L,2022Natur.609..915K,2024ApJ...976L..28M}. 
The second scenario is that these \hi~arms were created from inflowing material that could be feeding the 30~Doradus starburst region. \citet{2011ApJ...737...29O} discovered that the LMC has a population of AGB stars that are not only spatially distinct from the normal LMC AGB population in the bar, but are also either counter-rotating with the disk or co-rotating with the disk at a $54\pm2\arcdeg$ incline.  \citeauthor{2011ApJ...737...29O} suggest these uncharacteristic stars could have formed in the SMC and were tidally stripped into the LMC during a close encounter. The kinematics of these AGB stars agree with the \hi~arms \citep{2011ApJ...737...29O}, providing support for the infalling scenario.

Despite the dispute in their origin, arms~B and~E are traceable to similar supergiant shells near 30~Doradus and connect the LMC to different gaseous structures beyond the LMC's disk. Arm~B joins with the MS while arm~E links with LA~I \citep{2008ApJ...679..432N}---the LA is divided into four complexes known as LA~I, LA~II, LA~III \citep{1998Natur.394..752P,2005A&A...432...45B,2013ApJ...764...74F} and LA~IV \citep{2013ApJ...764...74F,2012A&A...547A..12V}. \citet{2008ApJ...679..432N} mention the differing paths of arms~B and~E could be related to their orientation relative to the LMC's disk. Assuming an outflow origin, \citet{2008ApJ...679..432N} postulate that the high-velocity gas in both arms would likely reside behind the LMC's disk. 
Because of the LMC's inclination relative to the MW, they reason that ejected material on the far side of the LMC's disk will experience a more direct headwind as it moves through the Milky Way's halo than the material on the nearside \citep{2008ApJ...679..432N}. However, the \citet{1994A&A...289..357D} study measured the velocity gradient and radial velocity offset of gas clouds from absorption features in the direction of compact continuum sources in and behind the LMC and concluded that the high-velocity gas likely resides on the nearside of the LMC. Given the uncertainty in the orientation of these arms, additional studies are needed to constrain their geometry.

One method for constraining the relative location of arms~B and~E along the line-of-sight relative to the LMC's disk is through an absorption-line study that uses stars embedded within the LMC's disk as background targets. Any associated absorbing material would therefore lie between the star and the observer. 
In this paper, we combine \hi~21-cm emission observations with UV absorption-line spectroscopy toward~8 O~and B~stars that are embedded in the LMC's disk and lie within $2\fdg1$ of 30~Doradus to search for signatures of these arms. Section~\ref{data} describes our \hi~21-cm emission line and UV absorption line datasets, their reduction, and analysis. We detail our procedure for identifying the kinematic bounds of the LMC's \hi~disk in Section~\ref{kinematic_bounds} and then outline the kinematic, line widths, and column density properties of the nearside gas in Section~\ref{UV_absorb}.  We explore whether outflows can create a filament on the nearside of the LMC in Section \ref{simulation} by incorporating a high-resolution simulation of the LMC and SMC as they orbit the MW. We discuss the complexity of the orientation of the Magellanic System in Section~\ref{discussion} and offer differing perspectives on the dynamics of the gas. Finally, we summarize our main findings in Section \ref{conclusion}.

\begin{figure*}[t]
  \centering

  \includegraphics[] {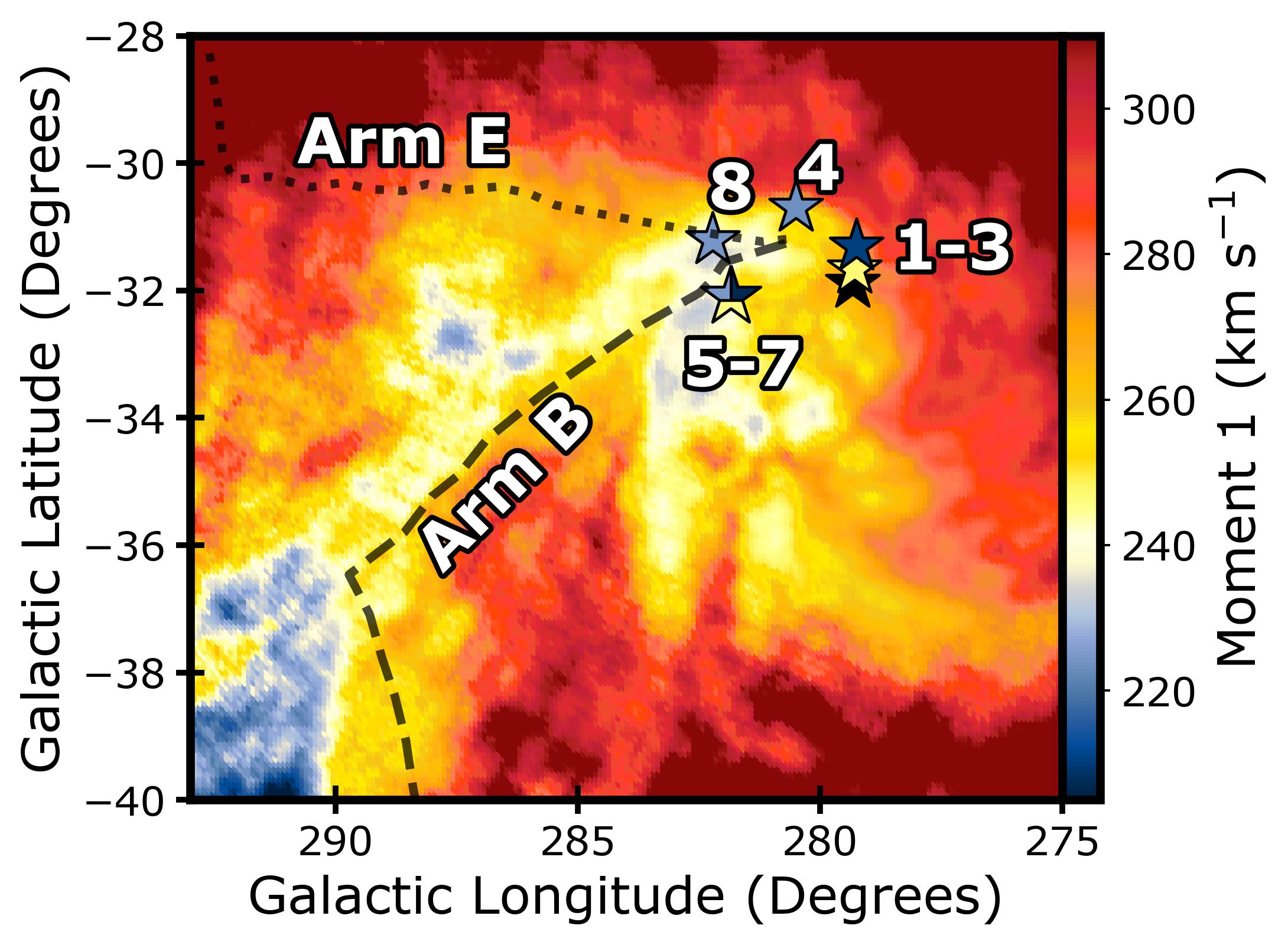}
\caption{A moment~1 (emission-weighted mean velocity; see Equation~\ref{Moment1}) map of the LMC from the GASS survey. The locations of the 8~sightlines are marked with a star symbol. Each sightline is color coded based on the star's radial velocity (see Table~\ref{tab:targets}). The star marker labeled as ``5--7" contains  three different colors that correspond to the radial velocities of sightlines~5--7. As sightline~2 does not have a measured radial velocity, we mark with a black star. The pathways of arm~B (dashed black line) and arm~E (dotted black line) from the \citet{2008ApJ...679..432N} Gaussian decomposition of the LAB survey are highlighted.}
\label{fig:hi_PV_global}
\end{figure*}

\section{Observations, Data Reduction, and Analysis}\label{data}

Considering the LMC's proximity ($d_\odot\approx50\,\rm{kpc}$; \citealp{1999ASSL..237..125W,2013IAUS..289..169P,2014AJ....147..122D}) and its nearly face-on inclination (22$\degree \lesssim i \lesssim$ 26$\degree$; \citealp{1998ApJ...503..674K, 2003MNRAS.339...87S, 2018ApJ...866...90C}), it is a prime candidate to explore the effects of galaxy interactions on galaxy evolution, specifically the formation of extended gaseous structures.  For this study, we use \hi~21-cm emission observations from both the Galactic All Sky Survey\footnote{\url{https://www.astro.uni-bonn.de/hisurvey/AllSky_gauss/}} (GASS; \citealp{2009ApJS..181..398M}, and the Galactic Australian Square Kilometre Array Pathfinder (GASKAP; \citealp{2013PASA...30....3D,2022PASA...39....5P}) survey and UV absorption-line spectroscopy from the HST's Ultraviolet Legacy Library of Young Stars as Essential Standards (ULLYSES) program to probe 8~sightlines near 30~Doradus (see Figure~\ref{fig:sightlines} and Table~\ref{tab:targets}). Below, we describe these observations and present our data processing. 

\begin{deluxetable*}{c c c c c c c c c c}
\tabletypesize{\scriptsize}
\tablecaption{Properties of the Sightlines in Our Sample\label{tab:targets}}
\tablewidth{0pt}
\tablehead{
\colhead{ID\tablenotemark{\footnotesize\rm{\color{blue}a}}}\vspace{-0.2cm} & \colhead{Background} & \colhead{Stellar Radial Velocity\tablenotemark{\footnotesize\rm{\color{blue}b}}} & \colhead{Stellar} & \colhead{\textit{l}} & \colhead{\textit{b}} &\colhead{$\Delta\theta_{\rm 30\,Dor}$\tablenotemark{\footnotesize\rm{\color{blue}c}}} & \colhead{Gratings \&}  \\
\colhead{} & \colhead{Target} & (\kms)&\colhead{Type} & \colhead{$(^\circ)$} & \colhead{$(^\circ)$} &\colhead{$(^\circ)$} &\colhead{Observing Modes} 
}
\startdata
(1) & Sk$-$68$^\circ$140  & $+246.5\pm 0.6$\tablenotemark{\scriptsize\rm{\color{blue}1}}&B0.7 Ib-Iab Nwk\tablenotemark{\scriptsize\rm{\color{blue}2}} & $279\fdg28$  &  $-31\fdg67$ & 0.15&HST/STIS E230M \& COS G130M \\
(2) & Sk$-$68$^\circ$129& -- & B1 I\tablenotemark{\scriptsize\rm{\color{blue}5}} & $279\fdg33$  & $-31\fdg89$ & 0.25&HST/STIS E230M \& COS G130M\\ 
(3) & Sk$-$68$^\circ$155  &$+210.7\pm0.9$\tablenotemark{\scriptsize\rm{\color{blue}1}}& B0.5 I\tablenotemark{\scriptsize\rm{\color{blue}6}} & $279\fdg24$ & $-31\fdg31$ & 0.40&HST/STIS E230M \& COS G130M \\
(4) & Sk$-$70$^\circ$115 &$+223.7 \pm 1.5$\tablenotemark{\scriptsize\rm{\color{blue}1}}& O6.5 If\tablenotemark{\scriptsize\rm{\color{blue}7}} & $280\fdg49$ & $-30\fdg70$ & 1.31&HST/STIS E230H E230M E140H E140M \\
(5) & BI184 &$+225.0$\tablenotemark{\scriptsize\rm{\color{blue}3}}& O8 (V)e\tablenotemark{\scriptsize\rm{\color{blue}3}} & $281\fdg84$ & $-32\fdg09$ & 2.07&HST/STIS E230M \& COS G130M + FUSE \\ 
(6) & Sk$-$71$^\circ$45  &$+207.0 \pm 1.5$\tablenotemark{\scriptsize\rm{\color{blue}1}}& O4-5 III(f)\tablenotemark{\scriptsize\rm{\color{blue}8}} & $281\fdg86$ & $-32\fdg02$ & 2.07&HST/STIS E230M E140M \\
(7) & Sk$-$71$^\circ$41  & $+245.9 \pm 0.5$\tablenotemark{\scriptsize\rm{\color{blue}1}}&O9.7 Iab\tablenotemark{\scriptsize\rm{\color{blue}3}}  & $281\fdg90$ & $-32\fdg07$ & 2.11&HST/STIS E140M + FUSE \\
(8) & Sk$-$71$^\circ$50  &$+224.5 \pm 1.1$\tablenotemark{\scriptsize\rm{\color{blue}1}}& O6.5 III\tablenotemark{\scriptsize\rm{\color{blue}9}} & $282\fdg21$ & $-31\fdg21$ & 2.39&HST/STIS E230M E140M\\ 
\enddata
\tablenotetext{\rm{\color{blue}a}}{We refer to our stellar sightlines by these ID numbers in Figure \ref{fig:sightlines} and throughout the text. }
\tablenotetext{\rm{\color{blue}b}}{The radial velocities are given in the Kinematic Local Standard of Rest (LSR) velocity frame. The global systemic velocity of the LMC's \hi\ disk is $v_{LSR}\approx+265~\kms$ and the localized systemic velocity near 30~Doradus is $v_{LSR}\approx232~\kms$.}
\tablenotetext{\rm{\color{blue}c}}{We calculated the angular offset from the center of 30~Doradus $ (l,\,b) = (279\fdg46,\,-31\fdg67)$ using an assumed distance $d_\odot=50\,\kpc$ for each of the stellar sightlines inside the LMC's disk. }
\tablenotetext{}{References: [\color{blue}1] \,\cite{2024AA...688A.104S}, [\color{blue}2]\, \cite{2015AA...574A..13E}, [\color{blue}3]\, \cite{2018AA...615A..40R}, 
 [\color{blue}4]\,\citet{2023AA...676A..85S}, 
[\color{blue}5] \,\cite{2000AJ....119.2214M},
[\color{blue}6] \,\cite{1978AAS...31..243R},
[\color{blue}7] \,\citet{2004NewAR..48..727N},
[\color{blue}8] \,\citet{1977ApJ...215...53W},
[\color{blue}9] \,\citet{1991PASP..103.1123F}}

{}

\end{deluxetable*}

\subsection{Radio Data}

GASS mapped the \hi~21-cm emission from the Southern Hemisphere at an angular resolution of 16$^{\prime}$, a velocity resolution of $1.0~\kms$, and 3$\sigma$ column density detection sensitivity of $\log{\left(N_{\rm H\textsc{~i},\,3\sigma}/\mathrm{cm}^{-2}\right)} = 18.2$ for a $30~\kms$ wide line \citep{2009ApJS..181..398M}. We used already reduced archival GASS observations that have been processed for bandpass correction and flux calibration. This reduction is detailed in \citet{2009ApJS..181..398M}. 

GASKAP is an ongoing survey that is mapping much of the Southern Hemisphere, concentrating on the Milky Way, LMC, SMC, and associated extended gaseous circumgalactic medium (CGM) structures. This \hi~21-cm survey has an angular resolution of 30$^{\prime \prime}$, velocity resolution of $0.24~\kms$, and 3$\sigma$ column density sensitivity of $\log{\left(N_{\rm H\textsc{~i},\,3\sigma}/\mathrm{cm}^{-2}\right)} = 20.0$ for a $30~\kms$ wide line \citep{2022PASA...39....5P}. The GASKAP observations are science ready and were processed with the ASKAPSoft pipeline \citep{2019ascl.soft12003G}, which applied a bandpass calibration to the data and flagged any unwanted signal. The GASKAP \hi~data were combined with observations from the 64\,m Parkes single-disk telescope \citep{2009ApJS..181..398M} to fill in short-spacings visibility and to enable a dust and molecular gas physical scale view \citep{2022PASA...39....5P}.

The GASS and GASKAP surveys both have different strengths. GASKAP's angular resolution is an order of magnitude finer than GASS's, enabling the investigation of the small-scale structure of the LMC and its gaseous arms. GASS is sensitive to fainter emission, allowing the diffuse structures to be resolved. We used GASS and GASKAP observations to estimate the kinematic bounds of the LMC's disk and the nearside gas for each sightline. However, with GASKAP's much smaller beam size that more closely resembles HST's, we used it as the primary dataset for the kinematic width determination of both features.

\subsection{UV Absorption Line Data}
Galaxies like the LMC that are inclined almost face-on work exceptionally well for ``down-the-barrel'' absorption-line studies in which background light sources embedded within galaxies are used to isolate absorbing material lying on their nearside. For this task, we utilized the 7th~data release (DR7) of the HST's ULLYSES program\footnote{\url{https://ullyses.stsci.edu/}}, which is a director's discretionary program that targets stars primarily located in the MCs to compile an ultraviolet spectroscopic library of high-quality spectra. By combining the Far Ultraviolet Spectroscopic Explorer (FUSE) with HST's Space Telescope Imaging Spectrograph (STIS) and Cosmic Origin Spectrograph (COS), the ULLYSES program spans a wide wavelength range of $937\le\lambda\le3{,}119\,\text{\AA}$. This survey provides science ready spectra with a continuum signal-to-noise ratio of $S/N > 20$ for STIS and $ S/N > 30$ for COS observations. The details of the ULLYSES reduction are outlined in \citet{2020RNAAS...4..205R}, which utilized the CalSTIS and CalCOS calibration pipelines to calibrate the spectrum and co-add all the observations. For this project, we used the spectral data products of individually co-added spectra observed using the same diffraction grating and instrument.

When selecting sightlines for our study, we had access to the target sample from the ULLYSES DR6, although we switched our analysis to DR7 once it became available. We imposed five conditions to determine our sightline sample size. First, we prioritized targets with HST/STIS observations over those that only have HST/COS or FUSE observations. The STIS spectra have a superior velocity resolution, which aids in characterizing the multi-component structure of the gas. 
With the COS observations, it is difficult  to kinematically disentangle the LMC's disk from the offset gas.
For example, 
the STIS observations have a velocity resolution of $\Delta v = 2.5~\kms$ for the high-resolution gratings (E140H and E230H) and $\Delta v=6.5~\kms$ and $\Delta v=10~\kms$ for the medium-resolution gratings, E140M and E230M, respectively. The COS and FUSE observations have velocity resolutions of $\Delta v\approx 17~\kms$ \citep{2012ApJ...744...60G, 2014cosi.book....6H} and  $\Delta v\approx 20~\kms$
\citep{2000ApJ...538L...7S}, respectively. We did, however, include sightlines with both HST/STIS E230M and HST/COS G130M coverage and sightlines with additional FUSE observations given the limited availability of targets with only STIS data in this region. We list the instruments and gratings used to observe each sightline in Table~\ref{tab:targets}. Second, we only considered sightlines in close proximity to either arm. Arm~B is spatially closer to 30~Doradus and has more sightlines potentially along its predicted path found from \citet{2008ApJ...679..432N}. Therefore, we limit our selection criteria to sightlines within ${\sim}1\fdg0$ of the predicted path of arm~B. However, arm~E only has 2~sightlines within ${\sim}0\fdg5$ degrees of its path. We extend our search window to ${\sim}1\fdg5$ degrees for arm~E. Third, we selected sightlines near 30~Doradus to explore the arms' extent and their possible connection to the starburst region. However, 30~Doradus is an active starburst region and has evidence of inflows from its stellar activity \citep{2025arXiv250305968P}. We survey near the \citet{2025arXiv250305968P} sightlines in order to discuss a possible origin of the arms in context with their findings. Fourth, we required that spectra contain redshifted material beyond the LMC's disk in the ULLYSES spectra that are kinematically near the arms (see Section~\ref{pv_plots_section} for discussion of disk boundary determination). Fifth, the emission along the sightlines exhibit filamentary, redshifted \hi~emission in their position-velocity map (see Section~\ref{pv_plots_section} for position-velocity map discussion) to strengthen the argument that we are detecting arm material in both the UV and \hi~emission. From the DR6 sample, these criteria narrowed our study to 8~sightlines (see Figure~\ref{fig:sightlines}).

\subsection{Normalization of UV Spectra}\label{norm_UV_absorb}
While the spectra from the ULLYSES program are science-ready, the flux is not normalized with respect to the continuum from the background target. For our normalization process, we split the UV spectra into smaller spectral segments that span $-1{,}500 \leq v_{LSR} \leq +1{,}500 ~\kms$ and are centered on the line transitions of interest in the rest frame velocity at a redshift $z=0$.  We truncated or shifted this velocity span for ion transitions with a complex background continuum to exclude problematic regions. Then, we selected all of the wavelength regions of the stellar continuum with no absorption features from the background star, which are usually much wider than the interstellar features, and without absorption features associated with intervening gas. We fit a best-fit polynomial to this selected continuum, typically spanning orders of $2 \leq n \leq 7$. For this task, we created an automatic polynomial fitter that selects the lowest-order polynomial, resulting in the smallest change in the reduced Chi-squared value between two consecutive orders. Additionally, this fitter measures the standard deviation of the continuum for each order and prioritizes the order with the smaller standard deviation value to model the continuum. 
\begin{figure*}[hbtp]
  \centering
  \includegraphics[height=0.32\textwidth]{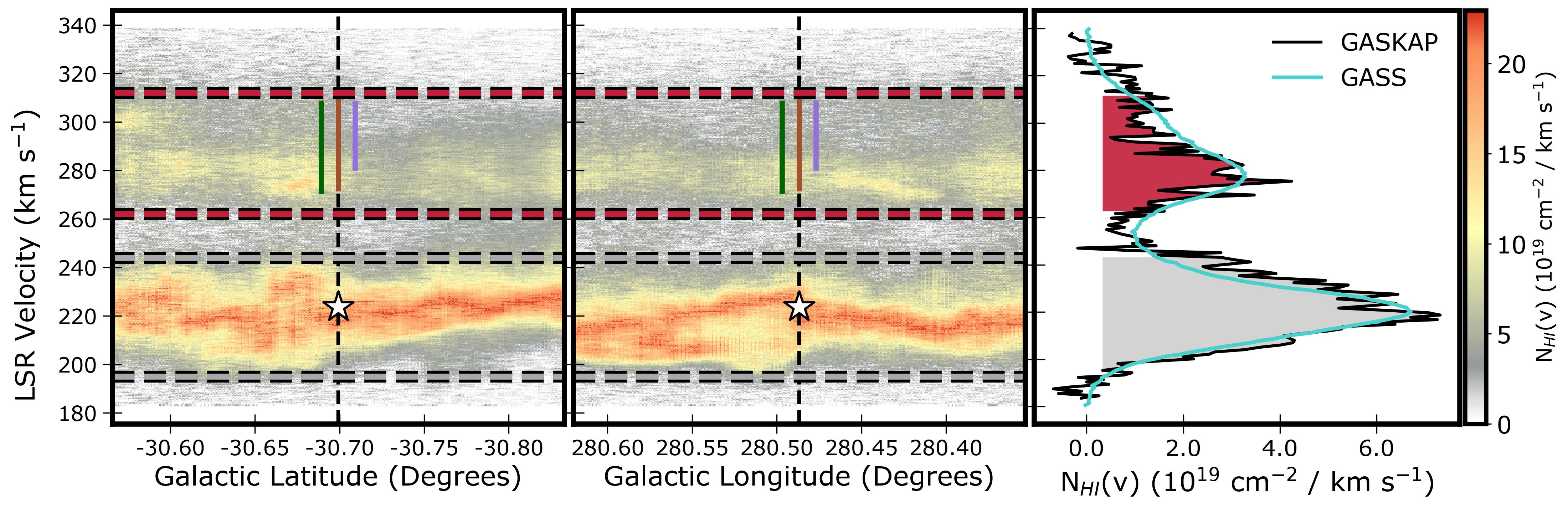}
  \caption{A position-velocity map for sightline 4 (Sk$-$70$^\circ$115) using \hi~21-cm emission data from GASKAP. The left and middle panels are sliced along a $0\fdg267$ region through the sightline's Galactic latitude and longitude, respectively. The LMC's disk is prominent between $ +190 \lesssim v_{\rm LSR} \lesssim +245\, \rm{km\, s^{-1}}$ (highlighted in gray in the right panel) while the offset material exterior to the disk is located around $ +260 \lesssim v_{\rm LSR} \lesssim +310\, \rm{km\, s^{-1}}$ (highlighted in red in the right panel). We caution that the separation between the two \hi~emission bands for sightline~4 is not the standard among the remaining sightlines. However, we choose to highlight sightline~4 as it is the strongest argument for arm material flowing in this region. We note that the offset gas's interior edge (side kinematically closer to the LMC's disk) has a more well-defined boundary than its exterior side which may suggest the material on leading edge experiences more compression than the trailing gas. The black dashed vertical line is the Galactic latitude or longitude of sightline~4. The vertical green, brown, and purple lines are the velocity ranges from the Voigt fitting analysis of the Fe\textsc{~ii}, Si\textsc{~ii}, and S\textsc{~ii} components, respectively (see Table~\ref{tab:Voigt_results_tab}). We intentionally offset the Galactic longitude and latitude of the Fe\textsc{~ii} and S\textsc{~ii} components by $+0\fdg01$ and $-0\fdg01$, respectively, for visual clarity. We include the radial velocity of the background star on the left and middle plots as indicated by the star symbol. The right panel has \hi~21-cm line emission expressed in terms of column density as a function of velocity from the GASKAP (black) and GASS (cyan) observations. The GASKAP data is rebinned by 1~\kms. The peaks in the emission-line profile align with the two bands of \hi \, emission in the position-velocity maps.}
  \label{fig:pv}
\end{figure*}

\subsection{Conversion to Local Standard of Rest Velocity Frame}\label{VLSR}

In order to aid in distinguishing the absorbers that belong to the MW and LMC, we convert the ULLYSES spectra from the heliocentric frame into the kinematic Local Standard of Rest (LSR) velocity frame. This calculation involves correcting for the motion of the Sun with respect to an idealized orbit about the Galactic center. In the equatorial direction (RA, DEC) = (18$h$, +30$\degree$), the Sun's motion is $\Delta v_{\rm{LSR, offset}}\approx 20$\kms. This conversion aligns the main component of the MW around 0\kms. In this velocity frame, the global systemic velocity of the LMC from its \hi\ emission is $v_{\rm{LSR}}\approx+265\,\kms$ at the center of the galaxy \citep{1992A&A...263...41L,1998ApJ...503..674K}. However, in the localized region near 30~Doradus, where we are investigating, the systemic velocity of the \hi\ gas is closer to $\vlsr\approx{\rm +232}\,\kms$.

\section{Kinematic Boundaries}\label{kinematic_bounds}
Before searching for absorption signatures of arms~B and~E in the UV spectra, we used the \hi~emission to characterize their spatial and kinematical distribution. The LMC's disk and arms~B and~E should be kinematically close together since the arms are traceable to \hi~supershells near 30~Doradus. \citet{2008ApJ...679..432N} tracked these gaseous arms by decomposing the \hi~emission-line observations using Gaussian fits; for this task, they used  Leiden/Argentine/Bonn (LAB) survey observations with a $0.6\arcdeg$ angular resolution. Combining the GASS and GASKAP \hi~observations with angular resolutions of 16$^{\prime}$ and 30$^{\prime \prime}$, respectively, we can better resolve and separate the LMC's disk and the gaseous arms. The \hi~emission observed with the LAB beam is diluted and spatially averaged over an area that is nearly 5{,}200 times larger than covered by the GASKAP beam (i.e., $\Omega_{\rm LAB}/\Omega_{\rm GASKAP}=\left(0.6\arcdeg/0.0083\arcdeg\right)^2$).  Below, we describe our approach to constrain the kinematic locations using GASS and GASKAP.

\subsection{\hi~Gas Kinematics and Small-Scale Structures}\label{pv_plots_section}
\
We used the GASKAP data to explore the kinematic range of both arms flowing through the LMC with a moment~1 (emission-weighted average velocity) map (see Figure \ref{fig:hi_PV_global}). We calculate the moment~1 values using the relationship:
\begin{equation}\label{Moment1}
       \frac{M_{1}}{\rm km^{-1} \s}= \frac{\sum v T_B(v)}{\sum T_B(v)},
\end{equation}
where $v$ is the velocity in~\kms~and $T_{B}(v)$ is the brightness temperature as a function of velocity. We find that arm~B is kinematically between $+235 \leq v_{LSR} \leq +285$\kms~while arm~E is at slightly higher velocities of $+240 \leq v_{LSR} \leq +300$\kms. These moment~1 values align very similarly to the predicted velocity ranges from \citet{2003ApJS..148..473K} and \citet{2003MNRAS.339...87S}. Therefore, along our 8~sightlines, we expect to observe \hi~emission associated with the arms around $+235 \lesssim v_{\rm LSR} \lesssim +300 ~\kms$.

To investigate small-scale variation, we created position-velocity maps centered at each sightline that slice 0\fdg267 (angular resolution of GASS) through the \hi~emission in that region for both Galactic longitude and latitude space (see Figures~\ref{fig:pv} and~\ref{fig:appendix_pv_plots}). These maps aid in separating \hi~emission that is associated with the disk with any material external to it. For sightlines~1 and~2 (Sk$-$68$^\circ$140 and Sk$-$68$^\circ$129), we expanded the search window to 0\fdg5 as little exterior \hi~emission was observed in the 0\fdg267 slice.

Sightline~4 (Sk$-$70$^\circ$115) is the only target with two \hi~emission regions; the LMC's disk and offset gas that are entirely separated kinematically in both the Galactic longitude and latitude position-velocity slices (see Figure~\ref{fig:pv}). Because of this distinction, we can implement a method of more tightly constraining the kinematic width of the LMC's disk and any exterior gas using the emission-weighted line centers (i.e., moment~1 values) and the emission weighted line widths (i.e., moment~2 values). We determined the moment~2 values using the relationship:
\begin{equation}\label{Moment2}
   \frac{M_{2}}{\rm km^{-1} \s} = \sqrt{\frac{\sum (v-M_{1})^2 T_B(v)}{\sum T_B(v)}}
\end{equation}

We find the moment~1 (see Equation~\ref{Moment1}) and moment~2 values for every spectral line in sightline~4's position-velocity map for both the LMC's disk and the high-velocity gas. We then determine a best-fit polynomial to all of the disk's moment~1 and moment~2 values, as well as, a separate polynomial to all of the moment~1 and moment~2 values for the exterior high-velocity gas. We then apply the results from these best fit polynomials to  estimate typical velocity and width of the \hi\ emission at the Galactic longitude and latitude coordinates of sightline~4. We report our determined widths for both the LMC's disk and offset gas from the position-velocity map of sightline~4 in Table~\ref{tab:boundaries_hi}.

\subsection{GASS Gaussian Decomposition}\label{Gass_Gauss_Decomp}
Sightline~4 is the only one where we can calculate moment~1 and moment~2 values to determine the kinematic width of the LMC's disk and the redshifted gas; the position-velocity maps of the 7 remaining sightlines exhibit kinematic blending between the \hi~emission of both structures (see Figure~\ref{fig:appendix_pv_plots}). In this situation, we cannot calculate moment~1 and moment~2 values of the separate features without prior knowledge of their kinematic extents. Therefore, we performed a Gaussian decomposition on the GASS observations for each sightline to separate the \hi~emission into individual Gaussian components for the disk and offset gas. We utilized the \textsc{GaussDecomp} Python package \citep{Gauss_decomp_citation}  which is based on the IDL software developed by \citet{2008ApJ...679..432N}. For more details on the automated Gaussian fitting algorithm, we refer the readers to \citet{2008ApJ...679..432N} and \citet{2000A&A...364...83H}. Using the center velocities as an estimate of the kinematic location of each high-velocity gas component, the Full Width at Half Maximum (FWHM) of the emission line provided an approximate boundary for the extent of the feature. We adopted these Gaussian decomposition results as a first approach estimate. 

With the center velocities and FWHM, we visually inspected each sightline's position-velocity map. We compared the first guess from the Gaussian decomposition with the small-scale \hi~variation displayed in the GASKAP data. The boundary values for both the LMC's disk and the high-velocity gas were adjusted (${\sim}$5--10~\kms\,beyond the upper FWHM limit) if needed to enclose the gaseous material of both structures accurately. Our \hi~emission boundary constraints for both the LMC's disk and the offset material from the Gaussian Decomposition technique are reported in Table~\ref{tab:boundaries_hi}.

\begin{table*}[hbtp]
\centering
\caption{\hi~Kinematic bounds of the LMC's Disk and Offset Gas}
\label{tab:boundaries_hi}
\begin{tabular}{ccccccc}
\hline
\hline 
ID & Sightline& \multicolumn{2}{c}{LMC Disk} & \colhead{} & \multicolumn{2}{c}{Offset Gas} \\
\cline{3-4}\cline{6-7}
 & & Left Boundary  & Right Boundary & & Left Boundary  & Right Boundary\\ 
   &       & ($v_{\mathrm{LSR}}$/km s$^{-1}$) & $(v_{\mathrm{LSR}}$/km s$^{-1}$) & & ($v_{\mathrm{LSR}}$/km s$^{-1}$) & $(v_{\mathrm{LSR}}$/km s$^{-1}$)\\
\hline
(1) & Sk$-$68$^\circ$140 & +230 & +280 && +280 & +300 \\
(2) & Sk$-$68$^\circ$129 & +235 & +269 && +269 & +285 \\
(3) & Sk$-$68$^\circ$155& +228 & +269 && +269 & +309 \\
(4) & Sk$-$70$^\circ$115\tablenotemark{\scriptsize\rm{\color{blue}1}}  & +195 & +244 && +262 & +312 \\
(5) & BI184& +200 & +239 && +239 & +285 \\
(6) & Sk$-$71$^\circ$45 & +214 & +242 && +242 & +280 \\
(7) & Sk$-$71$^\circ$41& +192 & +242 &&  +242 & +295 \\
(8) & Sk$-$71$^\circ$50 & +200 & +238 &&  +238 & +280 \\ \hline
\end{tabular}
\tablenotetext{\rm{\color{blue}1}}{The bounds of the disk and offset gas were determined through moment~1 and moment~2 analysis. }
\tablenotetext{}{We estimate the localized systemic velocity of the LMC's disk in this region to be $v_{LSR}\approx232~\kms$. To determine this, we find the average of the median velocity between the left and right boundary of the LMC's disk for each sightline.}
\end{table*}

\section{Characterization of High-Velocity Absorbers}\label{UV_absorb}
We explore the UV spectra for absorption signatures of the high-velocity gas relative to the LMC's disk that are roughly within the kinematic boundaries that we estimated from the \hi~emission in Section~\ref{kinematic_bounds}. We examine neutral, weakly\nobreakdash-, and moderately-ionized species covered by the \textit{HST}/STIS and COS observations that we traced through the following line transitions: $\rm{P\textsc{~ii}}\, \lambda 1152$, $\rm{S\textsc{~ii}}\,\lambda \lambda 1250,1253$, $\rm{C\textsc{~i}}\,\lambda\lambda 1277,1280,1328$, $\rm{Ni\textsc{~ii}}\, \lambda \lambda 1317,1370$, $\rm{C\textsc{~ii}^{*}}\,\lambda 1335$, $\rm{S\textsc{~i}}\, \lambda 1401$, $\rm{Al\textsc{~ii}}\, \lambda 1670$, $\rm{Si\textsc{~ii}\, \lambda 1808}$, $\rm{Al\textsc{~iii}}\, \lambda\lambda 1854,1862$, and $\rm{Fe\textsc{~ii}}$ $\lambda\lambda2249,2260$. We also incorporated the $\rm{Fe\textsc{~ii}}\, \lambda 1143$ transition for sightline~7  (Sk$-$71$^\circ$41), which is covered by the FUSE observations and is not covered in the E230M HST/STIS grating. We excluded $\rm{O\textsc{~i}}\, \lambda 1302$ from our analysis as the disk components are significantly saturated---the background light from our targets are completely absorbed---along all 8~sightlines making it unfeasible to distinguish the signatures of the offset absorbers. We elected to use the higher oscillator strength $\rm{Si\textsc{~ii}}\, \lambda 1260$ line over the $\rm{S\textsc{~ii}}\, \lambda 1259$ as the redshifted absorbers relative to the LMC's disk were too weak to detect in this latter line. We stress that in many cases the components of the ion species Si\textsc{~ii}, Al\textsc{~ii}, S\textsc{~ii}, and C\textsc{~ii*} are saturated (see Tables~\ref{tab:Voigt_results_tab} and~\ref{tab:Voigt_results_app}), which restricts our column density measurements to lower limits. While we considered investigating the chemical abundances and ion ratios along our 8~sightlines, the saturation proved difficult to provide any constraints to the properties of the offset absorbers. However, we explore their characteristics through kinematics and line widths below.

Below, we describe our techniques to characterize the physical properties of this redshifted, high-velocity gas.

\subsection{Techniques for Analyzing Absorber Properties}
Our study implements two different methods to measure the physical properties of the absorbers. The first is Voigt line fitting, which we used for unsaturated or weakly saturated absorption features. The second is the apparent optical depth (AOD) method, which we used in the case of strongly saturated absorbers. The results of the Voigt profile fitting and AOD method are provided in Tables~\ref{tab:Voigt_results_tab} and~\ref{tab:Voigt_results_app}.

\subsubsection{Voigt Profile Fitting}\label{voigt_fit_sec}

We fit Voigt line profiles to continuum normalized UV absorption lines using the \textsc{VoigtFit} \citep{2018arXiv180301187K} Python software. This line fitter determines a best-fit Voigt profile based on a non-linear least-squares minimization and employs user-supplied initial guesses of the absorption features' velocity centroids ($v$), Doppler parameters ($b$), and column densities ($N$). At first, we allowed all parameters to vary freely. If it could not converge on a reasonable total fit or if the absorption features were saturated, we either fixed the velocity centroids or the widths of the components. We set these fixed values to ones we measured for a different transition for the same or similar ionization species. 

We fit multiple transitions simultaneously for the same ionization species whenever possible as their resultant column densities, Doppler parameters, and centroids should all agree; however, there were cases where we found discrepancies in these parameters when fitting these transitions individually, which could have been associated with the noisy background continuum, unresolved saturation, or interloper absorption from a different line transition at a neighboring wavelength. For unresolved saturation, we used the best-fit Doppler parameters and velocity centroids for a component from a lower oscillator strength transition when fitting the higher oscillator strength transition. For the interloper case, we held the number of components consistent for the absorption associated with the same ionization species. If a best-fit Voigt profile could not be obtained, we exclusively used the transitions that provided the most constraint of the fit. 

We accounted for the broadening of the absorption lines associated with the instrument and grating using line spread functions (LSF), which we passed to the \textsc{VoigtFit} line fitter. For the STIS data, we assumed the LSF was Gaussian in shape with a FWHM value equal to the spectral resolution of the instrument configuration. \citet{2020ApJ...897...23F} demonstrated for high S/N and medium spectral resolution COS observations that the Gaussian LSF approximation does not significantly change the component parameters when compared to calculated parameters from the tabulated LSF.  \citet{2025arXiv250305968P} also tested this for the C\textsc{~iv} and Si\textsc{~iv} ions that utilize the tabulated LSFs for the STIS E140M grating and reported that the differences in the column density values were less than $0.1\, \rm dex$. The Doppler parameters determined from the two approximations were also found to be consistent within $<0.5\, \sigma$. The majority of our sightlines utilized both the STIS E140M and E230M gratings that have instrument profiles with $FWHM=6.5$ and $10.0~\kms$, respectively \citep{2023stii.book...22M}. Sightline~4 was observed using both the E230H and E140H gratings and we adopted a $FWHM=2.5~\kms$ for the width of the instrument profile \citep{2023stii.book...22M} for ion transitions covered by these gratings.

For each COS observation, we incorporated the line spread functions provided by Space Telescope Science Institute\footnote{\url{https://www.stsci.edu/hst/instrumentation/cos/performance/spectral-resolution}} with the appropriate central wavelength and lifetime position for each sightline. While \citet{2020ApJ...897...23F} noted no significant difference between a Gaussian and tabulated LSF for COS observations at lifetime position (LP)1, our COS observations include LP1, LP2, and LP3. The LP3 and LP2 spectra have lower velocity resolution than LP~1. For consistency, we utilized the tabulated LSFs for each LP across our COS sightlines.  Furthermore, we implemented a $FWHM=20~\kms$ Gaussian-shaped LSF for the FUSE observations as recommended in \citet{fuse_book_data}. 

From the total resultant fit, we measure the velocity centroids, Doppler parameters, and column densities of the absorbers. We convert the integrated normalized flux of the absorbers to column density using the rest frame wavelength and oscillator strength (Vienna Atomic Line Data Base: \citealt{1995A&AS..112..525P} and \citealt{2017ApJS..230....8C}) of the individual ion transition.

\begin{figure}[hbtp]
  \centering
  \includegraphics[width=0.35\textwidth]{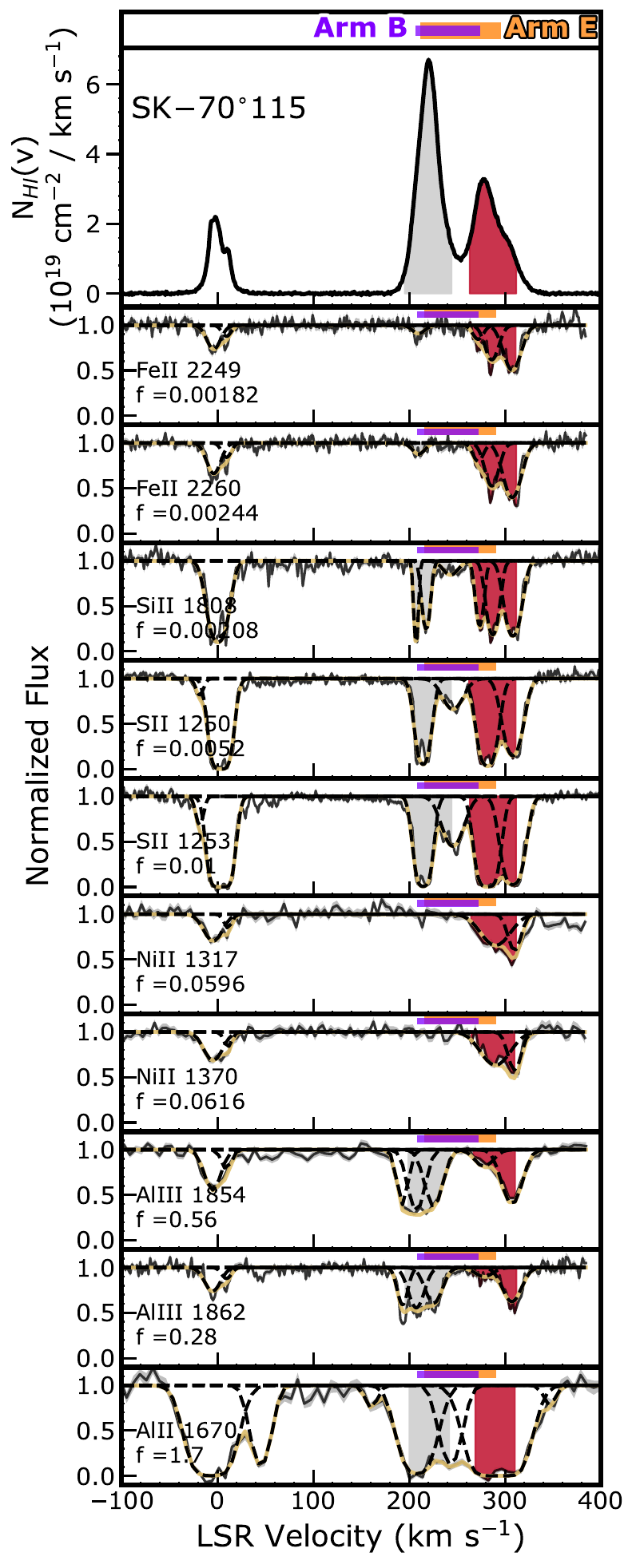}
  \caption{ A plotstack of various ion transitions for sightline~4. \textit{Top Panel}: GASS \hi~emission in terms of column density as a function of velocity. \textit{Lower Panels}: Low and intermediate ion transitions.  The Milky Way is located around $0~\kms$. The gray shaded region is the kinematic boundary of the LMC's disk and in red is the kinematic extent of the offset gas as determined from the \hi~emission (see Table~\ref{tab:boundaries_hi}). Both shaded regions were determined through assessing the reliability of the Gaussian fits from the Gaussian decomposition technique on the \hi~emission. We also label the predicted velocity range of arm~B and arm~E from the \hi~channel maps of \citet{2003ApJS..148..473K} and \citet{2003MNRAS.339...87S} by the purple and orange rectangles above the flux and \hi~emission, respectively. Each panel is labeled with the ion transition, wavelength in angstroms, and oscillator strength. The flux of the spectra is normalized and has a 1$\sigma$ error envelope. The light tan line outlines the total Voigt profile while the individual components are outlined by the black dashed lines. }

  \label{fig:plostacks}
\end{figure}

\subsubsection{Apparent Optical Depth (AOD)}\label{AOD_sec}
The Voigt fitting method does not determine robust column density measurements for saturated absorbers as their features are neither well-defined nor resolved. To determine a lower limit on the column density for saturated absorbers, we follow the AOD method outlined in \citet{1991ApJ...379..245S}. This procedure transforms the absorption-line spectrum into an apparent optical depth profile ($N_a(v)$) as a function of the line-of-sight velocity in~\kms. The conversion is given by:
\begin{equation}
 \frac{N_{a}}{\cm^{-2}}= 3.768 \times 10^{14} \frac{1}{\lambda_{0} f} \int_{v_{min}}^{v_{max}} \text{ln}\biggl (\frac{F_{c}(v)}{F_{obs}(v)} \biggr ) dv,
\end{equation}
where \textit{f} is the oscillator strength of the ion transition  \citep{1995A&AS..112..525P, 2017ApJS..230....8C}, $\lambda_0$ is the rest wavelength of the transition in angstroms, ${F}_{obs}(v)$ is the observed flux, and ${F}_{c}(v)$ is the continuum flux \citep{1991ApJ...379..245S}.  

The limits of integration for the apparent column density are the velocity bounds of the absorption feature. When the features are extremely saturated or blended, we define the velocity bounds from an unsaturated transition of the same species with a smaller oscillator strength or a species with a similar ionization potential. To estimate the lower limit of the column density for a saturated or blended feature, we follow the procedure outlined in \citet{2025arXiv250305968P} and assume the absorption-line profile is symmetric about its central velocity. For our Voigt-profile fitting of these lines, we fix the velocity centroid (\textit{v}) and Doppler parameter (\textit{b}) of a similar unsaturated transition and integrate between $v_{central}-\frac{b}{2}\leq v \leq v_{central} + 2b$.
For blending, we truncated the lower velocity limit to avoid over including LMC disk material to estimate a lower limit of the total apparent column density. About 25$\%$ of the area of the blended absorber falls between $v_{center}-0.71\sigma$. We set the upper integration limit to $2b$ because $2b \approx 3 \sigma$ ($b=\sqrt{2}~\sigma$) essentially encompasses the total area of the absorption feature on the non-blended side. In cases with blended neighboring absorbers, velocity boundaries were determined based on the clustering of absorption features to ensure that individual components were physically real. We tested this by varying the boundary limits and then evaluating the stability of the fits, similar to the approaches used in previous studies of the Magellanic System (e.g. \citealp{2018ApJ...865..145R}).

\subsection{Characterizing the Kinematics}\label{kinematic_distribution}
\begin{deluxetable*}{ccccc}[hbtp]
\centering
\caption{Summary of Line-Fitting Results.} 
\label{tab:Voigt_results_tab}
\tablewidth{0pt}
\tablehead{
\colhead{Ion} \vspace{-0.2cm} & \colhead{$v_{\rm LSR}$} & \colhead{$\log{\left(N_x/\cm^{-2}\right)}$} & \colhead{$b$} & \colhead{Notes}\\
\colhead{} & \colhead{(\kms)} & \colhead{(dex)} & \colhead{(\kms)} & \colhead{}\\
\hline \hline
\colhead{}& \colhead{}& \colhead{\textbf{Sk$-$70$^\circ$115}} }
\startdata
H\textsc{~i} &+220.8 $\pm$ 0.2  &21.34 $\pm$ 0.01&19.1 $\pm$ 0.3&Disk Component\\
&+280.8 $\pm$ 0.6 &21.04 $\pm$ 0.02&23.3 $\pm$ 0.9&Offset Component\\
S\textsc{~i}& $+265.4 \pm 2.4$ &  $14.08 \pm 0.05$ & $15.0 \pm 2.1$ & Offset Component \\
C\textsc{~i}& $ +205.4^{a} \pm 0.0$& $>$13.47 & $2.2 \pm 0.1$&Disk Component \\
&  $+266.1 \pm 0.3$ &  $13.63 \pm 0.01$ & $6.2 \pm 0.2$ &Offset Component \\
&  $+284.2 \pm 0.3$&  $12.91 \pm 0.04$ & $2.1 \pm 0.4$ &Offset Component \\
P\textsc{~ii} 
&  $+182.7 \pm 0.9$ &  $12.72 \pm 0.11$ & $ 2.9 \pm 2.2$& Disk Component \\
&  $+193.4 \pm 3.0$ & $13.31 \pm 0.05$ & $27.5 \pm 3.9$& Disk Component \\
&  $+259.1 \pm 3.0$& $13.48 \pm 0.08$ & $21.2 \pm 4.1$& Offset Component \\
&  $+282.0 \pm 1.2$ & $12.97 \pm 0.19$ &  $ 7.2 \pm 2.6$& Offset Component \\
& $+294.9 \pm 0.0$ & $12.26 \pm 0.27$ &  $ 5.1 \pm 0.0$&Offset Component \\   
S\textsc{~ii}   &  $+214.3^{a}\pm 0.0$ &  $>$15.34 & $ 7.7 \pm0.1$&Disk Component\\
&  $+245.3\pm 0.3$ & $ 14.78 \pm 0.01$ & $13.1 \pm 0.5$&Disk/Offset Component\\
&  $+281.3^{a}\pm 0.0$  & $>$15.47 & $ 10.0 \pm 0.2$&Offset Component \\
                       &  $+307.8^{a} \pm 0.0$ & $>$ 15.22  & $10.9 \pm 0.2$&Offset Component$^{*}$\\
Ni\textsc{~ii}&  $+289.7 \pm 0.0$ & $13.86 \pm 0.03$ & $19.6 \pm 1.6$ & Offset Component\\
&  $+309.6 \pm 0.0$ & $13.59 \pm 0.04$ &  $7.9 \pm 1.0$ &Offset Component\\
C\textsc{~ii}$^{*}$& $ +212.9^{a} \pm 0.0 $& $>$14.71 & $12.9 \pm 0.4$ &Disk Component\\
&  $+249.3 \pm 0.0$ &  $13.20 \pm 0.04$ &   $11.6 \pm 0.0$&Disk/Offset Component \\
&  $+272.3^{a} \pm 0.3$  & $>$ 13.66 &  $6.3 \pm 0.3$&Offset Component \\
                       &  $+284.8^{a}\pm0.3 $  & $>$13.42 &  $3.0 \pm 0.3$&Offset Component \\
                       &  $+302.8 \pm 1.5$&  $13.40 \pm 0.04$ &  $16.8 \pm 2.1$&Offset Component \\
Si\textsc{~ii} &  $+207.0^{a} \pm0.0$ & $>$15.01& $2.2 \pm 0.2$&Disk Component\\
&  $+217.4^{a} \pm0.3$& $>$15.02 & $5.0 \pm 0.3$&Disk Component \\
&  $+242.9 \pm 1.5$&  $14.51 \pm 0.07$ &  $10.6 \pm 0.0$&Disk Component\\
&  $+273.3^{a} \pm 0.6$ &  $>$15.12 & $ 5.0 \pm 0.5$ & Offset Component \\
&  $+286.9^{a} \pm 0.3$&  $>$ 15.30 &  $7.4 \pm 0.8$ & Offset Component\\
&  $+308.0^{a} \pm 0.3$& $>$ 15.33 &  $9.7 \pm 0.5$ & Offset Component$^{*}$\\
Fe\textsc{~ii} &  $+209.6 \pm 0.6$& $14.16 \pm 0.06$ &   $6.7 \pm 1.1$& Disk Component \\
&  $+271.6 \pm0.6$  & $14.34 \pm 0.07$ &  $5.0 \pm 0.8$& Offset Component \\
&  $+285.9 \pm0.3$ &  $14.90 \pm 0.03$ &  $7.5 \pm 0.6$ &Offset Component\\
&  $+307.2 \pm 0.3 $  & $15.13 \pm 0.01$ & $10.2 \pm 0.4$&Offset Component \\
Al\textsc{~ii}&  $+207.9^{a}\pm 3.0$ & $>$ 13.23 &  $17.4 \pm 2.9$&Disk Component \\
&  $+242.9^{a}\pm 3.0$ & $>$ 12.87 & $11.7 \pm 6.3$& Disk Component\\
&  $+291.8^{a} \pm 2.1$& $>$ 13.40 &   $23.0 \pm 0.0$&Offset Component$^{*}$ \\
&  $+346.2\pm3.9$  &  $11.69 \pm 0.2$ &    $6.0 \pm 0.0$ &Offset Component$^{*}$\\
Al\textsc{~iii} 
&  $+192.9  \pm 0.3$ & $12.75 \pm 0.05$ &  $4.3 \pm 0.4$& Disk Component\\
&  $+207.2  \pm 0.9$ & $12.92 \pm 0.09$ &  $9.0 \pm 1.8$& Disk Component\\
&  $+225.3  \pm 1.5$ & $12.88 \pm 0.08$ & $10.1 \pm 1.4$ & Disk Component\\
&  $+277.8 \pm 1.8$ & $12.24 \pm 0.08$ & $11.4 \pm 2.7$&Offset Component \\
&  $+306.8 \pm0.0$ & $12.92 \pm 0.02$ & $11.6 \pm 0.7$ &Offset Component$^{*}$\\             \enddata
\tablenotetext{a}{These components are saturated or could have unresolved saturation. The reported column densities are measured through the AOD method (See Section \ref{AOD_sec}) and are given as lower limits.}
\tablecomments{We present the Gaussian decomposition results of the GASKAP observations and the Voigt profile-fitting results for sightline~4. This target is observed through the E230H and E140H gratings of the STIS instrument (see Table~\ref{tab:targets}). We include components determined to be LMC disk material (``Disk Component") and note features that have blending issues between the disk and the offset features with ``Disk/Offset Component.'' For UV features significantly beyond the \hi~boundary of the offset gas, we classify it as a possible offset feature or an inflow from a galactic fountain and label it as ``Offset Component$^{*}$." In situations where we fixed the velocity center or Doppler parameter, the uncertainty is labeled with $\pm \,0.0~\kms$. The Voigt profile-fitting results for the remaining 7~sightlines are in Table~\ref{tab:Voigt_results_app} in the appendix. }
\end{deluxetable*}

\begin{figure*}[hbtp!]
  \centering
  \includegraphics[width=1\textwidth]{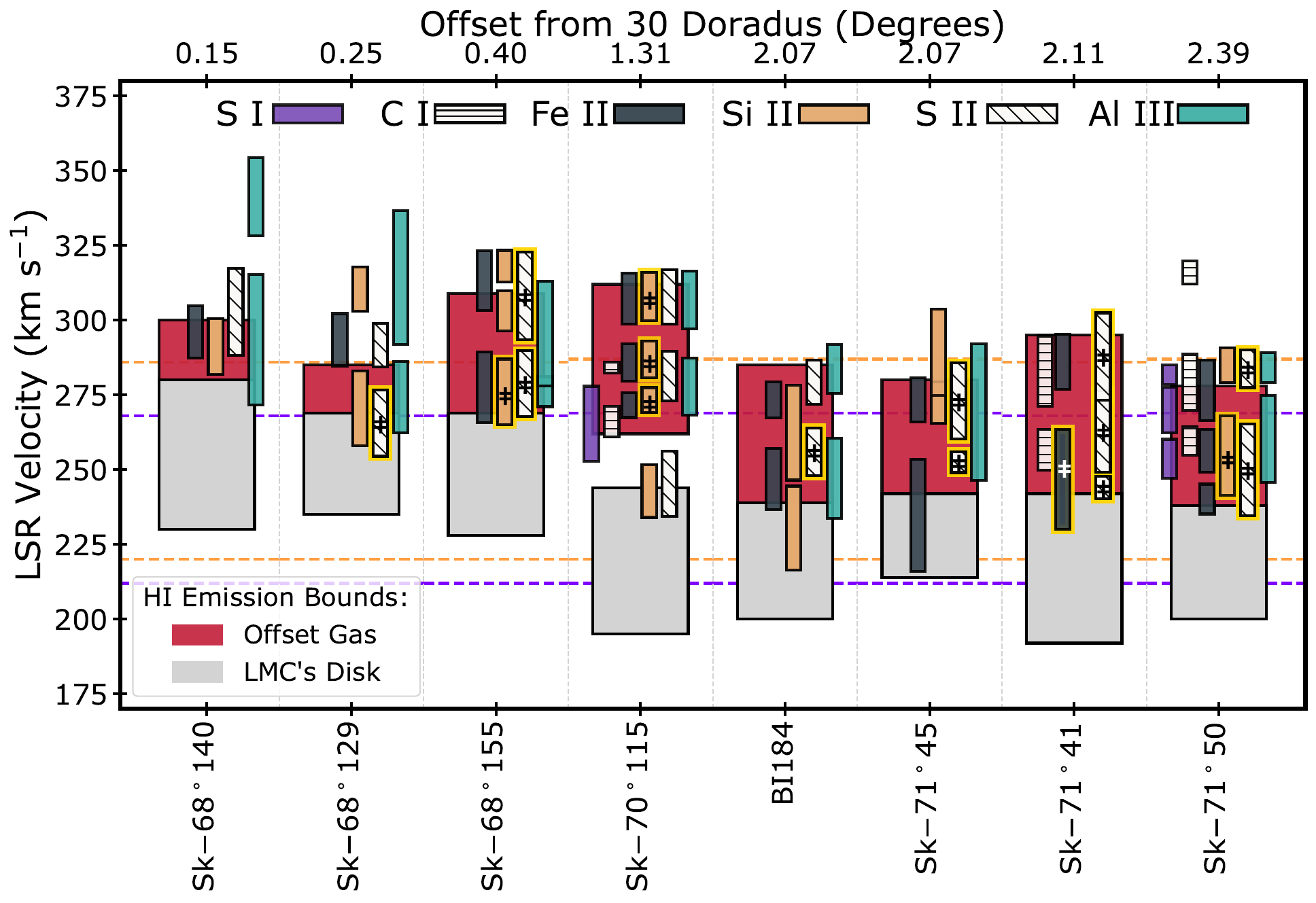}
  \caption{A comparison of the FWHM, center velocity, and offset from 30~Doradus for each sightline's absorption components. We choose to show each sightline's offset from 30~Doradus rather than display a linear increase on the top x-axis (all following offset plots are arranged in this layout). We prefer this labeling as sightlines~5--7 are located at similar angular offsets and a linear spacing would align the three sightlines on top of each other. The kinematic boundary of the LMC's disk  and offset gas along these sightlines are highlighted in gray and red, respectively. Six ion species are placed on this figure: S\textsc{~i} in purple, C\textsc{~i} in white with horizontal hatches, Fe\textsc{~ii} in dark gray, Si\textsc{~ii} in tan, S\textsc{~ii} in white with diagonal black hatches, and  Al\textsc{~iii} in green. The height of the rectangles corresponds to the FWHM of each component, where ${\rm FWHM}=2\sqrt{ln(2)}b$. Saturated components have a thicker yellow outline around them and a ``$\ddagger$'' symbol on top. We label the velocity ranges of arm~B and arm~E \citep{2003ApJS..148..473K, 2003MNRAS.339...87S} along each sightline by the horizontal purple and orange dotted lines, respectively. From left to right, the sightlines are arranged in order of increasing angular offset from 30~Doradus.  }
    \label{fig:FWHM}
\end{figure*}

To investigate the nature and structure of the high-velocity gas relative to the LMC's disk, we compare the velocity centers of their absorption components. Using S\textsc{~i} and C\textsc{~i} to trace the neutral gas (when available), Fe\textsc{~ii} and Si\textsc{~ii} to track the weakly-ionized gas, and S\textsc{~ii} and Al\textsc{~iii} to probe the moderately-ionized gas, we find that these ionization species kinematically overlap (see Figure~\ref{fig:FWHM}). The four sightlines within $1\fdg32$ of 30~Doradus have weakly-ionized components aligned around $+256\lesssim v_{\rm LSR} \lesssim +320~\kms$ and moderately-ionized components between $+265\lesssim v_{\rm LSR} \lesssim +345~\kms$. Sightline~4 has both neutral S\textsc{~i} and C\textsc{~i} components and these features are within $+263\lesssim v_{\rm LSR} \lesssim +285~\kms$. Therefore, close to 30~Doradus, the neutral, weakly-, and moderately-ionized gas occupy the same general kinematic space.

We observe a similar overlap for the four sightlines beyond $2\fdg0$ of 30~Doradus. Here, the majority of the 
weakly-ionized Fe\textsc{~ii} and Si\textsc{~ii} components and moderately-ionized S\textsc{~ii} and Al\textsc{~iii} components fall between $+230\lesssim v_{\rm LSR} \lesssim +290~\kms$ (see Figure~\ref{fig:FWHM}). Only sightlines~7 and~8 (Sk$-$71$^\circ$50) have detectable neutral absorption, which spans from around $+256 \lesssim v_{\rm LSR} \lesssim +316~\kms$. The remaining ion species we studied follow this kinematic overlap (see Tables~\ref{tab:Voigt_results_tab} and~\ref{tab:Voigt_results_app}). The observed overlap indicates that the high-velocity gas at all ionization stages probed in this study share the same kinematics.

We briefly highlight the kinematic similarity between the offset gas and the velocity ranges of arms~B and~E (see Section~\ref{connection}). From \citet{2003ApJS..148..473K}, \citet{2003MNRAS.339...87S}, and our velocity channel maps in Figure~\ref{fig:channel_maps}, we expect to find arm~B in the velocity range $+212 \leq v_{LSR} \leq +270$~\kms\, and arm~E around $+220 \leq v_{LSR} \leq +290$~\kms. We find that sightlines with angular offsets closer to 30~Doradus tend to have absorbers at the higher kinematic edge of arm~B with some extending beyond the \hi~boundaries of arm~E. This discrepancy is likely due to the difference in beam sizes between GASS/GASKAP and HST. Sightlines farther away have components that are situated deeper within the bounds of arm~B and approach the upper boundary of arm~E (see Figure~\ref{fig:FWHM}). Therefore, given the kinematic overlap between the UV absorbers and the velocity ranges of both arms, we suggest that we are likely probing the gaseous material of arm~B and/or arm~E. 

\begin{figure}[hbtp]
  \centering
\includegraphics[width=0.47\textwidth]{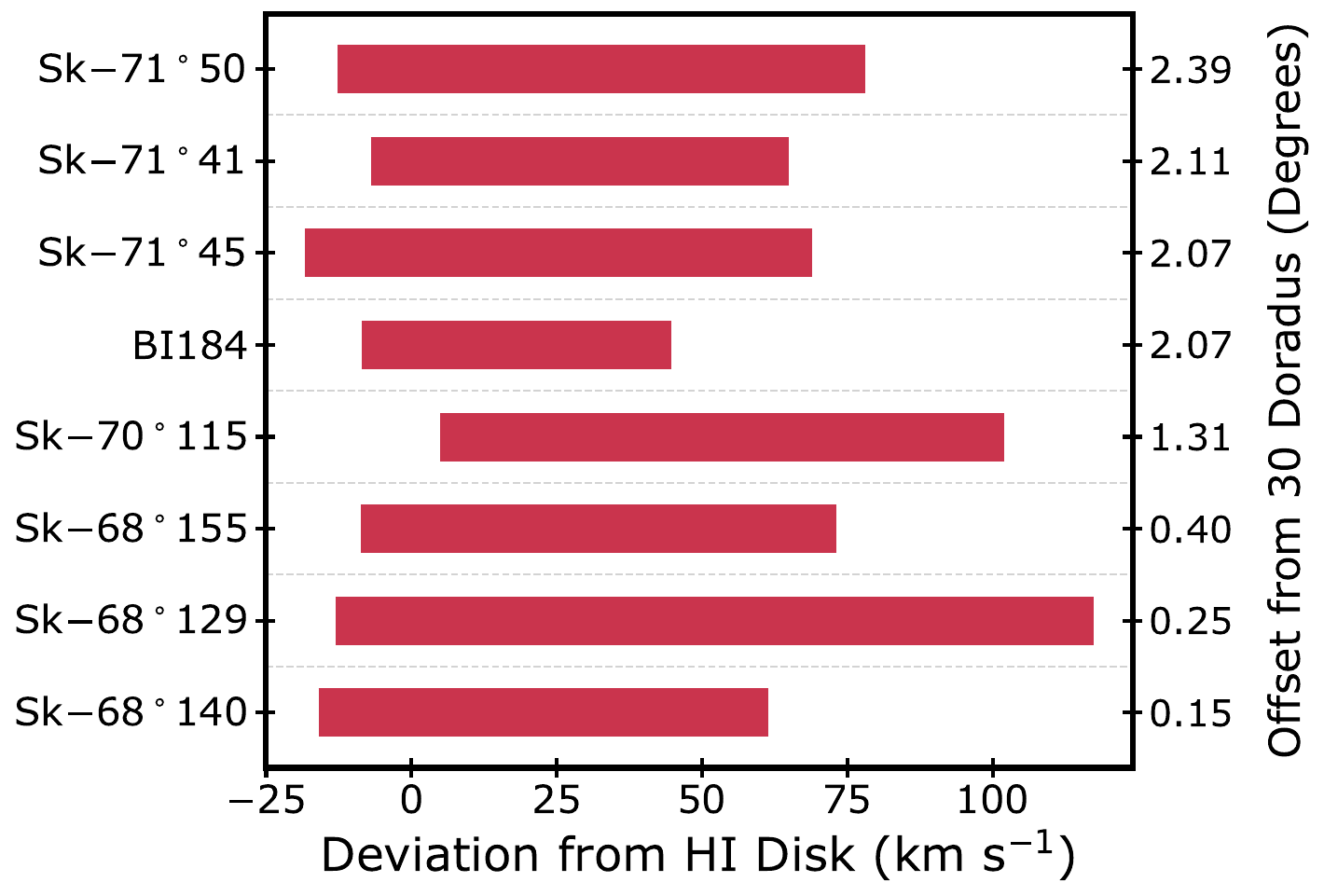}
  \caption{The deviation velocity range of the redshifted absorbers from the high-velocity edge of the LMC's \hi~disk as a function of offset from 30~Doradus (offset increasing from bottom to top). We emphasize that the right y-axis displaying the offset of each sightline from 30~Doradus is not linearly spaced. Rather, it labels the measured offset of each sightline recorded in Table~\ref{tab:targets}. The deviation ($v_{\rm Dev}$) is found from  $v_{\rm Dev}=v_{\rm offset\, gas}-v_{\rm LMC\,right\,bound}$ where $v_{\rm offset\, gas}$ is the center velocity of the redshifted absorbers and $v_{\rm LMC\,right\,bound}$ is the red velocity boundary of the LMC's disk as determined from the \hi~emission (see Table~\ref{tab:boundaries_hi}). We perform the subtraction for each absorber and then identify the minimum and maximum values as the deviation range. We note the deviation can be below zero for components blended with the LMC's disk.}\label{fig:deviation_from_disk}
\end{figure}

We note previous studies have observed a velocity gradient along the LMC's \hi~disk in the LSR frame \citep{1994A&A...289..357D,2003MNRAS.339...87S,2005A&A...432...45B,2008ApJ...679..432N}. This gradient is evident from our \hi~kinematic boundaries (see Table~\ref{tab:boundaries_hi} and Figure~\ref{fig:FWHM}), which shift with distance from 30~Doradus. We further explore the \hi~velocity gradient by calculating the kinematic deviation of the absorbers from the kinematic edge of the LMC's disk. We subtract the determined \hi~kinematic boundary of the LMC's disk from the velocity centroids of the UV features. We find that each sightline covers a similar deviation range (see Figure~\ref{fig:deviation_from_disk}). We note the deviations for sightlines~4 and~2 are larger than the other 6~sightlines with features extending past $+75$~\kms. Sightline~4 has a redshifted Al\textsc{~ii} feature found at $+346.2\pm3.9~\kms$, which is beyond its saturated disk component at $v_{\rm LSR}=+291.8\pm2.1~\kms$ (see Table~\ref{tab:Voigt_results_tab} and Figure~\ref{fig:plostacks}). Further, sightline~2 contains three redshifted  Al\textsc{~iii} absorbers at $+314.3 \pm 0.0$~\kms, $+57.3 \pm 4.8$~\kms, and $+386.3 \pm 0.0$~\kms. Overall, we find no relationship between absorber offset from the disk and the sightline's proximity to 30~Doradus. 

Additionally, we investigated the line widths of the offset gas. The Doppler parameter is impacted by both thermal and non-thermal broadening processes. The observed line width is given by the sum in quadrature of the thermal and non-thermal broadening contributions (i.e., $b^2 = b^2_{thermal} + b^2_{non-thermal}$). Gas that is thermally broadened through photoionization generally contains narrow features. Assuming only thermal motions, the physical line width is given by: 
\begin{equation}\label{thermal_line_widths}
    b = \left(\frac{2 k T}{M}\right)^{1/2} =12.90~\left(\frac{T_{4}}{M/amu}\right)^{1/2}~\kms,
\end{equation}
where $k$ is the Boltzmann constant, $M$ is the mass of the ionization species, and $T_{4} = T/10^{4} $~K. For the ionized offset gas in our project, we adopt a temperature of $10{,}000$~K, corresponding to the temperature in the warm ionized medium \citep{1999ApJ...523..223H}. For the neutral S\textsc{~i} and C\textsc{~i} components, we assume a temperature of $1,000 - 5,000$~K. Absorption features with line widths much larger than predicted by Equation~\ref{thermal_line_widths} are more likely also affected by non-thermal broadening processes like collisions.

When comparing the Doppler parameters of the ionization species across the 8~sightlines, we have to consider the velocity resolution of the instrument involved; COS and FUSE have much larger velocity resolutions than STIS (see Section~\ref{voigt_fit_sec}) and cannot physically resolve line widths on the same scale. Therefore, we focus on the E140H ($\Delta v = 2.5~\kms$), E230H ($\Delta v = 2.5~\kms$), E140M ($\Delta v=6.5~\kms$), and E230M ($\Delta v=10~\kms$) STIS observations and use them to compare to the line widths from Equation~\ref{thermal_line_widths}.

Generally, across the low and neutral ionization species for the majority of our sightlines, we observe a combination of medium ($\sim4.0-15.0$~\kms) and broad line widths ($\geq\sim19.0$~\kms). We do find a few narrow features that may be thermally broadened that are detected along sightline~6 in S\textsc{~ii}, sightlines~2 and~3 in Al\textsc{~ii}, and sightlines~4 and~8 in C\textsc{~i}. We note that some of the components of these species that are kinematically closer to the LMC's disk and/or that are saturated tend to have Doppler parameters that are broader and could indicate more non-thermal broadening processes like collisions are influencing the gas, however, there may also be unresolved blending. To compare to the ionization processes occurring in the disk, we analyzed the widths of the disk absorbers along all 8~sightlines. We find mainly medium width absorbers for the LMC's ISM indicating non-thermal broadening. Interestingly, along sightline~4, there are narrow disk features for Si\textsc{~ii} and P\textsc{~ii} which suggests contributions from both thermal and non-thermal mechanisms present in the disk. For the lower and neutrally ionized species, we conclude that they likely experience both thermal and non-thermal broadening effects given the range of their line widths.

Then, we investigated the kinematic width of the moderately-ionized offset gas using the Al\textsc{~iii} absorbers. Sightlines~1 and~8 have very broad absorbers kinematically near the disk and more medium width components at higher velocities. Sightlines~2 and~3 are similar in that they have a mix of broad and medium width components at varying velocities. Sightlines~4 and 5 contain only slightly moderate absorbers while sightline~6 has a singular very broad feature. We suggest that the moderately ionized phase of the gas, probed by  Al\textsc{~iii},  experiences more non-thermal contributions.

\subsection{Electron Density and Cooling Rate Calculations}\label{electron_density}

To estimate the density of the offset gas, we can use collisionally ionized species like C\textsc{~ii$^{*}$}, which has the $2(p)^{2} P_{3/2}$ and $2(p)^{2} P_{1/2}$ fine-structure levels for its ground state. The $2(p)^{2} P_{3/2}$ level is populated through collisional excitation as electrons, hydrogen atoms and molecules, and helium collide with ions in the $2(p)^{2} P_{1/2}$ level in the surrounding gas. Once collisionally ionized into the $2(p)^{2} P_{3/2}$ level, an electron can undergo spontaneous emission which leads to cooling. However, electrons in the $ P_{3/2}$ level can also transition to a higher energy state through the absorption we observe at the C\textsc{~ii$^{*}$}$\lambda$~1335.7 line. When calculating densities, we compare collisionally excited doublets to eliminate any dependencies on abundances and ionization fractions. Previous studies (e.g., \citealp{2004ApJ...615..767L,2005ApJ...623..767J,2008ApJ...679..460Z}) have estimated electron densities in the interstellar and circumgalactic mediums with the column-density ratios $N(C\textsc{~ii}^{*})/N(C\textsc{~ii})$. We calculate the electron density following the procedure outlined in \citet{2004ApJ...615..767L}, where 
\begin{equation}\label{electron_den_c}
    n_{e} = 0.531 \frac{\sqrt(T)}{\Omega_{12}}e^{91.7K/T} \frac{N(\rm{CII^{*}})}{N(\rm{CII})}
\end{equation}
Here, $\Omega_{12}$ is the collisional strength. 
We make similar assumptions as in \citet{2004ApJ...615..767L} and \citet{2013ApJ...772..111R} and use S\textsc{~ii} as a proxy for C\textsc{~ii} as C is lightly depleted onto dust grains while S is not \citep{2009ApJ...700.1299J}. Therefore, \citet{2004ApJ...615..767L} rewrites Equation~\ref{electron_den_c} above as
\begin{equation}\label{electron_density_S}
    n_{e} = 0.531 \frac{\sqrt(T)}{\Omega_{12}}e^{91.7K/T} 10^{0.2}\frac{N(\rm{CII^{*}})}{N(\rm{SII})} (\rm{S}/\rm{C})_{\odot}
\end{equation}
We assume a temperature of ${\sim}24,000\,\rm K$ as this is where the $\rm C^+/C$ ionization fraction species is expected to be highest based on the \citet{2007ApJS..168..213G} computational models (see their Figures~2 and~8). For the collisional strength, we use $\Omega_{12} = 2.82$ at $T=20,000$~K from Table~2 of \citet{1984A&A...134..193H}. We also adopt the $(S/C)_{\odot}$ value from \citet{2021A&A...653A.141A}. We emphasize that Equations~\ref{electron_den_c} and~\ref{electron_density_S} assume that the cold gas along the line-of-sight is negligible, the \hi~emission and UV ionization species are probing the warm neutral and ionized gas phases, and only spontaneous emission is occurring. In the case of cold gas where collisions with hydrogen atoms cannot be neglected, electron densities calculated using Equations~5 and~6 become upper limits.

\begin{figure*}[h]
  \centering
  \includegraphics[width=0.80\textwidth]{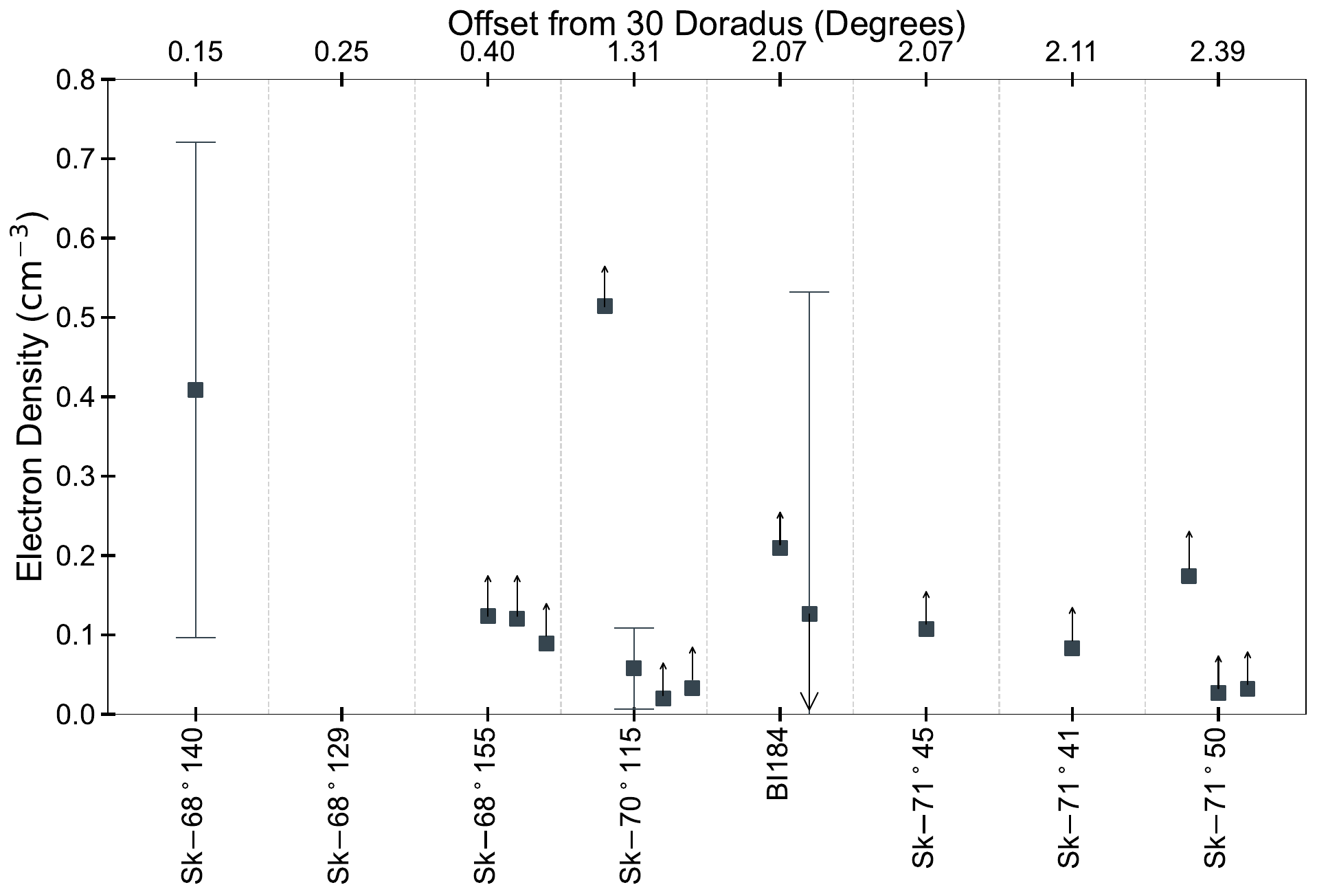}
  \caption{The electron density measurements across all our sightlines. Data points with upward arrows are lower limits. The error bar for the data point associated with BI184 extends below $0~\rm{cm^{-3}}$ and we indicate this with a downward arrow. We order the sightlines from left to right with increasing angular offset from 30~Doradus.  
}\label{fig:electron_density}

\end{figure*}

With these values, we calculate the electron density across 7~of the 8~sightlines. We exclude sightline~2 as it does not have S\textsc{~ii} and C\textsc{~ii}$^{*}$ components that align kinematically. Additionally, we note that across the majority of our sightlines, the S\textsc{~ii} components are saturated which leads to only lower limit estimates of the electron density. From our results, we find components along the same sightline that are kinematically closer to the disk tend to have larger electron densities than their more kinematically offset counterparts (see Figure~\ref{fig:electron_density}). We also report a lower limit of $n_{e}\gtrsim 0.1\,\rm{cm^{-3}}$ for the electron density of the nearside gas. We can compare our measurements to the \citet{2013ApJ...772..111R} study, where they found a electron density of $n_{e} \leqslant 5.0\times 10^{-2}~\rm{cm^{-3}}$ along one sightline near the body of the MS. We expect and find the electron density to be higher near the 30~Doradus region rather than in the extended structure of the MS.  Additionally, the \citet{2011ApJ...735....6P} study characterizing the dust emission in three different regimes in the LMC---diffuse ionized gas, typical \hii~regions, and very bright \hii~regions---reports electron densities of $n_{e}=0.055~\rm{cm}^{-3}$, $n_{e}=1.52 \pm 0.44~\rm{cm}^{-3}$, and $n_{e}=3.98 \pm 2.05~\rm{cm}^{-3}$, respectively. Our lower limit indicates the nearside gas has a larger electron density than the diffuse ionized gas in the LMC.

\begin{figure*}[h]
  \centering
  \includegraphics[width=0.80\textwidth]{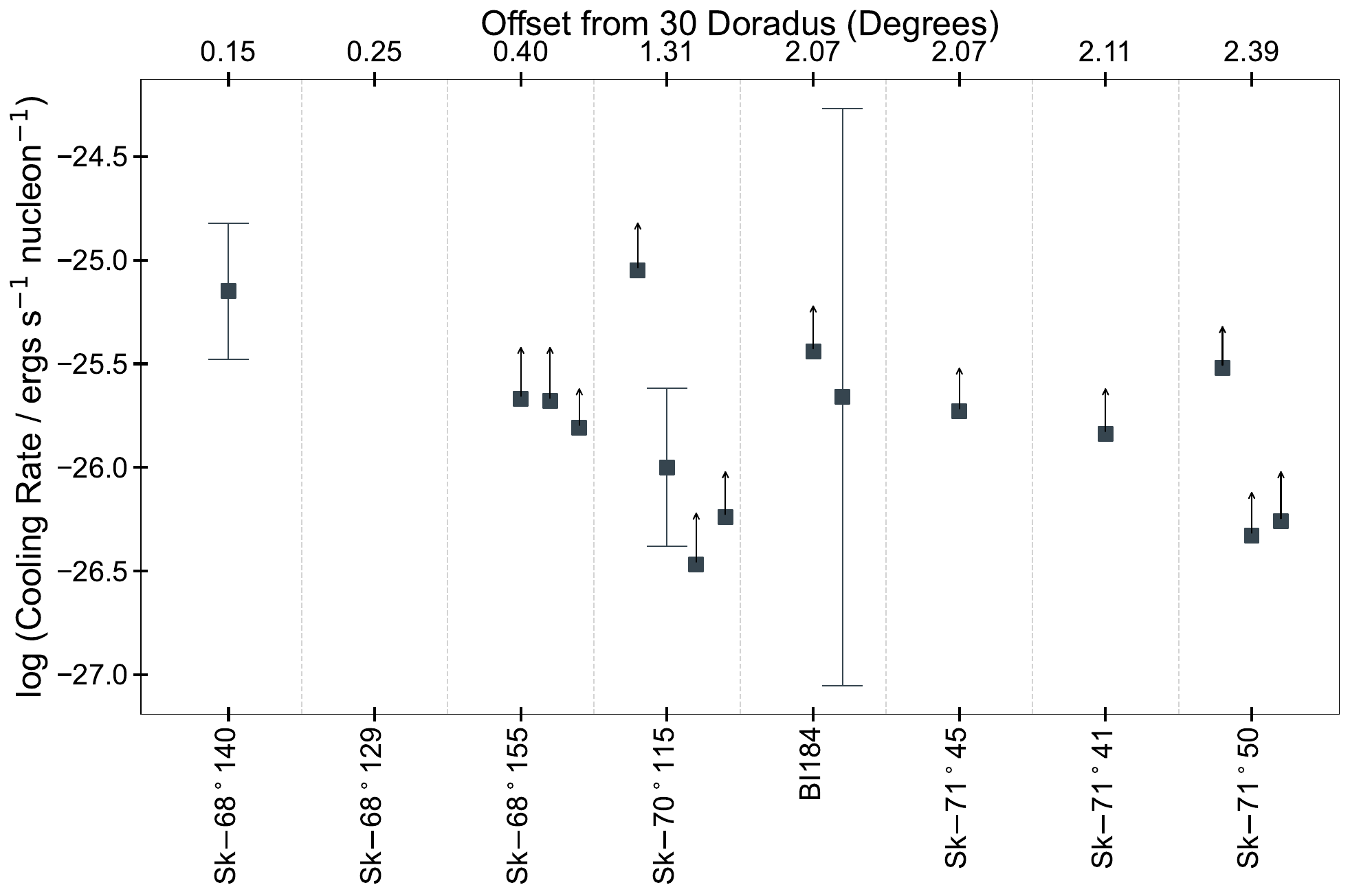}
  \caption{The cooling rate per nucelon measurements across all our sightlines. Data points with upward arrows are lower limits. Sightlines are arranged from left to right with increasing angular offset from 30~Doradus.
}\label{fig:cooling_rate}

\end{figure*}

Further, \citet{2004ApJ...615..767L} derived an expression for the cooling rate per nucleon given in the equation below. 

\begin{equation}\label{cooling_rate}
    L_{c} = 2.89 \times 10^{-20} \frac{N(\rm{CII^{*}})}{N(\rm{SII})} (\rm{S}/\rm{H})_{\odot}\, \rm{ergs} \,s^{-1}\, nucleon^{-1}
\end{equation}
We estimate the lower limit for the offset absorbers given the saturation of S\textsc{~ii} along most of our sightlines of $\rm{log}(L_{c}/ ergs~s^{-1}~nucleon^{-1})\gtrsim -25.8$ (see Figure~\ref{fig:cooling_rate}). Comparing to the work of \citet{2004ApJ...615..767L} studying the \mbox{low-,} intermediate-, and high-velocity clouds belonging to the Milky Way, we find that our cooling rate limit aligns fairly well with the values determined for the Milky Way ISM material. 

\subsection{Total Integrated Column Densities}\label{total_cd}

As an assessment of the ionization conditions of the redshifted absorbers, we compare the total integrated column density values of Fe\textsc{~ii}, Si\textsc{~ii}, Ni\textsc{~ii}, Al\textsc{~iii}, and P\textsc{~ii} as these species are found across the majority of the sightlines and there is limited saturation. We determine the total integrated column densities using the AOD method. We integrate between the left boundary (closest to the disk) of the offset gas as determined by the \hi\, emission (see Table~\ref{tab:boundaries_hi}) and $v_{central}+2b$ of the last offset absorber in the UV spectra. For sightline~4, there is a kinematic separation between the LMC's disk and the offset gas between $+244\leq v_{LSR}\leq +262$~\kms. The UV spectra contains Si\textsc{~ii} and P\textsc{~ii} absorbers that fall within this velocity range (see Figure~\ref{fig:plostacks}). For those two ionization species, we adjusted the lower limit of integration to  $+244$~\kms, the right edge of the LMC's disk.

\begin{figure*}[h]
  \centering
\includegraphics[width=0.80\textwidth]{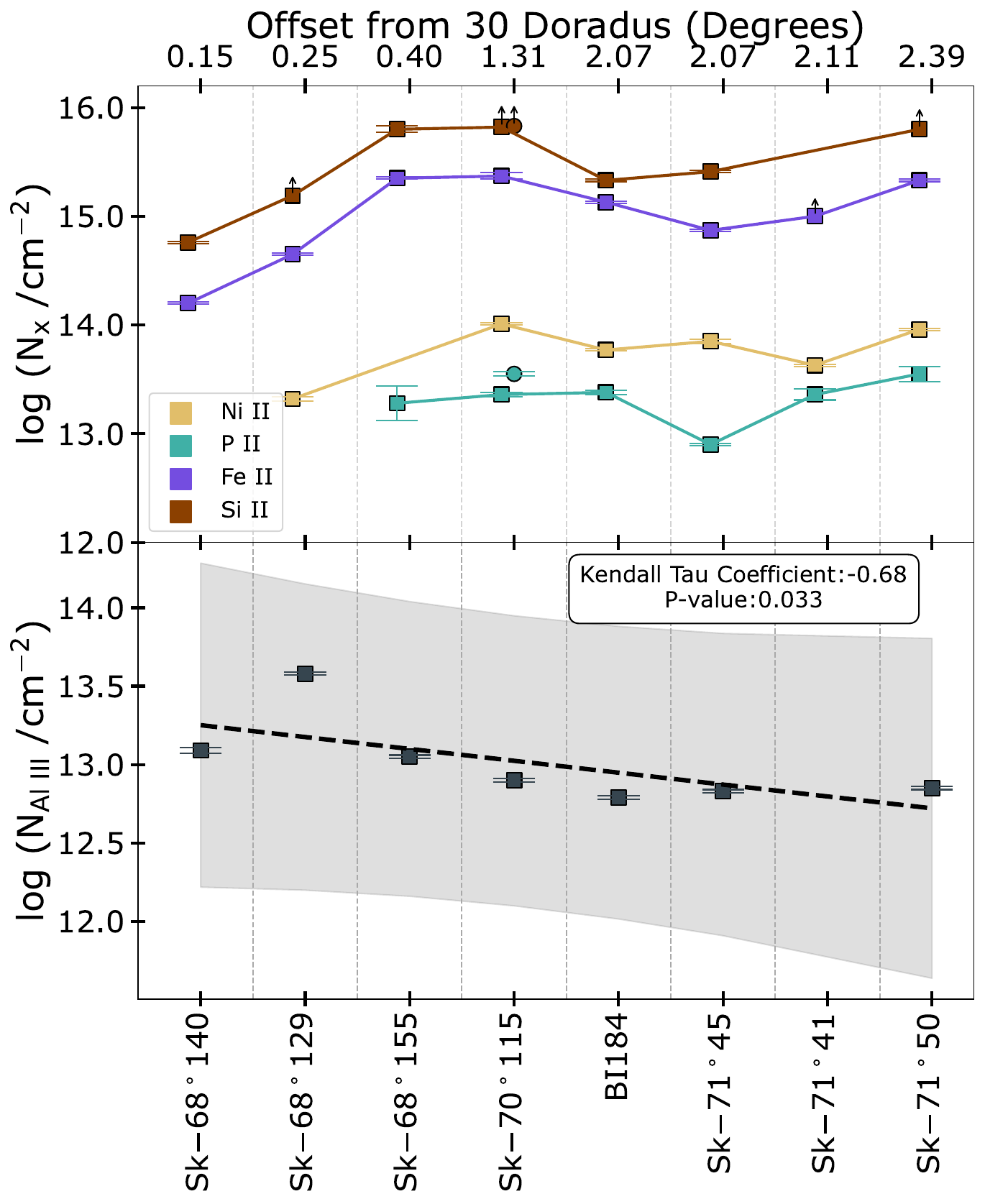}
  \caption{\textit{Top Panel:} Total integrated column densities for the offset absorbers of Ni\textsc{~ii} (tan), P\textsc{~ii} (green), Fe\textsc{~ii} (purple), and Si\textsc{~ii} (brown) for each of the sightlines. The upward arrows on the data points indicate saturated components that are lower limits. The two circle data points for Si\textsc{~ii} and P\textsc{~ii} along sightline~4 are integrated between the velocity of the right edge of the disk boundary and the last offset UV absorber of the offset gas. We include these data points because there are UV absorbers kinematically located between the determined \hi~boundaries of the disk and offset gas. \textit{Bottom Panel:} Total integrated column densities for the offset absorbers of Al\textsc{~iii}. The gray envelope is the 3$\sigma$ prediction interval of the linear regression fit. We hightlight the Kendall Tau Coefficient and p-value of the fit in the top right corner. The sightlines are arranged in order of increasing angular offset from 30~Doradus from left to right.} \label{tot_int_CD}

\end{figure*}

When interpreting the results from the total integrated column density plots, we emphasize that arms~B and~E do not contain stars and the 8~sightlines are at some arbitrary depth inside the disk of the LMC. We do not know the entire 3D~geometry of the arms or physically how much of the arms are in front of our sightlines. Therefore, the path along the line-of-sight from which we measure our total integrated column densities will vary among the sightlines contributing to some of the scatter in the values. This effect is specifically seen in nearby sightlines~5--7 which are probing the same general region but likely are at different depths relative to the arms. With that in consideration, we point out a global trend for Fe\textsc{~ii}, Si\textsc{~ii}, and Ni\textsc{~ii} where the two closest sightlines to 30~Doradus typically have smaller values than sightlines at a greater angular offset (see Figure~\ref{tot_int_CD}). However, the total integrated column density of Al\textsc{~iii} decreases as a function of increased angular offset for the Voigt fit limits. 

To measure the strength and direction of the relationship between the total integrated column densities of Al\textsc{~iii} and offset from 30~Doradus, we performed a Kendall's Tau correlation test. The Kendall rank coefficient $\tau$ describes how well the two variables agree and whether there is a positive or negative correlation. The statistical significance of the correlation is given by the p-value and we assume a significance level of $p =0.05$. For the Al\textsc{~iii} column density values, we find $\tau =-0.68$ and $p=0.033$ (see Figure~\ref{tot_int_CD}). Our results indicate that there is a statistically strong negative correlation between the  Al\textsc{~iii} values and the offset from 30~Doradus. We anticipate these findings as Al\textsc{~iii} has a moderate ionization potential (28.45 eV). We expect to find larger column densities for moderate species closer to 30~Doradus given that its enhanced stellar activity ionizes the surrounding gas. Considering that the two sightlines closest to 30~Doradus have lower total integrated column densities for singly-ionized species and that there is a strong decreasing trend for the Al\textsc{~iii},  we suggest that sightlines~1 and~2 may probe different gaseous material than the other sightlines. Given their proximity to 30~Doradus, these absorbers could be ejected material that is outflowing or falling back into the disk rather than part of either arm.

\section{Simulation of the Magellanic System}\label{simulation}

The existence of redshifted material on the nearside of the LMC is intriguing and merits a more thorough investigation into its formation. Therefore, it is useful to compare our findings against simulations. Specifically, we will focus on constraining the likelihood of gaseous debris existing on the nearside and how it gets there. The simulation used in this study is originally published in \citet{2021ApJ...921L..36L}. We utilize the meshless finite-mass method implemented in the GIZMO massively parallel multiphysics code \citep{2015MNRAS.450...53H,2005MNRAS.364.1105S}.
This simulation includes star formation and basic gas cooling and heating; however metal-line cooling and radiation (from stars and the UV background) are not included. Our simulated LMC galaxy includes a very active star-forming region that is analogous to the 30 Doradus star burst region, though this simulated star cluster is less active and has a gentler history with forming $1.3\times10^5\,M_\odot$ of stars in the past 2~million years and $2.4\times10^5\,M_\odot$ in past 20 million years (see \citealp{2025arXiv250305968P}). Due to the Lagrangian nature of the meshless finite-mass method, we can separate the gas that belongs to the MW and the MCs at late times based on the initial locations of the particles. We define material that originates from within the disks of the MCs or from within the Magellanic Corona as Magellanic material. 

For each of the three galaxies, the MCs and MW, we include a live dark matter halo (following a Hernquist profile; \citealt{1990ApJ...356..359H}) and stellar and gaseous exponential disks. The dark matter masses for the LMC, SMC, and MW are $1.8\times10^{11}$, $1.9\times10^{10}$, and $10^{12}$~$\mathrm{M}_\odot$, respectively. Additionally, we include gaseous halos around the MW and LMC \citep{2020Natur.585..203L}. We excised the circumgalactic medium from an LMC analog in the Auriga simulations \citep{2017MNRAS.467..179G} and overlaid it on top of our analytic LMC disk and dark matter halo. Its total mass was scaled to account for the difference in the total mass between the Auriga galaxy and our LMC. Its temperature was uniformly set to the LMC's virial temperature ($2.4\times10^5$~K). For the MW, we used a beta profile halo following \citet{2015ApJ...815...77S} and set its virial temperature to $2.4\times10^6$~K. To match the velocity gradient along the MS, the total mass of the MW's gaseous halo was increased by a factor of two. More details can be found in  Table~1 of \citet{2021ApJ...921L..36L}.

To compare the kinematic and spatial location of the gas with the ULLYSES observations, we generate mock velocity distribution profiles that extend through the disk plane of the LMC.
We select sightlines originating from the Solar position that pass through the LMC's disk and map out the gas velocities. The benefits of using the simulation are that we can identify the gas velocities and connect them to the positions of the gas. Therefore, we can isolate the redshifted or blueshifted material on the near side of the disk and investigate its properties. To create these maps, we reorient the simulation such that the $z'$ axis is aligned with the angular momentum of the LMC's stellar disk, and the origin aligns with the 3D position of the LMC. 
Then, we extract line-of-sight velocity profiles from the Sun through the leading, middle, and trailing regions of the LMC's disk. We take a 4~kpc thick slice of the simulation and bin the gaseous material along each sightline. These slices are large enough for us to probe over homogeneous vs. small scale stochastic conditions so that we could reliably trace the bulk material and associated motions of the gas that extends beyond the LMC's disk. We measure the mean radial velocity for the gas within $1\times1\times4$~kpc boxes (with the 4~kpc length along the $x'$ direction, perpendicular to the line-of-sight) at 100~locations through the line-of-sight for each sightline (see Figure~\ref{fig:los_vel}).

\begin{figure*}[hbtp]
    \centering
\includegraphics[width=\textwidth]{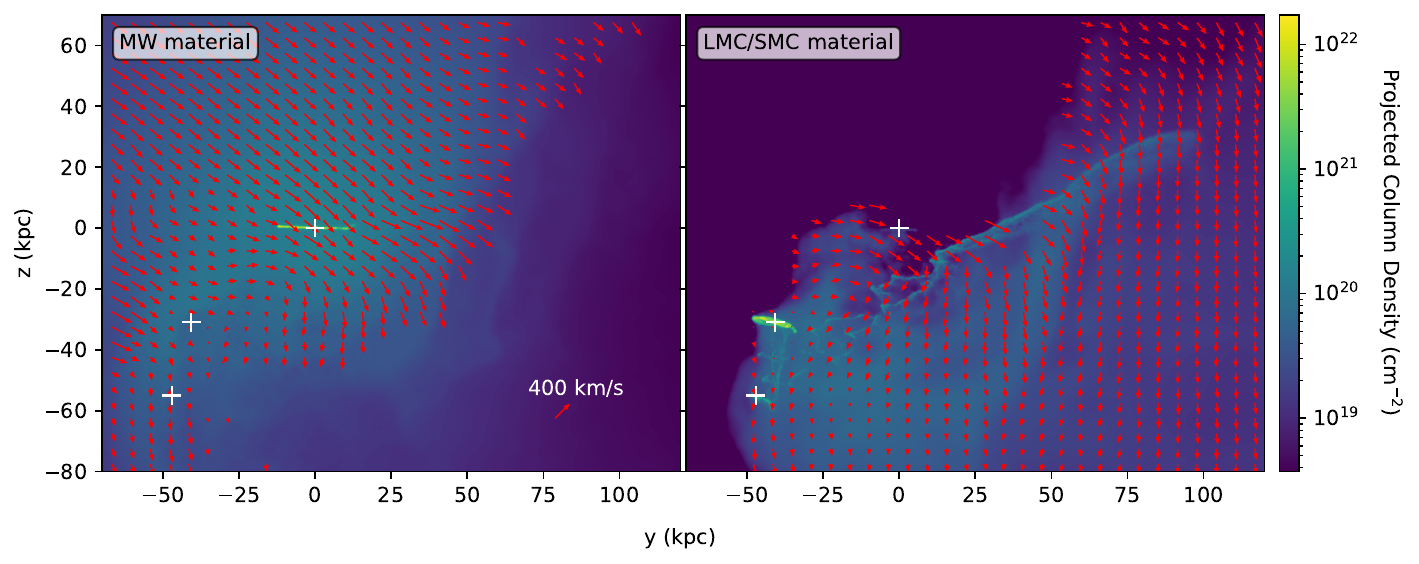}
    \caption{View of entire simulated Milky Way and MC system shown in Cartesian coordinates. The sun is located at $(x,y,z)=(-8.122,0,0.023)$~kpc, and the $z$-axis is aligned with the Galactic north pole. \textit{Left}: Background showing the projected gas density of the Milky Way material with velocity field overlaid with red arrows. The velocities are shown in the LMC center-of-mass frame. The red arrow in the bottom right corner of the left panel is shown for scale. White plus marks denote the centers of the MW, LMC, and SMC. \textit{Right}: As left, except showing the LMC material, again with the velocity field shown with respect to the LMC's center-of-mass motion.}
    \label{fig:zoom_out_system}
\end{figure*}

\begin{figure*}[hbtp]
    \centering\includegraphics[width=\textwidth]{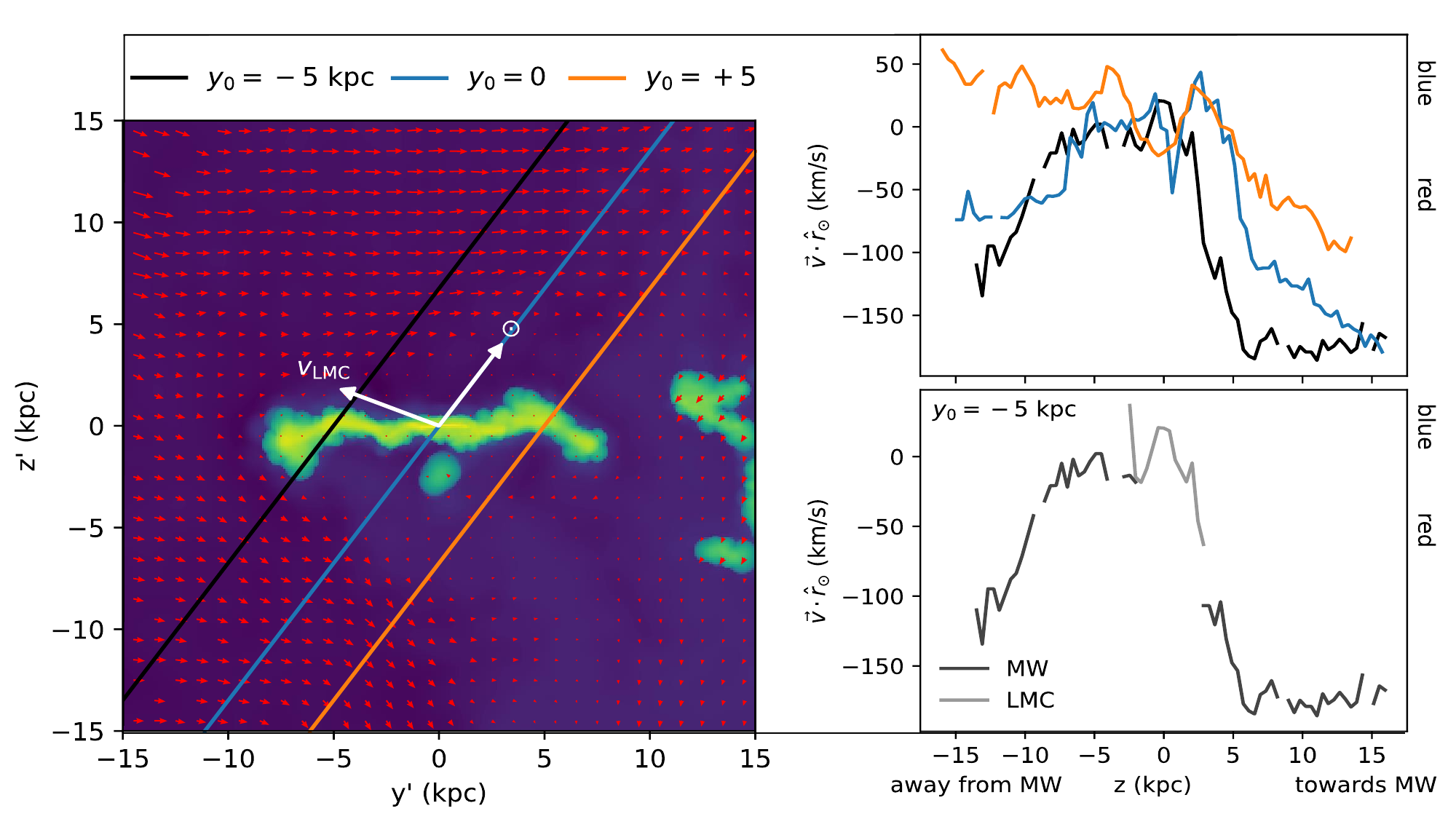}
    \caption{\textit{Left}: Projected gas density of a zoomed-in region around the LMC 4~kpc thick overlaid with the gas velocity field in the LMC's rest frame. This image is rotated with respect to Figure~\ref{fig:zoom_out_system} in order to align the angular momentum of the LMC's disk with the $z$-axis. The white arrows denote the direction of the LMC's velocity (labeled $v_{\rm LMC}$), and the direction toward the sun (labeled $\odot$). The three lines (black, blue, and orange) mark the trajectories for three sightlines along which we measure the radial velocities of the gas in the simulation. \textit{Upper right}: The radial velocity profiles of the three sightlines. Positive velocities indicate material moving toward the sun (blueshifted material), while negative velocities indicate redshifted gas. \textit{Lower right}: The radial velocity profile of the $y_0=-5$~kpc sightline (black in left and upper right panels) split up based on MW (dark gray) and LMC (light gray) material. The LMC-originating gas is spatially isolated from the MW gas due to the protection of the Magellanic Corona and extends up to $\sim3$~kpc above and below the plane of the LMC's disk.}
    \label{fig:los_vel}
\end{figure*}

The Magellanic Corona that we include in our simulations affects the flow of the gas near the LMC's disk. In Figures~\ref{fig:zoom_out_system} and Figure~\ref{fig:los_vel} (left panel), we display the velocity vectors of the gas surrounding the LMC's disk. The small vectors on the near side of the LMC's disk are indicators that the Magellanic Corona provides a strong barrier toward the leading edge of the LMC that significantly lessens the ram-pressure forces affecting the position of the gas near this galaxy's disk. Without this barrier, the gas flows on the leading edge of the LMC would be suppressed (see \citealp{2021ApJ...908...62C}).
We also observe a distinct spatial separation between the MW and LMC material (see bottom right panel of Figure \ref{fig:los_vel}). The Magellanic Corona guards the gaseous debris out to about ${\sim}$3 kpc above the LMC's disk, limiting mixing between the Milky Way's gas and the LMC. We find the \hi~column density of the redshifted, nearside gas ranges from $18.0\leq \rm{log(N_{\hi}/cm^{-2})} \leq 21.2$. Closer to the disk, we observe higher \hi~column densities. The \hi\ column density of the gas drops away from the disk, with redshifted material on the nearside of the LMC dropping more rapidly. 

\section{Discussion}\label{discussion}

Below, we discuss the kinematic overlap between the GASS/GASKAP \hi~emission, the offset absorbers in the UV, and the arms to offer support that we are probing continuous gaseous structures, likely arm~B and/or arm~E, flowing in front of the LMC. Additionally, we provide implications for environmental processes in order to form the arms on the nearside of the LMC.  We place into context the properties of the redshifted absorbers with previous studies of the 30~Doradus region. Finally, we include additional arguments on the explanation of the possible nature of the redshifted gas relative to the LMC's disk.

\subsection{Complementary Perspectives from Neutral Hydrogen and UV Observations}\label{connection}

\begin{figure*}[h!]
  \centering
  \includegraphics[width=1\textwidth]{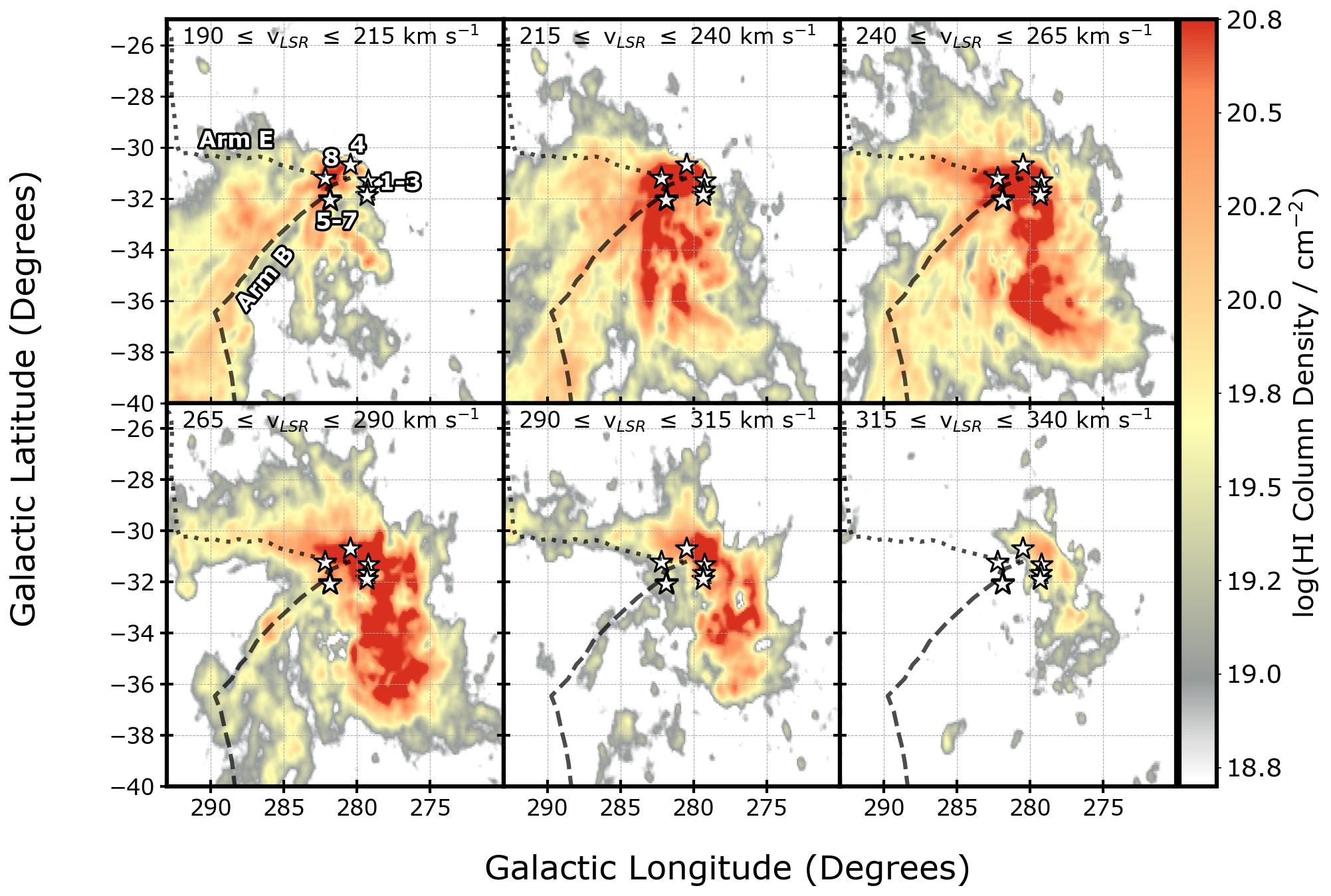}
  \caption{Individual integrated \hi~emission velocity channel maps over a velocity range of $+190\le\  v_{\rm LSR}\le+340~\kms$ from the GASS survey. We label the locations of the 8~sightlines in our project with stars and their numbers correspond to the IDs in Table~\ref{tab:targets}. We label the velocity range of each panel at the top of each plot. The predicted locations of arm~B and arm~E from the Gaussian decomposition of the LAB survey in \citet{2008ApJ...679..432N} are highlighted by the dashed black line and dotted black line, respectively. }
  \label{fig:channel_maps}
\end{figure*}

By examining both the \hi~and UV data simultaneously, we can focus on the small-scale structure of the \hi~gas while comparing it to the kinematics of the UV features. We compared the \hi~gas motion in both the LMC's disk and its exterior by creating individual position-velocity maps for each sightline (see Section~\ref{pv_plots_section} and Figures~\ref{fig:pv} and~\ref{fig:appendix_pv_plots}). In all of our sightlines, we detect coherent, redshifted \hi~emission that is offset from the LMC's disk by roughly $20-68~\kms$. In this region, arms~B and~E contain redshifted gas consistent with what we are observing.  Then, we overlaid the velocity extent of the Fe\textsc{~ii}, Si\textsc{~ii}, and S\textsc{~ii} components from the Voigt profile fitting on each sightlines' position-velocity map (see Figures~\ref{fig:pv} and ~\ref{fig:appendix_pv_plots}). We find that the kinematic spread of the Fe\textsc{~ii}, Si\textsc{~ii}, and S\textsc{~ii} absorbers align with the filamentary \hi~emission, suggesting that we are likely investigating the same material in the radio and UV data. We created velocity channel maps from the GASS observations to examine the kinematical and spatial distribution of the \hi~emission in this region (see Figure~\ref{fig:channel_maps}). Sightlines~5, 6, and 7 lie spatially in the same predicted route as arm~B. Similarly, sightline~8 lies within the \hi~emission of arm~E. We suggest that we are likely detecting one or both arms given their spatial and kinematical proximity to our sightlines.

\begin{figure*}
  \centering
  \includegraphics[width=1\textwidth]{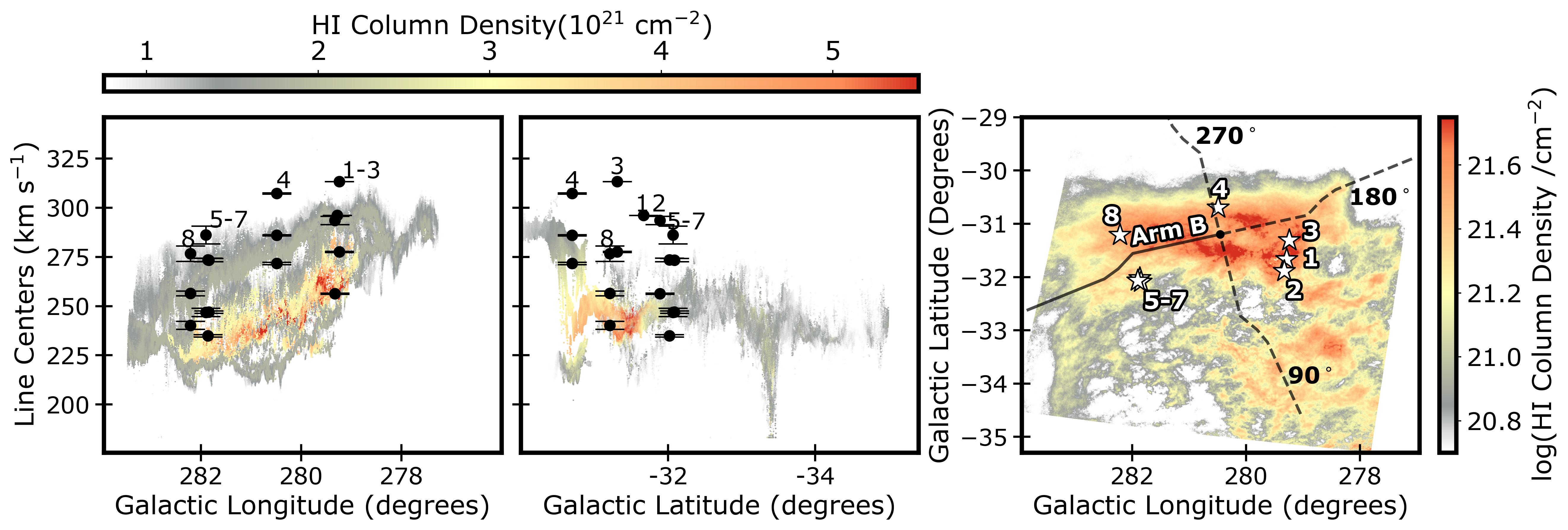}
  \caption{\textit{Left}: A position-velocity longitude slice through the path of arm~B and its rotated 180\degree\, trajectory. \textit{Middle}: A position-velocity latitude slice through arm~B's 90\degree\, and 270\degree\,rotations. In both the left and middle panels, the black dots are the velocity centroids from the Voigt profile fitting of each sightline's Fe\textsc{~ii} components. The dots are located at the sightlines' Galactic longitudes (left) and Galactic latitudes (middle). \textit{Right}: An integrated \hi~column density map from GASKAP that outlines the path of arm~B and its 90\degree, 180\degree, and 270\degree\, rotation (each labeled in bold) around arm~B's starting point (marked by the solid black point) as determined from the \citet{2008ApJ...679..432N} Gaussian decomposition of the LAB data: $(l,b)=(280\fdg50,-31\fdg20)$. Each of the 8~sightlines are labeled. }
  \label{fig:arm_b_long_lat}
\end{figure*}

\begin{figure*}
  \centering
  \includegraphics[width=1\textwidth]{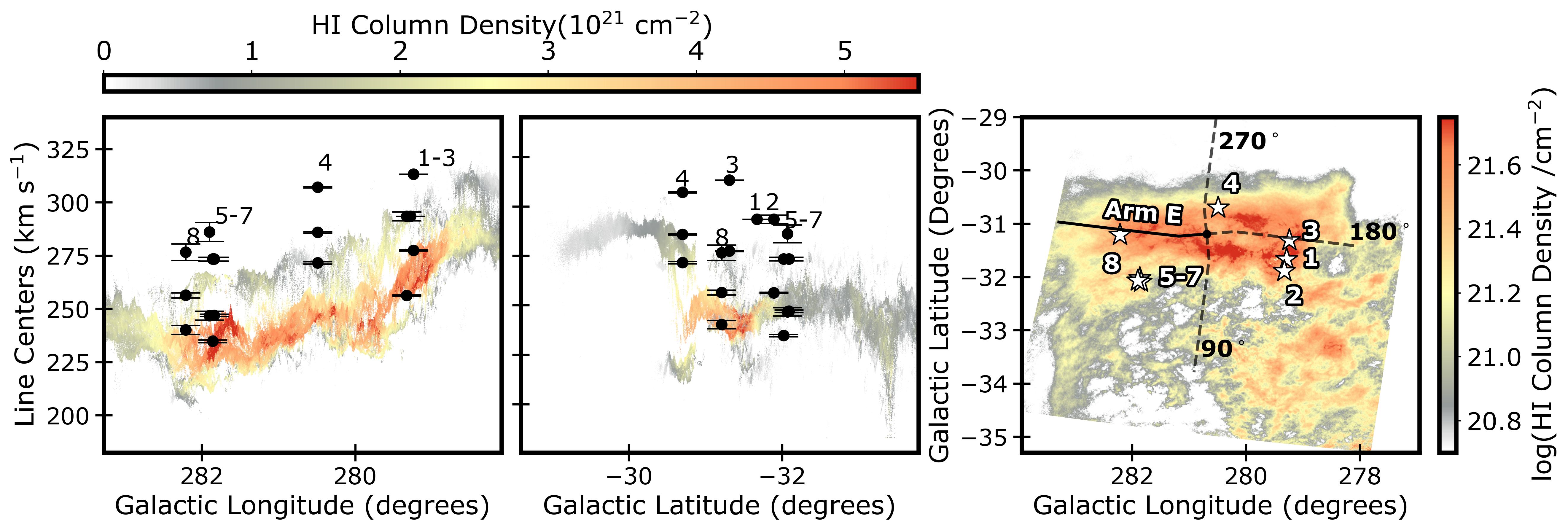}
  \caption{Similar position-velocity plots as in Figure ~\ref{fig:arm_b_long_lat}. \textit{Left}: A Galactic longitude slice through the path of arm~E and its rotated 180\degree\, trajectory. \textit{Middle}: A Galactic latitude slice through arm~E's 90\degree\, and 270\degree\,rotations. Again, the black dots are the velocity centroids from the Voigt profile fitting analysis for the Fe\textsc{~ii} components. \textit{Right}: An integrated \hi~column density map from GASKAP that outlines the path of arm~E and its 90\degree, 180\degree, and 270\degree\, rotation (labeled in bold)  around $(l,b)=(280\fdg69,-31\fdg19)$, the starting point of arm~E as determined by \citet{2008ApJ...679..432N}.}
\label{fig:arm_e_long_lat}
\end{figure*}

To help clarify which sightlines could contain arm~B and~E material, we performed the same Gaussian decomposition technique as in \citet{2008ApJ...679..432N} on the new GASKAP observations to trace the kinematic behavior along the arms. While \textsc{GaussDecomp} works well automatically over a range of longitudes and latitudes, we manually performed Gaussian decomposition for sightline~8. 
We rotated the predicted path of the arms from the Gaussian decomposition of the LAB survey in \citet{2008ApJ...679..432N} by $180\arcdeg$, creating a new position-position cut through the LMC's disk (see Figures~\ref{fig:arm_b_long_lat} and~\ref{fig:arm_e_long_lat}). Similarly, we also rotated the predicted LAB survey path by $90\arcdeg$ and $270\arcdeg$ to create a perpendicular position-position slice through the region to ensure we could isolate the arms' kinematic behavior from the material in the LMC's disk (see Figures~\ref{fig:arm_b_long_lat} and~\ref{fig:arm_e_long_lat}). The high-velocity \hi~emission along the 180\degree\, rotated paths of arm~E and arm~B connect seamlessly through the \hi~gas at Galactic longitudes near sightline~4 ($l = 280\fdg49$) and even extend further toward 30~Doradus near sightline~3 ($l = 279\fdg24$). Therefore, we suggest that arm~B and arm~E may extend about ${\sim} 1\fdg0$ further into the \hi~overdensity region that houses 30~Doradus than previously thought. Furthermore, both arms flow cohesively through the measured velocity centroids of the Fe\textsc{~ii} UV absorbers from the Voigt fitting procedure for sightlines~1--4 (see Figure~\ref{fig:arm_e_long_lat}), supporting that our \hi~and UV observations are probing the same material.

 If both arms extend farther toward lower Galactic longitudes, the arms likely converge spatially and kinematically. The predicted path of arm~B as traced by \citet{2008ApJ...679..432N} cuts diagonally through the upper portion of the \hi~overdensity region, spanning $\sim 5$ degrees in Galactic longitude and $\sim 8$ degrees in Galactic latitude. The predicted path of arm~E spatially converges with arm~B around $(l,b) =(280\fdg5, -31\fdg2) $ and extends to Galactic longitudes of $279\fdg2\leq l \leq 280\fdg7 $.

\citet{2008ApJ...679..432N} determined that arm~B and arm~E diverge in different directions at higher Galactic longitudes because of their position relative to the disk. As the LMC is inclined with the 30~Doradus overdensity region on its leading edge, that area is targeted with the most direct headwinds as it moves through the Milky Way's halo (as first discussed in \citealp{1998A&A...329L..49D}). If supergiant shells near 30~Doradus are feeding the gas in these arms, the gas will experience more direct ram-pressure forces sweeping it to the far side. However, an opposing view by \citet{1994A&A...289..357D} suggests that the high-velocity gas is on the nearside of the LMC. They calculated the line-of-sight velocity gradients of gas clouds along various background targets in and behind the LMC and found that the low-velocity absorbers have~0 or negative velocity gradients placing them on the far side. In contrast, the high-velocity gas has positive velocity gradients positioning them in front of the disk. With two possible interpretations and the potential extension of the arms toward the 30~Doradus region, arm~B and arm~E could diverge at lower Galactic longitudes, creating complexity in determining how much of the arms are in front of and/or behind the LMC's disk. 

However, we caution that since the \hi~emission for the arms could extend toward the 30~Doradus region, there may be a blending of both arms' material near the center of 30~Doradus and we are not necessarily only detecting one arm along sightlines~1-4. We also reemphasize from our findings in Section~\ref{total_cd}, that sightlines~1 and~2 may be probing material unassociated with the remaining 6~sightlines. We found that the ionization conditions seem to change as the angular offset to 30~Doradus increases; sightlines~1 and~2, which are within $0\fdg25$ of 30~Doradus, have larger Al\textsc{~iii} total integrated column density values than their more distant counterparts. As 30~Doradus is a starburst region with stellar-driven galactic winds \citep{1984MNRAS.211..521M,2003MNRAS.344..741R,2021ApJ...908...62C,2025arXiv250305968P}, we expect the nearby sightlines to contain higher ionized species due to the heightened activity occurring in their proximity. Given the accelerated stellar activity near 30~Doradus and the potential crossing of the two arms, we cannot definitively determine which arms each sightline probes. We can, however, conclude that there is coherency between the \hi~emission of the arms and the velocity centroids of the UV absorbers, strengthening our argument that we are probing the same material in both datasets.

\subsection{Nature of the High-Velocity Gas}

Below, we discuss explanations for the nature of the redshifted material. One possibility is that the redshifted absorbers could be disk components moving at higher velocities. Our detection of C\textsc{~i} along 3~sightlines and S\textsc{~i} along 2~sightlines indicates that a phase of the gas is cold and dense, resembling the conditions of the disk. However, \citet{2018ApJ...865..145R} found one component of C\textsc{~i} in LA~II, demonstrating that cold, dense gas can be found beyond the extent of the disk. An absorption-line study of approximately 182~sightlines across the LMC found no signatures of redshifted multi-disk components for the majority of their targets \citep{poudel_2025}. We re-emphasize that 1) the redshifted absorbers are predominantly found in a localized area around the arms, 2) the kinematics of the offset gas and the arms align, and 3) the sightlines lie spatially near the arms. Therefore, we concentrate on two other explanations: outflows and inflows.

In an outflowing scenario, the \hi~overdensity region housing 30~Doradus could be funneling material into these arms through stellar activity that ejects material outside the disk. This dislodged material could then be swept away into a filament by a combination of ram-pressure and tidal forces. During inflows, this region may receive material from these arms as a fuel source for its stellar activity. The origin of these arms is an open area of debate, leaving many questions about the formation and chemical enrichment history of the Magellanic System. We cannot verify which of these theories is occurring or, more likely, whether both are simultaneously taking place. We offer that both outflows and inflows could be detected along our 8~sightlines through our Fe\textsc{~ii} and Ni\textsc{~ii} observations, since they are produced similarly. 

For the ions that we observed, both Fe\textsc{~ii} and Ne\textsc{~ii} are produced similarly. These elements further have similar 1st ($\Delta IP_{{\rm Fe}^+-{\rm Ne}^+} \approx 0.26\,\rm eV$) and 2nd ($\Delta IP_{{\rm Fe}^{2+}-{\rm Ne}^{2+}} \approx -2.0\,\rm eV$) ionization potentials and are depleted similarly onto dust grains. Therefore, their column density ratio should be sensitive to the relative production of these elements in the gas. If they followed solar abundance patterns, then this would yield $\left({\rm Ni}/{\rm H}\right)_{\odot}/\left({\rm Fe}/{\rm H}\right)_{\odot}\approx0.064$ when correcting for warm ionized medium like depletion \citep{2009ApJ...700.1299J}. Along sightline~4, we measured a $N_{\rm Ni\textsc{~ii}}/N_{\rm Fe\textsc{~ii}}=0.091$ for the component at $\vlsr\approx+287\,\kms$ and $0.029$ for the component at $\vlsr\approx+308\,\kms$, which are respectively a factor of ${\sim}1.4$ and ${\sim}0.45$ from solar. The differences in these two components could indicate that they originated from two different sources, such as outflows and inflows. 

However, we can explore whether or not gas on the nearside of the LMC is possible through a purely outflowing or inflowing scenario. We offer that future observations could provide additional constraints on the dynamics of these arms. For example, UV absorption-line metallicity measurements of the nearside gas could distinguish between gas stripped from the LMC and gas accreted from the intergalactic medium or the SMC. Furthermore, from the position-velocity map of sightline~4, our GASS and GASKAP observations may hint that the interior edge of the offset gas along this sightline is more compressed than its trailing components (see Figure~\ref{fig:pv}). Therefore, deeper \hi~emission mappings with the next generation Square Kilometre Array radio telescopes could reveal more small-scale structure to decipher whether the arms are dynamically coherent or disrupted by environmental forces. Furthermore, better angular resolution \hi~absorption-line observations could reveal a more complete orientation of the small-scale structure of the gaseous arms relative to the LMC's disk adding to the work of \citet{1994A&A...289..357D}.

\subsubsection{Outflowing Scenario}

Multiple star-forming regions throughout the LMC's disk are powering wide spread galactic winds that are apparent across the entire face of the LMC (e.g. \citealp{2007MNRAS.377..687L}). This wind has been characterized by multiple UV absorption-line (e.g. \citealp{2016ApJ...817...91B,2024ApJ...974...22Z,2025arXiv250305968P}), and radio and emission-line (e.g \citealp{2003MNRAS.344..741R, 2021ApJ...908...62C}) studies. The 30~Doradus starburst region has distorted the surrounding ISM as evident by the several supergiant \hi~shells \citep{1999AJ....118.2797K}. The extreme stellar activity in this region could have propelled material outside the disk, feeding a galactic wind \citep{1984MNRAS.211..521M,2003MNRAS.344..741R,2004A&A...423..895O,2021ApJ...908...62C,2025arXiv250305968P}. Arms~B and~E could be this outflowing material due to their connections to this starbursting region.

As further evidence for a wide-spread outflow, \citet{1993A&A...271..402K} compared the large-scale structure of the magnetic field in the LMC with its global gas kinematics. They find two linear polarized filaments near the southern region of 30~Doradus which link with high-density, low-velocity \hi~gas discussed in \citet{1992A&A...263...41L}. \citet{1993A&A...271..402K} conclude that polarized structures bend toward the Milky Way and away from the LMC suggesting that as the low-velocity gas got disrupted by environmental forces, the magnetic field got dragged along. The 3D structure of the magnetic field in the LMC points to an outflowing scenario where gas is swept away taking the field with it.

While there is evidence for outflowing gas across the entire LMC, it is unclear whether material is being swept into arm~B, and/or arm~E directly. To explore this possibility, \citet{2008ApJ...679..432N} characterized the kinematic and spatial patterns in the MS using a Gaussian decomposition technique on the \hi~emission from the LAB survey; They traced arm~B near the 30~Doradus region, specifically to supergiant shells 18 and 20, and arm~E to multiple supergiant shells in the 30~Doradus region (see Figures 21, 22, 23, 25, and 26 in \citealt{2008ApJ...679..432N}). The \citet{2008ApJ...679..432N} study found: (1) arm~E shares similar spatial and kinematic progressions as LA~I suggesting that the two are physically connected, and (2) the radial velocities of LA~I align more closely with an LMC origin. Additionally, \citet{2008ApJ...679..432N} confirmed that the MS contains two spatially and kinematically separate filaments that trace to two different regions of the MCs. The first has a higher metallicity (0.5 $Z_{\odot}$; \citealp{2013ApJ...772..111R}) than the present-day values of the SMC or the LMC ($\sim 0.3$ and $\sim 0.4$ $Z_{\odot}$, respectively; \citealp{1992ApJ...384..508R}). \citet{2013ApJ...772..111R} explain that Type~II supernovae from an active star formation region could have homogeneously enriched the gas before it was removed from the host galaxy. The second filament has a lower metallicity (0.1 $Z_{\odot}$; \citealp{2010ApJ...718.1046F,2013ApJ...772..110F}) and was kinematically tracked back to the SMC. 

These two filaments of the MS also have a sinusoidal pattern in both velocity and position space with an amplitude of ${\sim}2\degree$---similar to the angular offset of ${\sim}2.5\degree$ between the center of the LMC and the 30~Doradus \hi~overdensity region \citep{2008ApJ...679..432N}. The pattern could have been generated as the \hi~overdensity region rotated around the center of the LMC. A smoothed particle hydrodynamical simulation by \citet{2015ApJ...813..110H}---which included gas cooling, stellar feedback, and star formation---suggests that the oscillatory behavior of the two filaments is produced through a ``ram-pressure plus collision scenario"; the SMC passes by the LMC and then collides with it. They recreated the overall shape and extension of the MS along with two twisting tails, each originating from one of the MCs as expected as a result of ram pressure stripping. They suggest that gas was stripped during the collision between the SMC and LMC, and that supernovae explosions are crucial for expelling gas from the disk of the LMC.

Further, \citet{2018ApJ...863...49B} created a 3D simulation of clustered supernova outflows for the LMC to test whether material from thermally driven supernova explosions could be swept away by ram-pressure stripping to form the extended gaseous structures around the MCs. The simulated galaxy contained a starburst region that engaged in 10~successive outflow events spaced $60\,\rm Myr$ apart, ultimately forming a gaseous filament that trailed behind the galaxy. However, neither the \citet{2018ApJ...863...49B} or \citet{2015ApJ...813..110H} study included a corona around the MCs, which could hinder or even prohibit outflowing material from being swept into such structures. 

Recently, \citet{2022Natur.609..915K} found evidence of a warm corona surrounding the LMC. Therefore, for a thorough investigation into outflow-driven winds, we utilized a GIZMO hydrodynamic simulation of the LMC and Milky Way that incorporated a corona around the LMC and SMC (described in Section~\ref{simulation}; \citealt{2021ApJ...921L..36L}). We find that with the protective shield offered by the Magellanic Corona in our simulation, suppression of the outflows due to ram-pressure interactions with the MW's halo is reduced. The corona offers the most shielding on the leading edge, near where the 30~Doradus analog is located. Our findings are in agreement with the bow shock proposition suggested by \citet{2023ApJ...959L..11S} and further explored in \citet{2024ApJ...974...22Z}. \citet{2024ApJ...974...22Z} argues that while there is no direct observational evidence of a bow shock around the LMC, the galactic winds are not significantly impacted by ram-pressure stripping. A bow shock or a shielding halo could explain the source of protection for the winds as the LMC orbits the MW. However, despite this protection, we find the outflows can be ultimately swept away by tidal interactions, which results in both blueshifted and redshifted material on the nearside of the LMC's disk along the leading, middle, and trailing edge of the galaxy (see Figure~\ref{fig:los_vel}). Consequently, outflows from 30~Doradus are a reasonable explanation for the material in extended gaseous structures surrounding the LMC like the arms.

The results of our study indicate that outflows from the 30~Doradus region are a viable origin for the gaseous material of arms~E and~B. Given that our sightlines are within $2\fdg1$ of 30~Doradus, the offset gas likely contains material from this region. We note that our redshifted absorbers share similar Doppler parameters as the blueshifted galactic wind absorbers in \citet{2025arXiv250305968P}. To quantify the ionization conditions, we compared our average Fe\textsc{~ii} and Al\textsc{~iii} Doppler parameters to the values measured in \citet{2025arXiv250305968P}. In the case that sightlines~1 and~2 are probing material unassociated with sightlines~3--8 (the 2~sightlines closest to 30~Doradus appear separate from the remaining~6 when examining the total integrated column densities; see Section \ref{total_cd}), we divide the sightlines into those two groupings. For sightlines~1--2, we find average values of $\langle b_{Fe\textsc{~ii}}\rangle = 11.8 \pm 1.4~\kms$  and $\langle b_{Al\textsc{~iii}}\rangle = 20.8 \pm 1.9~\kms$. For sightlines~3--8, we report
$\langle b_{Fe\textsc{~ii}}\rangle = 11.3 \pm 4.5~\kms$ and $\langle b_{Al\textsc{~iii}}\rangle = 14.1 \pm 6.7~\kms$. \citet{2025arXiv250305968P} found $\langle b_{Fe\textsc{~ii}}\rangle  = 10.0 \pm 6.3~\kms$ and $\langle b_{Al\textsc{~iii}}\rangle  = 13.9 \pm 8.0~\kms$ along sightlines within $1\fdg7$ of 30~Doradus for the galactic wind material. As our average Doppler parameter values are within the uncertainty values of \citet{2025arXiv250305968P}, we conclude the material detected along each sightline could share similar ionization processes. With the additional property of coherency observed in the \hi~emission structure in the position-velocity maps, we conclude that the detected redshifted absorbers could be ejected debris from the 30~Doradus area shaped into a filamentary structure over an extended period of time by ram-pressure stripping and tidal forces in the environment.

\subsubsection{Inflowing Scenario}

The most straightforward interpretation of redshifted material beyond the disk of the LMC is inflowing material. Previous absorption-line studies have detected material at positive velocities beyond the LMC's disk \citep{2002ApJ...569..214H, 2009ApJ...702..940L,2024ApJ...974...22Z,2025arXiv250305968P}. \citet{2002ApJ...569..214H} used FUSE observations to detect redshifted $\rm{O\textsc{~vi}}$ absorbers along two of the twelve sightlines in their study. They conclude that highly-ionized inflows are uncommon across the LMC. \citet{2009ApJ...702..940L} examined the absorption profile of $\rm{Fe\textsc{~ii}}$ from FUSE observations and found only a small subset of their 139~sightlines show signatures of infalling gas. While they state there is no strong evidence of inflows, they found absorbers around $+50-70 ~\kms$ beyond the LMC's disk aligning kinematically similarly to where our components are measured. A more recent study by \citet{2024ApJ...974...22Z} also detected possible inflows through $\rm{Si\textsc{~iv}}$, $\rm{C\textsc{~iv}}$, and $\rm{S\textsc{~ii}}$ in regions of the LMC that are relatively quiet. They suggest that while inflows exist, they are weaker to detect or in areas not probed by DR5 of the ULLYSES program utilized in their paper. With DR6 of ULLYSES, \citet{2025arXiv250305968P} probed within $1\fdg7$ of 30~Doradus and detected inflowing low-ionization species
along three of their eight sightlines, suggesting that the inflows could be supplying material into that region. Additionally, an optical survey of diffuse interstellar bands (DIBs) detect infalling gas through the $\rm{Na\textsc{~i}~D}$ doublet \citep{2013A&A...550A.108V}. Using spectra of early type stars from the Very Large Telescope Flames Tarantula Survey, \citet{2013A&A...550A.108V} found that the area surrounding R136---a compact concentration of stars inside NGC2070 (e.g.\citealp{1991IAUS..148..145W,2018A&A...614A.147C})--- lacks gas except for a high-speed gas cloud beyond the disk of the LMC. Therefore, given the detection of redshifted material in both quiet and active regions around the LMC, it is plausible that arm~B and~E are infalling into this region.

Another observational piece of evidence supporting inflows is a group of AGB stars that have line-of-sight velocities counter-rotating to the LMC's disk \citep{2011ApJ...737...29O}. This AGB population has Ca~metallicity measurements similar to the abundance values of periphery SMC red giants as determined by \citet{2010ApJ...714L.249D}. The stars also kinematically overlap with the \hi~arms \citep{2011ApJ...737...29O}. These characteristics hint toward the AGB stars being tidally stripped from the SMC and then accreted onto the LMC. \citet{2011ApJ...737...29O} further surmise that the \hi~overdensity of gas surrounding the 30~Doradus region could have been formed as the LMC's disk and infalling gas interacted creating a shock which potentially elevated the star formation. A comparison, conducted by \citet{1998AJ....116.1275D}, between the 30~Doradus and Shapley Constellation~III star-forming region supports this inflowing situation. Both 30~Doradus and Constellation~III are star-forming regions in the LMC that formed from roughly the same amount of gas; However, 30~Doradus was much more efficient at creating stars and in a more densely concentrated area \citep{1998AJ....116.1275D}. \citeauthor{1998AJ....116.1275D} explains that the resulting stellar mass and spatial distribution of 30~Doradus is possible with an inflow depositing a compact gas reservoir in that area.

Furthermore, various studies have demonstrated through the age-metallicity relation of the LMC that younger star clusters, specifically formed in the last $\sim$2~Gyr, have on average a lower metallicity than their intermediate-age cluster counterparts (e.g. \citealp{2005A&A...442..597V,2009AJ....138.1243H,2012A&A...537A.106R}). With an inflowing scenario, relatively metal-poor gas from the SMC is accreted onto the LMC which then mixes with the ISM. Then, the gaseous material available to form new stars is diluted by the lower-metallicity inflows. Likely, the younger star clusters formed $\sim$2~Gyr ago are from a lower metallicity gas reservoir supplied by the inflowing SMC material.

While there are currently no abundance measurements near where arm~B and arm~E connect to the MS and LA~I, respectively, we include these following two metallcity arguments to emphasize the larger gaseous structures do support a scenario with inflowing gas from the SMC. \citet{2018ApJ...865..145R} investigated the abundances around LA~II and found the elemental abundance values almost identical to the stellar body of the SMC. \citet{2018ApJ...854..142F} measured the oxygen abundance values in LA~III and found they are comparable to the values measured in the MS filament that traces kinematically to the SMC (see \citealt{2008ApJ...679..432N}).

To verify that an inflowing scenario can reproduce the morphology, kinematics, and general characteristics of the entire MC system, \citet{2012ApJ...750...36D} performed an N-body simulation of the MCs using a multiple passage orbit of the MW. They found that with two close interactions between the MCs, around ${\sim}2$~Gyr ago and  ${\sim}250$~Myr ago, the LMC's tidal forces ripped the MS and LA material out from the SMC.  They demonstrate that stars stripped from the SMC's disk are incorporated into the LMC, providing support to the origin of the AGB stars found in \citet{2011ApJ...737...29O}. However, the simulations estimated the total amount of stars transferred to the LMC is comparable to the assumed initial mass of their SMC disk ($2.7 \times 10^{8} \, \rm{M\odot}$). This estimate is significantly larger than what was observed by \citet{2011ApJ...737...29O}.
This caveat can be explained by the SMC having a larger initial mass before any interactions (e.g., \citealt{2009MNRAS.395..342B}).

The results of our study could also imply an inflowing scenario. We note a simple interpretation is that these redshifted absorbers could be an inflowing filament created through tidal interactions between the LMC and SMC (e.g. \citealt{2011ApJ...737...29O}). The continuous nature of the \hi~emission in our position-velocity maps, offers support toward this theory. However, these high-velocity absorbers could also be associated with galactic fountains as our sightlines are very close to 30~Doradus and contain similar properties as the galactic wind material. \citet{1976ApJ...205..762S} proposed the ``galactic fountain" model where supernovae explosions will launch hot gas out of the galaxy's disk and then after reaching a certain height, the gas will cool down and cascade back onto the disk in the form of clouds that move at more intermediate and high-velocities. We note we could be detecting galactic fountain material given the results of our total integrated column density values for Al\textsc{~iii}. We find the values decrease with offset from 30~Doradus, hinting toward an origin near this region. Additionally, we find in Section~\ref{kinematic_distribution} that the sightlines closer to 30~Doradus do not have absorbers at a greater offset from the disk as we would assume near an active starburst region ejecting a violent outflow.

\section{Summary \& Conclusion}\label{conclusion}
In this project, we analyzed \hi~21-cm emission from GASS and GASKAP and UV absorption-line data from the HST's ULLYSES program across 8 stellar sightlines in the LMC to study high-velocity gas flowing through the 30~Doradus starburst region. We argue that the detected absorbers could be arm~B and/or arm~E given their spatial and kinematic proximity to the sightlines in our study.  We provide evidence that the neutral, low- and moderately-ionized gas in this work all share the same kinematics and is at least partially located on the nearside of the LMC, adding clarity to the 3D structure of the Magellanic System. We present the main results from our work below.

\begin{enumerate}
    \item {\bf Nearside Redshifted Absorbers:} There are redshifted absorbers beyond the extent of the LMC's disk which overlap kinematically with observed \hi~emission structures at positive velocities beyond the disk. The absorbers and \hi~emission kinematically correspond to the gaseous structures arm~B and arm~E. Given the physical placement of our background targets, we provide evidence that likely both of these arms, which trace to the MS and LA~I, are at least partially located on the nearside of the LMC. 
    
  \item {\bf Orientation of the Arms:} 
  With the results of the Gaussian decomposition of the GASKAP data, we suggest that arm~B and arm~E likely track further into the \hi~overdensity region near 30~Doradus than previously assumed. Arm~B and arm~E could cross and connect to gaseous material near a Galactic longitude of $l = 279\fdg0$, increasing their horizontal extent by $\sim 1\fdg0$.

  \item{\bf Inflows vs. Outflow Scenario:}
A natural explanation for the redshifted gas is that it is flowing away from us and toward the disk of the LMC. These inflows could funnel material into 30~Doradus fueling its star formation. However, this starburst region is generating intense outflows; through hydrodynamical simulations, we find that with a warm coronal around the LMC, outflows in this region could be protected out to $3~\kpc$. Interestingly, we find both redshifted and blueshifted nearside winds. Therefore, our results coincide with either scenario. Unfortunately, the saturation of the UV ion lines limit our ability to measure the chemical composition of the gas; this missing information could aid in resolving this origin debate.
    \item {\bf Kinematics, and Velocity Spread:} The offset gas consists of multiple phases, with three distinct ionization states existing within the same general kinematic space. We confirm a velocity gradient is present in the \hi~disk and do not observe sightlines closer to 30~Doradus containing absorbers that are more significantly offset from the disk.  The high-velocity absorbers, for neutral, and weakly-ionized species are likely both thermally and non-thermally broadened while the moderately-ionized gas is affected by more non-thermal broadening processes. We find average Doppler parameter values that align with the galactic wind results of \citet{2025arXiv250305968P}. We suggest the offset and galactic wind material could undergo similar ionization processes.

    \item{\bf{Electron Density and Cooling Rate:}} Under the assumption that the gas along the line-of-sight for our 8~sightlines is warm, we determined the electron density and cooling rate per nucleon using the C\textsc{~ii}$^{*}$ and S\textsc{~ii} absorption lines. We measure a lower limit for the electron density to $n_{e}\gtrsim 0.1~\rm{cm^{-3}}$. We find this value is larger than the body of the MS as expected given the starburst activity near our sightlines. The limit also is larger than expected for the diffuse ionized gas in the LMC. We also estimate the cooling rate to be $\rm{log}(L_{c}/ ergs~s^{-1}~nucleon^{-1})\gtrsim -25.8$, similar to the cooling rate for Milky Way ISM material. 

    \item {\bf Total Integrated Column Densities:} We observe a decrease in the total integrated column density values for the Al\textsc{~iii} components with an increase in angular offset from 30~Doradus. However, we note the two sightlines closest to 30~Doradus have smaller total integrated column density values for the singly-ionized species. We suggest there is more moderately ionized gas within 0.25\degree of 30~Doradus than in the regions beyond.

\end{enumerate}

With future spectroscopic surveys, higher-resolution observations of arm~B and arm~E can provide new insights into their formation and evolution. By combining these observations with high-resolution hydrodynamical simulations, we can better differentiate between various gas stripping mechanisms and enhance our understanding of the LMC's interaction history.

\clearpage

\acknowledgments
Support for this program was provided by NASA through the grant HST-AR-16602.001-A from the Space Telescope Science Institute, which is operated by the Association of Universities for Research in Astronomy, Incorporated, under NASA contract NAS5-26555. 
Horton received additional support provided by NSF grant 2334434. This material is based upon work supported by the National Science Foundation Graduate Research Fellowship Program under Grant No. 2334434. Any opinions, findings, and conclusions or recommendations expressed in this material are those of the author(s) and do not necessarily reflect the views of the National Science Foundation.
N.L. acknowledges support by NASA through grants HST-AR-16602 and HST-AR-17051 from the Space Telescope Science Institute. This project used ULLYSES observations obtained with the NASA/ESA Hubble Space Telescope \citep{2020RNAAS...4..205R}, retrieved from the Mikulski Archive for Space Telescopes (MAST: http://archive.stsci.edu) at the Space Telescope Science Institute (STScI). STScI is operated by the Association of Universities for Research in Astronomy, Inc. under NASA contract NAS 5-26555. 
This study used archived \hi~LMC data obtained through the Australia Telescope Online Archive (http://atoa.atnf.csiro.au) and the Galactic All Sky Survey Archive (http://www.atnf.csiro.au/research/HI4PI). 
\\
\software{
\textsc{Astropy} \citep{astropy:2013,astropy:2018,astropy:2022},
\textsc{VoigtFit} \citep{2018arXiv180301187K},
\textsc{GaussDecomp} \citep{Gauss_decomp_citation}
}

\bibliographystyle{aasjournal} 
\bibliography{References}

\begin{thebibliography}{}
\expandafter\ifx\csname natexlab\endcsname\relax\def\natexlab#1{#1}\fi
\providecommand{\url}[1]{\href{#1}{#1}}
\providecommand{\dodoi}[1]{doi:~\href{http://doi.org/#1}{\nolinkurl{#1}}}
\providecommand{\doeprint}[1]{\href{http://ascl.net/#1}{\nolinkurl{http://ascl.net/#1}}}
\providecommand{\doarXiv}[1]{\href{https://arxiv.org/abs/#1}{\nolinkurl{https://arxiv.org/abs/#1}}}

\bibitem[{{Asplund} {et~al.}(2021){Asplund}, {Amarsi}, \& {Grevesse}}]{2021A&A...653A.141A}
{Asplund}, M., {Amarsi}, A.~M., \& {Grevesse}, N. 2021, \aap, 653, A141, \dodoi{10.1051/0004-6361/202140445}

\bibitem[{{Astropy Collaboration} {et~al.}(2013){Astropy Collaboration}, {Robitaille}, {Tollerud}, {Greenfield}, {Droettboom}, {Bray}, {Aldcroft}, {Davis}, {Ginsburg}, {Price-Whelan}, {Kerzendorf}, {Conley}, {Crighton}, {Barbary}, {Muna}, {Ferguson}, {Grollier}, {Parikh}, {Nair}, {Unther}, {Deil}, {Woillez}, {Conseil}, {Kramer}, {Turner}, {Singer}, {Fox}, {Weaver}, {Zabalza}, {Edwards}, {Azalee Bostroem}, {Burke}, {Casey}, {Crawford}, {Dencheva}, {Ely}, {Jenness}, {Labrie}, {Lim}, {Pierfederici}, {Pontzen}, {Ptak}, {Refsdal}, {Servillat}, \& {Streicher}}]{astropy:2013}
{Astropy Collaboration}, {Robitaille}, T.~P., {Tollerud}, E.~J., {et~al.} 2013, \aap, 558, A33, \dodoi{10.1051/0004-6361/201322068}

\bibitem[{{Astropy Collaboration} {et~al.}(2018){Astropy Collaboration}, {Price-Whelan}, {Sip{\H{o}}cz}, {G{\"u}nther}, {Lim}, {Crawford}, {Conseil}, {Shupe}, {Craig}, {Dencheva}, {Ginsburg}, {Vand erPlas}, {Bradley}, {P{\'e}rez-Su{\'a}rez}, {de Val-Borro}, {Aldcroft}, {Cruz}, {Robitaille}, {Tollerud}, {Ardelean}, {Babej}, {Bach}, {Bachetti}, {Bakanov}, {Bamford}, {Barentsen}, {Barmby}, {Baumbach}, {Berry}, {Biscani}, {Boquien}, {Bostroem}, {Bouma}, {Brammer}, {Bray}, {Breytenbach}, {Buddelmeijer}, {Burke}, {Calderone}, {Cano Rodr{\'\i}guez}, {Cara}, {Cardoso}, {Cheedella}, {Copin}, {Corrales}, {Crichton}, {D'Avella}, {Deil}, {Depagne}, {Dietrich}, {Donath}, {Droettboom}, {Earl}, {Erben}, {Fabbro}, {Ferreira}, {Finethy}, {Fox}, {Garrison}, {Gibbons}, {Goldstein}, {Gommers}, {Greco}, {Greenfield}, {Groener}, {Grollier}, {Hagen}, {Hirst}, {Homeier}, {Horton}, {Hosseinzadeh}, {Hu}, {Hunkeler}, {Ivezi{\'c}}, {Jain}, {Jenness}, {Kanarek}, {Kendrew}, {Kern}, {Kerzendorf}, {Khvalko}, {King}, {Kirkby}, {Kulkarni},
  {Kumar}, {Lee}, {Lenz}, {Littlefair}, {Ma}, {Macleod}, {Mastropietro}, {McCully}, {Montagnac}, {Morris}, {Mueller}, {Mumford}, {Muna}, {Murphy}, {Nelson}, {Nguyen}, {Ninan}, {N{\"o}the}, {Ogaz}, {Oh}, {Parejko}, {Parley}, {Pascual}, {Patil}, {Patil}, {Plunkett}, {Prochaska}, {Rastogi}, {Reddy Janga}, {Sabater}, {Sakurikar}, {Seifert}, {Sherbert}, {Sherwood-Taylor}, {Shih}, {Sick}, {Silbiger}, {Singanamalla}, {Singer}, {Sladen}, {Sooley}, {Sornarajah}, {Streicher}, {Teuben}, {Thomas}, {Tremblay}, {Turner}, {Terr{\'o}n}, {van Kerkwijk}, {de la Vega}, {Watkins}, {Weaver}, {Whitmore}, {Woillez}, {Zabalza}, \& {Astropy Contributors}}]{astropy:2018}
{Astropy Collaboration}, {Price-Whelan}, A.~M., {Sip{\H{o}}cz}, B.~M., {et~al.} 2018, \aj, 156, 123, \dodoi{10.3847/1538-3881/aabc4f}

\bibitem[{{Astropy Collaboration} {et~al.}(2022){Astropy Collaboration}, {Price-Whelan}, {Lim}, {Earl}, {Starkman}, {Bradley}, {Shupe}, {Patil}, {Corrales}, {Brasseur}, {N{"o}the}, {Donath}, {Tollerud}, {Morris}, {Ginsburg}, {Vaher}, {Weaver}, {Tocknell}, {Jamieson}, {van Kerkwijk}, {Robitaille}, {Merry}, {Bachetti}, {G{"u}nther}, {Aldcroft}, {Alvarado-Montes}, {Archibald}, {B{'o}di}, {Bapat}, {Barentsen}, {Baz{'a}n}, {Biswas}, {Boquien}, {Burke}, {Cara}, {Cara}, {Conroy}, {Conseil}, {Craig}, {Cross}, {Cruz}, {D'Eugenio}, {Dencheva}, {Devillepoix}, {Dietrich}, {Eigenbrot}, {Erben}, {Ferreira}, {Foreman-Mackey}, {Fox}, {Freij}, {Garg}, {Geda}, {Glattly}, {Gondhalekar}, {Gordon}, {Grant}, {Greenfield}, {Groener}, {Guest}, {Gurovich}, {Handberg}, {Hart}, {Hatfield-Dodds}, {Homeier}, {Hosseinzadeh}, {Jenness}, {Jones}, {Joseph}, {Kalmbach}, {Karamehmetoglu}, {Ka{l}uszy{'n}ski}, {Kelley}, {Kern}, {Kerzendorf}, {Koch}, {Kulumani}, {Lee}, {Ly}, {Ma}, {MacBride}, {Maljaars}, {Muna}, {Murphy}, {Norman}, {O'Steen},
  {Oman}, {Pacifici}, {Pascual}, {Pascual-Granado}, {Patil}, {Perren}, {Pickering}, {Rastogi}, {Roulston}, {Ryan}, {Rykoff}, {Sabater}, {Sakurikar}, {Salgado}, {Sanghi}, {Saunders}, {Savchenko}, {Schwardt}, {Seifert-Eckert}, {Shih}, {Jain}, {Shukla}, {Sick}, {Simpson}, {Singanamalla}, {Singer}, {Singhal}, {Sinha}, {Sip{H{o}}cz}, {Spitler}, {Stansby}, {Streicher}, {{{S}}umak}, {Swinbank}, {Taranu}, {Tewary}, {Tremblay}, {Val-Borro}, {Van Kooten}, {Vasovi{'c}}, {Verma}, {de Miranda Cardoso}, {Williams}, {Wilson}, {Winkel}, {Wood-Vasey}, {Xue}, {Yoachim}, {Zhang}, {Zonca}, \& {Astropy Project Contributors}}]{astropy:2022}
{Astropy Collaboration}, {Price-Whelan}, A.~M., {Lim}, P.~L., {et~al.} 2022, \apj, 935, 167, \dodoi{10.3847/1538-4357/ac7c74}

\bibitem[{{Barger} {et~al.}(2016){Barger}, {Lehner}, \& {Howk}}]{2016ApJ...817...91B}
{Barger}, K.~A., {Lehner}, N., \& {Howk}, J.~C. 2016, \apj, 817, 91, \dodoi{10.3847/0004-637X/817/2/91}

\bibitem[{{Bekki} \& {Stanimirovi{\'c}}(2009)}]{2009MNRAS.395..342B}
{Bekki}, K., \& {Stanimirovi{\'c}}, S. 2009, \mnras, 395, 342, \dodoi{10.1111/j.1365-2966.2009.14514.x}

\bibitem[{{Besla} {et~al.}(2012){Besla}, {Kallivayalil}, {Hernquist}, {van der Marel}, {Cox}, \& {Kere{\v{s}}}}]{2012MNRAS.421.2109B}
{Besla}, G., {Kallivayalil}, N., {Hernquist}, L., {et~al.} 2012, \mnras, 421, 2109, \dodoi{10.1111/j.1365-2966.2012.20466.x}

\bibitem[{{Bestenlehner} {et~al.}(2020){Bestenlehner}, {Crowther}, {Caballero-Nieves}, {Schneider}, {Sim{\'o}n-D{\'\i}az}, {Brands}, {de Koter}, {Gr{\"a}fener}, {Herrero}, {Langer}, {Lennon}, {Ma{\'\i}z Apell{\'a}niz}, {Puls}, \& {Vink}}]{2020MNRAS.499.1918B}
{Bestenlehner}, J.~M., {Crowther}, P.~A., {Caballero-Nieves}, S.~M., {et~al.} 2020, \mnras, 499, 1918, \dodoi{10.1093/mnras/staa2801}

\bibitem[{{Brandl} {et~al.}(1996){Brandl}, {Sams}, {Bertoldi}, {Eckart}, {Genzel}, {Drapatz}, {Hofmann}, {Loewe}, \& {Quirrenbach}}]{1996ApJ...466..254B}
{Brandl}, B., {Sams}, B.~J., {Bertoldi}, F., {et~al.} 1996, \apj, 466, 254, \dodoi{10.1086/177507}

\bibitem[{{Br{\"u}ns} {et~al.}(2005){Br{\"u}ns}, {Kerp}, {Staveley-Smith}, {Mebold}, {Putman}, {Haynes}, {Kalberla}, {Muller}, \& {Filipovic}}]{2005A&A...432...45B}
{Br{\"u}ns}, C., {Kerp}, J., {Staveley-Smith}, L., {et~al.} 2005, \aap, 432, 45, \dodoi{10.1051/0004-6361:20040321}

\bibitem[{{Bustard} {et~al.}(2018){Bustard}, {Pardy}, {D'Onghia}, {Zweibel}, \& {Gallagher}}]{2018ApJ...863...49B}
{Bustard}, C., {Pardy}, S.~A., {D'Onghia}, E., {Zweibel}, E.~G., \& {Gallagher}, J.~S., I. 2018, \apj, 863, 49, \dodoi{10.3847/1538-4357/aad08f}

\bibitem[{{Cashman} {et~al.}(2017){Cashman}, {Kulkarni}, {Kisielius}, {Ferland}, \& {Bogdanovich}}]{2017ApJS..230....8C}
{Cashman}, F.~H., {Kulkarni}, V.~P., {Kisielius}, R., {Ferland}, G.~J., \& {Bogdanovich}, P. 2017, \apjs, 230, 8, \dodoi{10.3847/1538-4365/aa6d84}

\bibitem[{{Castro} {et~al.}(2018){Castro}, {Crowther}, {Evans}, {Mackey}, {Castro-Rodriguez}, {Vink}, {Melnick}, \& {Selman}}]{2018A&A...614A.147C}
{Castro}, N., {Crowther}, P.~A., {Evans}, C.~J., {et~al.} 2018, \aap, 614, A147, \dodoi{10.1051/0004-6361/201732084}

\bibitem[{{Choi} {et~al.}(2018){Choi}, {Nidever}, {Olsen}, {Blum}, {Besla}, {Zaritsky}, {van der Marel}, {Bell}, {Gallart}, {Cioni}, {Johnson}, {Vivas}, {Saha}, {de Boer}, {No{\"e}l}, {Monachesi}, {Massana}, {Conn}, {Martinez-Delgado}, {Mu{\~n}oz}, \& {Stringfellow}}]{2018ApJ...866...90C}
{Choi}, Y., {Nidever}, D.~L., {Olsen}, K., {et~al.} 2018, \apj, 866, 90, \dodoi{10.3847/1538-4357/aae083}

\bibitem[{{Ciampa} {et~al.}(2021){Ciampa}, {Barger}, {Lehner}, {Horn}, {Hernandez}, {Haffner}, {Smart}, {Bustard}, {Barber}, \& {Boot}}]{2021ApJ...908...62C}
{Ciampa}, D.~A., {Barger}, K.~A., {Lehner}, N., {et~al.} 2021, \apj, 908, 62, \dodoi{10.3847/1538-4357/abd320}

\bibitem[{{Cignoni} {et~al.}(2015){Cignoni}, {Sabbi}, {van der Marel}, {Tosi}, {Zaritsky}, {Anderson}, {Lennon}, {Aloisi}, {de Marchi}, {Gouliermis}, {Grebel}, {Smith}, \& {Zeidler}}]{2015ApJ...811...76C}
{Cignoni}, M., {Sabbi}, E., {van der Marel}, R.~P., {et~al.} 2015, \apj, 811, 76, \dodoi{10.1088/0004-637X/811/2/76}

\bibitem[{{Crowther} {et~al.}(2016){Crowther}, {Caballero-Nieves}, {Bostroem}, {Ma{\'\i}z Apell{\'a}niz}, {Schneider}, {Walborn}, {Angus}, {Brott}, {Bonanos}, {de Koter}, {de Mink}, {Evans}, {Gr{\"a}fener}, {Herrero}, {Howarth}, {Langer}, {Lennon}, {Puls}, {Sana}, \& {Vink}}]{2016MNRAS.458..624C}
{Crowther}, P.~A., {Caballero-Nieves}, S.~M., {Bostroem}, K.~A., {et~al.} 2016, \mnras, 458, 624, \dodoi{10.1093/mnras/stw273}

\bibitem[{{de Boer} {et~al.}(1998){de Boer}, {Braun}, {Vallenari}, \& {Mebold}}]{1998A&A...329L..49D}
{de Boer}, K.~S., {Braun}, J.~M., {Vallenari}, A., \& {Mebold}, U. 1998, \aap, 329, L49, \dodoi{10.48550/arXiv.astro-ph/9711052}

\bibitem[{{de Grijs} {et~al.}(2014){de Grijs}, {Wicker}, \& {Bono}}]{2014AJ....147..122D}
{de Grijs}, R., {Wicker}, J.~E., \& {Bono}, G. 2014, \aj, 147, 122, \dodoi{10.1088/0004-6256/147/5/122}

\bibitem[{{De Propris} {et~al.}(2010){De Propris}, {Rich}, {Mallery}, \& {Howard}}]{2010ApJ...714L.249D}
{De Propris}, R., {Rich}, R.~M., {Mallery}, R.~C., \& {Howard}, C.~D. 2010, \apjl, 714, L249, \dodoi{10.1088/2041-8205/714/2/L249}

\bibitem[{{Diaz} \& {Bekki}(2012)}]{2012ApJ...750...36D}
{Diaz}, J.~D., \& {Bekki}, K. 2012, \apj, 750, 36, \dodoi{10.1088/0004-637X/750/1/36}

\bibitem[{{Dickey} {et~al.}(1994){Dickey}, {Mebold}, {Marx}, {Amy}, {Haynes}, \& {Wilson}}]{1994A&A...289..357D}
{Dickey}, J.~M., {Mebold}, U., {Marx}, M., {et~al.} 1994, \aap, 289, 357

\bibitem[{{Dickey} {et~al.}(2013){Dickey}, {McClure-Griffiths}, {Gibson}, {G{\'o}mez}, {Imai}, {Jones}, {Stanimirovi{\'c}}, {Van Loon}, {Walsh}, {Alberdi}, {Anglada}, {Uscanga}, {Arce}, {Bailey}, {Begum}, {Wakker}, {Bekhti}, {Kalberla}, {Winkel}, {Bekki}, {For}, {Staveley-Smith}, {Westmeier}, {Burton}, {Cunningham}, {Dawson}, {Ellingsen}, {Diamond}, {Green}, {Hill}, {Koribalski}, {McConnell}, {Rathborne}, {Voronkov}, {Douglas}, {English}, {Ford}, {Lockman}, {Foster}, {Gomez}, {Green}, {Bland-Hawthorn}, {Gulyaev}, {Hoare}, {Joncas}, {Kang}, {Kerton}, {Koo}, {Leahy}, {Lo}, {Migenes}, {Nakashima}, {Zhang}, {Nidever}, {Peek}, {Tafoya}, {Tian}, \& {Wu}}]{2013PASA...30....3D}
{Dickey}, J.~M., {McClure-Griffiths}, N., {Gibson}, S.~J., {et~al.} 2013, \pasa, 30, e003, \dodoi{10.1017/pasa.2012.003}

\bibitem[{{Dolphin} \& {Hunter}(1998)}]{1998AJ....116.1275D}
{Dolphin}, A.~E., \& {Hunter}, D.~A. 1998, \aj, 116, 1275, \dodoi{10.1086/300493}

\bibitem[{{Evans} {et~al.}(2015){Evans}, {Kennedy}, {Dufton}, {Howarth}, {Walborn}, {Markova}, {Clark}, {de Mink}, {de Koter}, {Dunstall}, {H{\'e}nault-Brunet}, {Ma{\'\i}z Apell{\'a}niz}, {McEvoy}, {Sana}, {Sim{\'o}n-D{\'\i}az}, {Taylor}, \& {Vink}}]{2015AA...574A..13E}
{Evans}, C.~J., {Kennedy}, M.~B., {Dufton}, P.~L., {et~al.} 2015, \aap, 574, A13, \dodoi{10.1051/0004-6361/201424414}

\bibitem[{{Fitzpatrick}(1991)}]{1991PASP..103.1123F}
{Fitzpatrick}, E.~L. 1991, \pasp, 103, 1123, \dodoi{10.1086/132934}

\bibitem[{{For} {et~al.}(2013){For}, {Staveley-Smith}, \& {McClure-Griffiths}}]{2013ApJ...764...74F}
{For}, B.-Q., {Staveley-Smith}, L., \& {McClure-Griffiths}, N.~M. 2013, \apj, 764, 74, \dodoi{10.1088/0004-637X/764/1/74}

\bibitem[{{Fox} {et~al.}(2020){Fox}, {Frazer}, {Bland-Hawthorn}, {Wakker}, {Barger}, \& {Richter}}]{2020ApJ...897...23F}
{Fox}, A.~J., {Frazer}, E.~M., {Bland-Hawthorn}, J., {et~al.} 2020, \apj, 897, 23, \dodoi{10.3847/1538-4357/ab92a3}

\bibitem[{{Fox} {et~al.}(2013){Fox}, {Richter}, {Wakker}, {Lehner}, {Howk}, {Ben Bekhti}, {Bland-Hawthorn}, \& {Lucas}}]{2013ApJ...772..110F}
{Fox}, A.~J., {Richter}, P., {Wakker}, B.~P., {et~al.} 2013, \apj, 772, 110, \dodoi{10.1088/0004-637X/772/2/110}

\bibitem[{{Fox} {et~al.}(2010){Fox}, {Wakker}, {Smoker}, {Richter}, {Savage}, \& {Sembach}}]{2010ApJ...718.1046F}
{Fox}, A.~J., {Wakker}, B.~P., {Smoker}, J.~V., {et~al.} 2010, \apj, 718, 1046, \dodoi{10.1088/0004-637X/718/2/1046}

\bibitem[{{Fox} {et~al.}(2014){Fox}, {Wakker}, {Barger}, {Hernandez}, {Richter}, {Lehner}, {Bland-Hawthorn}, {Charlton}, {Westmeier}, {Thom}, {Tumlinson}, {Misawa}, {Howk}, {Haffner}, {Ely}, {Rodriguez-Hidalgo}, \& {Kumari}}]{2014ApJ...787..147F}
{Fox}, A.~J., {Wakker}, B.~P., {Barger}, K.~A., {et~al.} 2014, \apj, 787, 147, \dodoi{10.1088/0004-637X/787/2/147}

\bibitem[{{Fox} {et~al.}(2018){Fox}, {Barger}, {Wakker}, {Richter}, {Antwi-Danso}, {Casetti-Dinescu}, {Howk}, {Lehner}, {D'Onghia}, {Crowther}, \& {Lockman}}]{2018ApJ...854..142F}
{Fox}, A.~J., {Barger}, K.~A., {Wakker}, B.~P., {et~al.} 2018, \apj, 854, 142, \dodoi{10.3847/1538-4357/aaa9bb}

\bibitem[{{Gardiner} {et~al.}(1998){Gardiner}, {Turfus}, \& {Putman}}]{1998ApJ...507L..35G}
{Gardiner}, L.~T., {Turfus}, C., \& {Putman}, M.~E. 1998, \apjl, 507, L35, \dodoi{10.1086/311668}

\bibitem[{{Gnat} \& {Sternberg}(2007)}]{2007ApJS..168..213G}
{Gnat}, O., \& {Sternberg}, A. 2007, \apjs, 168, 213, \dodoi{10.1086/509786}

\bibitem[{{Grand} {et~al.}(2017){Grand}, {G{\'o}mez}, {Marinacci}, {Pakmor}, {Springel}, {Campbell}, {Frenk}, {Jenkins}, \& {White}}]{2017MNRAS.467..179G}
{Grand}, R. J.~J., {G{\'o}mez}, F.~A., {Marinacci}, F., {et~al.} 2017, \mnras, 467, 179, \dodoi{10.1093/mnras/stx071}

\bibitem[{{Green} {et~al.}(2012){Green}, {Froning}, {Osterman}, {Ebbets}, {Heap}, {Leitherer}, {Linsky}, {Savage}, {Sembach}, {Shull}, {Siegmund}, {Snow}, {Spencer}, {Stern}, {Stocke}, {Welsh}, {B{\'e}land}, {Burgh}, {Danforth}, {France}, {Keeney}, {McPhate}, {Penton}, {Andrews}, {Brownsberger}, {Morse}, \& {Wilkinson}}]{2012ApJ...744...60G}
{Green}, J.~C., {Froning}, C.~S., {Osterman}, S., {et~al.} 2012, \apj, 744, 60, \dodoi{10.1088/0004-637X/744/1/6010.1086/141956}

\bibitem[{{Guzman} {et~al.}(2019){Guzman}, {Whiting}, {Voronkov}, {Mitchell}, {Ord}, {Collins}, {Marquarding}, {Lahur}, {Maher}, {Van Diepen}, {Bannister}, {Wu}, {Lenc}, {Khoo}, \& {Bastholm}}]{2019ascl.soft12003G}
{Guzman}, J., {Whiting}, M., {Voronkov}, M., {et~al.} 2019, {ASKAPsoft: ASKAP science data processor software}, Astrophysics Source Code Library, record ascl:1912.003

\bibitem[{{Haffner} {et~al.}(1999){Haffner}, {Reynolds}, \& {Tufte}}]{1999ApJ...523..223H}
{Haffner}, L.~M., {Reynolds}, R.~J., \& {Tufte}, S.~L. 1999, \apj, 523, 223, \dodoi{10.1086/307734}

\bibitem[{{Hammer} {et~al.}(2015){Hammer}, {Yang}, {Flores}, {Puech}, \& {Fouquet}}]{2015ApJ...813..110H}
{Hammer}, F., {Yang}, Y.~B., {Flores}, H., {Puech}, M., \& {Fouquet}, S. 2015, \apj, 813, 110, \dodoi{10.1088/0004-637X/813/2/110}

\bibitem[{{Harris} \& {Zaritsky}(2009)}]{2009AJ....138.1243H}
{Harris}, J., \& {Zaritsky}, D. 2009, \aj, 138, 1243, \dodoi{10.1088/0004-6256/138/5/1243}

\bibitem[{{Haud}(2000)}]{2000A&A...364...83H}
{Haud}, U. 2000, \aap, 364, 83

\bibitem[{{Hayes} \& {Nussbaumer}(1984)}]{1984A&A...134..193H}
{Hayes}, M.~A., \& {Nussbaumer}, H. 1984, \aap, 134, 193

\bibitem[{{Hernquist}(1990)}]{1990ApJ...356..359H}
{Hernquist}, L. 1990, \apj, 356, 359, \dodoi{10.1086/168845}

\bibitem[{{Holland}(2014)}]{2014cosi.book....6H}
{Holland}, S.~T. 2014, {COS Instrument Handbook v. 6.0}

\bibitem[{{Hopkins}(2015)}]{2015MNRAS.450...53H}
{Hopkins}, P.~F. 2015, \mnras, 450, 53, \dodoi{10.1093/mnras/stv195}

\bibitem[{{Howk} {et~al.}(2002){Howk}, {Sembach}, {Savage}, {Massa}, {Friedman}, \& {Fullerton}}]{2002ApJ...569..214H}
{Howk}, J.~C., {Sembach}, K.~R., {Savage}, B.~D., {et~al.} 2002, \apj, 569, 214, \dodoi{10.1086/339322}

\bibitem[{{Hyland} {et~al.}(1992){Hyland}, {Straw}, {Jones}, \& {Gatley}}]{1992MNRAS.257..391H}
{Hyland}, A.~R., {Straw}, S., {Jones}, T.~J., \& {Gatley}, I. 1992, \mnras, 257, 391, \dodoi{10.1093/mnras/257.3.391}

\bibitem[{{Jenkins}(2009)}]{2009ApJ...700.1299J}
{Jenkins}, E.~B. 2009, \apj, 700, 1299, \dodoi{10.1088/0004-637X/700/2/1299}

\bibitem[{{Jenkins} {et~al.}(2005){Jenkins}, {Bowen}, {Tripp}, \& {Sembach}}]{2005ApJ...623..767J}
{Jenkins}, E.~B., {Bowen}, D.~V., {Tripp}, T.~M., \& {Sembach}, K.~R. 2005, \apj, 623, 767, \dodoi{10.1086/428878}

\bibitem[{{Kim} {et~al.}(1999){Kim}, {Dopita}, {Staveley-Smith}, \& {Bessell}}]{1999AJ....118.2797K}
{Kim}, S., {Dopita}, M.~A., {Staveley-Smith}, L., \& {Bessell}, M.~S. 1999, \aj, 118, 2797, \dodoi{10.1086/301116}

\bibitem[{{Kim} {et~al.}(1998){Kim}, {Staveley-Smith}, {Dopita}, {Freeman}, {Sault}, {Kesteven}, \& {McConnell}}]{1998ApJ...503..674K}
{Kim}, S., {Staveley-Smith}, L., {Dopita}, M.~A., {et~al.} 1998, \apj, 503, 674, \dodoi{10.1086/306030}

\bibitem[{{Kim} {et~al.}(2003){Kim}, {Staveley-Smith}, {Dopita}, {Sault}, {Freeman}, {Lee}, \& {Chu}}]{2003ApJS..148..473K}
---. 2003, \apjs, 148, 473, \dodoi{10.1086/376980}

\bibitem[{{Klein} {et~al.}(1993){Klein}, {Haynes}, {Wielebinski}, \& {Meinert}}]{1993A&A...271..402K}
{Klein}, U., {Haynes}, R.~F., {Wielebinski}, R., \& {Meinert}, D. 1993, \aap, 271, 402

\bibitem[{{Krishnarao} {et~al.}(2022){Krishnarao}, {Fox}, {D'Onghia}, {Wakker}, {Cashman}, {Howk}, {Lucchini}, {French}, \& {Lehner}}]{2022Natur.609..915K}
{Krishnarao}, D., {Fox}, A.~J., {D'Onghia}, E., {et~al.} 2022, \nat, 609, 915, \dodoi{10.1038/s41586-022-05090-5}

\bibitem[{{Krogager}(2018)}]{2018arXiv180301187K}
{Krogager}, J.-K. 2018, arXiv e-prints, arXiv:1803.01187, \dodoi{10.48550/arXiv.1803.01187}

\bibitem[{{Lehner} \& {Howk}(2007)}]{2007MNRAS.377..687L}
{Lehner}, N., \& {Howk}, J.~C. 2007, \mnras, 377, 687, \dodoi{10.1111/j.1365-2966.2007.11631.x}

\bibitem[{{Lehner} {et~al.}(2009){Lehner}, {Staveley-Smith}, \& {Howk}}]{2009ApJ...702..940L}
{Lehner}, N., {Staveley-Smith}, L., \& {Howk}, J.~C. 2009, \apj, 702, 940, \dodoi{10.1088/0004-637X/702/2/940}

\bibitem[{{Lehner} {et~al.}(2004){Lehner}, {Wakker}, \& {Savage}}]{2004ApJ...615..767L}
{Lehner}, N., {Wakker}, B.~P., \& {Savage}, B.~D. 2004, \apj, 615, 767, \dodoi{10.1086/424730}

\bibitem[{{Lucchini} {et~al.}(2021){Lucchini}, {D'Onghia}, \& {Fox}}]{2021ApJ...921L..36L}
{Lucchini}, S., {D'Onghia}, E., \& {Fox}, A.~J. 2021, \apjl, 921, L36, \dodoi{10.3847/2041-8213/ac3338}

\bibitem[{{Lucchini} {et~al.}(2020){Lucchini}, {D'Onghia}, {Fox}, {Bustard}, {Bland-Hawthorn}, \& {Zweibel}}]{2020Natur.585..203L}
{Lucchini}, S., {D'Onghia}, E., {Fox}, A.~J., {et~al.} 2020, \nat, 585, 203, \dodoi{10.1038/s41586-020-2663-4}

\bibitem[{{Luks} \& {Rohlfs}(1992)}]{1992A&A...263...41L}
{Luks}, T., \& {Rohlfs}, K. 1992, \aap, 263, 41

\bibitem[{{Massey} {et~al.}(2000){Massey}, {Waterhouse}, \& {DeGioia-Eastwood}}]{2000AJ....119.2214M}
{Massey}, P., {Waterhouse}, E., \& {DeGioia-Eastwood}, K. 2000, \aj, 119, 2214, \dodoi{10.1086/301345}

\bibitem[{{McClure-Griffiths} {et~al.}(2009){McClure-Griffiths}, {Pisano}, {Calabretta}, {Ford}, {Lockman}, {Staveley-Smith}, {Kalberla}, {Bailin}, {Dedes}, {Janowiecki}, {Gibson}, {Murphy}, {Nakanishi}, \& {Newton-McGee}}]{2009ApJS..181..398M}
{McClure-Griffiths}, N.~M., {Pisano}, D.~J., {Calabretta}, M.~R., {et~al.} 2009, \apjs, 181, 398, \dodoi{10.1088/0067-0049/181/2/398}

\bibitem[{{McGee} \& {Newton}(1986)}]{1986PASA....6..471M}
{McGee}, R.~X., \& {Newton}, L.~M. 1986, \pasa, 6, 471, \dodoi{10.1017/S1323358000018415}

\bibitem[{{Meaburn}(1984)}]{1984MNRAS.211..521M}
{Meaburn}, J. 1984, \mnras, 211, 521, \dodoi{10.1093/mnras/211.3.521}

\bibitem[{{Medallon} \& {Welty}(2023)}]{2023stii.book...22M}
{Medallon}, S., \& {Welty}, D. 2023, {STIS Instrument Handbook for Cycle 31 v. 22.0}

\bibitem[{{Mishra} {et~al.}(2024){Mishra}, {Fox}, {Krishnarao}, {Lucchini}, {D'Onghia}, {Cashman}, {Barger}, {Lehner}, \& {Tumlinson}}]{2024ApJ...976L..28M}
{Mishra}, S., {Fox}, A.~J., {Krishnarao}, D., {et~al.} 2024, \apjl, 976, L28, \dodoi{10.3847/2041-8213/ad8b9d}

\bibitem[{{Nidever} \& {Chwalik}(2022)}]{Gauss_decomp_citation}
{Nidever}, D.~L., \& {Chwalik}, E. 2022, {GaussDecomp}, 1.0.1, \dodoi{10.5281/zenodo.227955549}

\bibitem[{{Nidever} {et~al.}(2008){Nidever}, {Majewski}, \& {Butler Burton}}]{2008ApJ...679..432N}
{Nidever}, D.~L., {Majewski}, S.~R., \& {Butler Burton}, W. 2008, \apj, 679, 432, \dodoi{10.1086/587042}

\bibitem[{{Niemela} \& {Gamen}(2004)}]{2004NewAR..48..727N}
{Niemela}, V., \& {Gamen}, R. 2004, \nar, 48, 727, \dodoi{10.1016/j.newar.2004.03.008}

\bibitem[{{Olano}(2004)}]{2004A&A...423..895O}
{Olano}, C.~A. 2004, \aap, 423, 895, \dodoi{10.1051/0004-6361:20040177}

\bibitem[{{Olsen} {et~al.}(2011){Olsen}, {Zaritsky}, {Blum}, {Boyer}, \& {Gordon}}]{2011ApJ...737...29O}
{Olsen}, K. A.~G., {Zaritsky}, D., {Blum}, R.~D., {Boyer}, M.~L., \& {Gordon}, K.~D. 2011, \apj, 737, 29, \dodoi{10.1088/0004-637X/737/1/29}

\bibitem[{{Paradis} {et~al.}(2011){Paradis}, {Paladini}, {Noriega-Crespo}, {Lagache}, {Kawamura}, {Onishi}, \& {Fukui}}]{2011ApJ...735....6P}
{Paradis}, D., {Paladini}, R., {Noriega-Crespo}, A., {et~al.} 2011, \apj, 735, 6, \dodoi{10.1088/0004-637X/735/1/6}

\bibitem[{{Pardy} {et~al.}(2018){Pardy}, {D'Onghia}, \& {Fox}}]{2018ApJ...857..101P}
{Pardy}, S.~A., {D'Onghia}, E., \& {Fox}, A.~J. 2018, \apj, 857, 101, \dodoi{10.3847/1538-4357/aab95b}

\bibitem[{{Pietrzy{\'n}ski} {et~al.}(2013){Pietrzy{\'n}ski}, {Gieren}, {Graczyk}, {Thompson}, {Pilecki}, {Nardetto}, {Kudritzki}, {Bresolin}, {Bono}, {Moroni}, {Konorski}, {Gorski}, {Storm}, {Smolec}, \& {Karczmarek}}]{2013IAUS..289..169P}
{Pietrzy{\'n}ski}, G., {Gieren}, W., {Graczyk}, D., {et~al.} 2013, in Advancing the Physics of Cosmic Distances, ed. R.~{de Grijs}, Vol. 289, 169--172, \dodoi{10.1017/S174392131202131X}

\bibitem[{{Pingel} {et~al.}(2022){Pingel}, {Dempsey}, {McClure-Griffiths}, {Dickey}, {Jameson}, {Arce}, {Anglada}, {Bland-Hawthorn}, {Breen}, {Buckland-Willis}, {Clark}, {Dawson}, {D{\'e}nes}, {Di Teodoro}, {For}, {Foster}, {G{\'o}mez}, {Imai}, {Joncas}, {Kim}, {Lee}, {Lynn}, {Leahy}, {Ma}, {Marchal}, {McConnell}, {Miville-Desch{\`e}nes}, {Moss}, {Murray}, {Nidever}, {Peek}, {Stanimirovi{\'c}}, {Staveley-Smith}, {Tepper-Garcia}, {Tremblay}, {Uscanga}, {van Loon}, {V{\'a}zquez-Semadeni}, {Allison}, {Anderson}, {Ball}, {Bell}, {Bock}, {Bunton}, {Cooray}, {Cornwell}, {Koribalski}, {Gupta}, {Hayman}, {Harvey-Smith}, {Lee-Waddell}, {Ng}, {Phillips}, {Voronkov}, {Westmeier}, \& {Whiting}}]{2022PASA...39....5P}
{Pingel}, N.~M., {Dempsey}, J., {McClure-Griffiths}, N.~M., {et~al.} 2022, \pasa, 39, e005, \dodoi{10.1017/pasa.2021.59}

\bibitem[{{Piskunov} {et~al.}(1995){Piskunov}, {Kupka}, {Ryabchikova}, {Weiss}, \& {Jeffery}}]{1995A&AS..112..525P}
{Piskunov}, N.~E., {Kupka}, F., {Ryabchikova}, T.~A., {Weiss}, W.~W., \& {Jeffery}, C.~S. 1995, \aaps, 112, 525

\bibitem[{{Poudel} {et~al.}(2025){Poudel}, {Horton}, {Vazquez}, {Barger}, {Cashman}, {Fox}, {Lehner}, {Lucchini}, {Krishnarao}, {McClure-Griffiths}, {D'Onghia}, {Tumlinson}, {Tuli}, {Sdun}, {Gebhart}, {Anthony}, {Cole}, {van Loon}, {Roman-Duval}, {Ma}, {Lynn}, {Lee}, \& {Leahy}}]{2025arXiv250305968P}
{Poudel}, S., {Horton}, A., {Vazquez}, J., {et~al.} 2025, \apj, 984, 161, \dodoi{10.3847/1538-4357/adc099}

\bibitem[{{Poudel} {et~al.}(in prep){Poudel}, {Horton}, {Vazquez}, {Barger}, {Cashman}, {Fox}, {Lehner}, {Lucchini}, {Krishnarao}, {McClure-Griffiths}, {D'Onghia}, {Tumlinson}, {Goon Tull}, {Sdun}, {Gebhart}, {Anthony}, {Cole}, {Van Loon}, {Roman-Duval}, {Ma}, {Lynn}, {Lee}, \& {Leahy}}]{poudel_2025}
---. in prep, \apj

\bibitem[{{Putman} {et~al.}(1998){Putman}, {Gibson}, {Staveley-Smith}, {Banks}, {Barnes}, {Bhatal}, {Disney}, {Ekers}, {Freeman}, {Haynes}, {Henning}, {Jerjen}, {Kilborn}, {Koribalski}, {Knezek}, {Malin}, {Mould}, {Oosterloo}, {Price}, {Ryder}, {Sadler}, {Stewart}, {Stootman}, {Vaile}, {Webster}, \& {Wright}}]{1998Natur.394..752P}
{Putman}, M.~E., {Gibson}, B.~K., {Staveley-Smith}, L., {et~al.} 1998, \nat, 394, 752, \dodoi{10.1038/29466}

\bibitem[{{Ramachandran} {et~al.}(2018){Ramachandran}, {Hamann}, {Hainich}, {Oskinova}, {Shenar}, {Sander}, {Todt}, \& {Gallagher}}]{2018AA...615A..40R}
{Ramachandran}, V., {Hamann}, W.~R., {Hainich}, R., {et~al.} 2018, \aap, 615, A40, \dodoi{10.1051/0004-6361/201832816}

\bibitem[{{Redman} {et~al.}(2003){Redman}, {Al-Mostafa}, {Meaburn}, \& {Bryce}}]{2003MNRAS.344..741R}
{Redman}, M.~P., {Al-Mostafa}, Z.~A., {Meaburn}, J., \& {Bryce}, M. 2003, \mnras, 344, 741, \dodoi{10.1046/j.1365-8711.2003.06865.x}

\bibitem[{{Richter} {et~al.}(2018){Richter}, {Fox}, {Wakker}, {Howk}, {Lehner}, {Barger}, {D'Onghia}, \& {Lockman}}]{2018ApJ...865..145R}
{Richter}, P., {Fox}, A.~J., {Wakker}, B.~P., {et~al.} 2018, \apj, 865, 145, \dodoi{10.3847/1538-4357/aadd0f}

\bibitem[{{Richter} {et~al.}(2013){Richter}, {Fox}, {Wakker}, {Lehner}, {Howk}, {Bland-Hawthorn}, {Ben Bekhti}, \& {Fechner}}]{2013ApJ...772..111R}
---. 2013, \apj, 772, 111, \dodoi{10.1088/0004-637X/772/2/111}

\bibitem[{{Roman-Duval} {et~al.}(2020){Roman-Duval}, {Proffitt}, {Taylor}, {Monroe}, {Fischer}, {Fischer}, {Fullerton}, {Aloisi}, {Britt}, {Busko}, {Carlberg}, {De Rosa}, {Jedrzejewski}, {Lockwood}, {Frazer}, {Hernandez}, {James}, {Oliveira}, {Plesha}, {Riedel}, {Riley}, {Sahnow}, {Sankrit}, {Shaw}, {Smith}, {Sohn}, {Som}, {Ubeda}, \& {Welty}}]{2020RNAAS...4..205R}
{Roman-Duval}, J., {Proffitt}, C.~R., {Taylor}, J.~M., {et~al.} 2020, Research Notes of the American Astronomical Society, 4, 205, \dodoi{10.3847/2515-5172/abca2f}

\bibitem[{{Rousseau} {et~al.}(1978){Rousseau}, {Martin}, {Pr{\'e}vot}, {Rebeirot}, {Robin}, \& {Brunet}}]{1978AAS...31..243R}
{Rousseau}, J., {Martin}, N., {Pr{\'e}vot}, L., {et~al.} 1978, \aaps, 31, 243

\bibitem[{{Rubele} {et~al.}(2012){Rubele}, {Kerber}, {Girardi}, {Cioni}, {Marigo}, {Zaggia}, {Bekki}, {de Grijs}, {Emerson}, {Groenewegen}, {Gullieuszik}, {Ivanov}, {Miszalski}, {Oliveira}, {Tatton}, \& {van Loon}}]{2012A&A...537A.106R}
{Rubele}, S., {Kerber}, L., {Girardi}, L., {et~al.} 2012, \aap, 537, A106, \dodoi{10.1051/0004-6361/201117863}

\bibitem[{{Rubio} {et~al.}(1992){Rubio}, {Roth}, \& {Garcia}}]{1992A&A...261L..29R}
{Rubio}, M., {Roth}, M., \& {Garcia}, J. 1992, \aap, 261, L29

\bibitem[{{Russell} \& {Dopita}(1992)}]{1992ApJ...384..508R}
{Russell}, S.~C., \& {Dopita}, M.~A. 1992, \apj, 384, 508, \dodoi{10.1086/170893}

\bibitem[{{Sahnow} {et~al.}(2000){Sahnow}, {Moos}, {Ake}, {Andersen}, {Andersson}, {Andre}, {Artis}, {Berman}, {Blair}, {Brownsberger}, {Calvani}, {Chayer}, {Conard}, {Feldman}, {Friedman}, {Fullerton}, {Gaines}, {Gawne}, {Green}, {Gummin}, {Jennings}, {Joyce}, {Kaiser}, {Kruk}, {Lindler}, {Massa}, {Murphy}, {Oegerle}, {Ohl}, {Roberts}, {Romelfanger}, {Roth}, {Sankrit}, {Sembach}, {Shelton}, {Siegmund}, {Silva}, {Sonneborn}, {Vaclavik}, {Weaver}, \& {Wilkinson}}]{2000ApJ...538L...7S}
{Sahnow}, D.~J., {Moos}, H.~W., {Ake}, T.~B., {et~al.} 2000, \apjl, 538, L7, \dodoi{10.1086/312794}

\bibitem[{{Salem} {et~al.}(2015){Salem}, {Besla}, {Bryan}, {Putman}, {van der Marel}, \& {Tonnesen}}]{2015ApJ...815...77S}
{Salem}, M., {Besla}, G., {Bryan}, G., {et~al.} 2015, \apj, 815, 77, \dodoi{10.1088/0004-637X/815/1/77}

\bibitem[{{Sana} {et~al.}(2024){Sana}, {Tramper}, {Abdul-Masih}, {Blomme}, {Dsilva}, {Maravelias}, {Martins}, {Mehner}, {Wofford}, {Banyard}, {Barbosa}, {Bestenlehner}, {Hawcroft}, {John Hillier}, {Todt}, {Larkin}, {Mahy}, {Najarro}, {Ramachandran}, {Ram{\'\i}rez-Tannus}, {Rubio-D{\'\i}ez}, {Sander}, {Shenar}, {Vink}, {Backs}, {Brands}, {Crowther}, {Decin}, {de Koter}, {Hamann}, {Kehrig}, {Kuiper}, {Oskinova}, {Pauli}, {Sundqvist}, {Verhamme}, \& {Xshoot-U Collaboration}}]{2024AA...688A.104S}
{Sana}, H., {Tramper}, F., {Abdul-Masih}, M., {et~al.} 2024, \aap, 688, A104, \dodoi{10.1051/0004-6361/202347479}

\bibitem[{{Savage} \& {Sembach}(1991)}]{1991ApJ...379..245S}
{Savage}, B.~D., \& {Sembach}, K.~R. 1991, \apj, 379, 245, \dodoi{10.1086/170498}

\bibitem[{{Selman} {et~al.}(1999){Selman}, {Melnick}, {Bosch}, \& {Terlevich}}]{1999A&A...341...98S}
{Selman}, F., {Melnick}, J., {Bosch}, G., \& {Terlevich}, R. 1999, \aap, 341, 98

\bibitem[{{Serebriakova} {et~al.}(2023){Serebriakova}, {Tkachenko}, {Gebruers}, {Bowman}, {Van Reeth}, {Mahy}, {Burssens}, {IJspeert}, {Sana}, \& {Aerts}}]{2023AA...676A..85S}
{Serebriakova}, N., {Tkachenko}, A., {Gebruers}, S., {et~al.} 2023, \aap, 676, A85, \dodoi{10.1051/0004-6361/202346108}

\bibitem[{{Setton} {et~al.}(2023){Setton}, {Besla}, {Patel}, {Hummels}, {Zheng}, {Schneider}, \& {Salem}}]{2023ApJ...959L..11S}
{Setton}, D.~J., {Besla}, G., {Patel}, E., {et~al.} 2023, \apjl, 959, L11, \dodoi{10.3847/2041-8213/ad0da6}

\bibitem[{{Shapiro} \& {Field}(1976)}]{1976ApJ...205..762S}
{Shapiro}, P.~R., \& {Field}, G.~B. 1976, \apj, 205, 762, \dodoi{10.1086/154332}

\bibitem[{{Sonnentrucker} {et~al.}(2009){Sonnentrucker}, {Massa}, {Kruk}, {Blair}, {Ake}, {Andersson}, {Chayer}, {Dixon}, {Fullerton}, Elizabeth, {Moos}, \& {Sahnow}}]{fuse_book_data}
{Sonnentrucker}, P., {Massa}, D., {Kruk}, J., {et~al.} 2009, {The FUSE Archival Data Handbook}

\bibitem[{{Springel}(2005)}]{2005MNRAS.364.1105S}
{Springel}, V. 2005, \mnras, 364, 1105, \dodoi{10.1111/j.1365-2966.2005.09655.x}

\bibitem[{{Staveley-Smith} {et~al.}(2003){Staveley-Smith}, {Kim}, {Calabretta}, {Haynes}, \& {Kesteven}}]{2003MNRAS.339...87S}
{Staveley-Smith}, L., {Kim}, S., {Calabretta}, M.~R., {Haynes}, R.~F., \& {Kesteven}, M.~J. 2003, \mnras, 339, 87, \dodoi{10.1046/j.1365-8711.2003.06146.x}

\bibitem[{{van Loon} {et~al.}(2005){van Loon}, {Marshall}, \& {Zijlstra}}]{2005A&A...442..597V}
{van Loon}, J.~T., {Marshall}, J.~R., \& {Zijlstra}, A.~A. 2005, \aap, 442, 597, \dodoi{10.1051/0004-6361:20053528}

\bibitem[{{van Loon} {et~al.}(2013){van Loon}, {Bailey}, {Tatton}, {Ma{\'\i}z Apell{\'a}niz}, {Crowther}, {de Koter}, {Evans}, {H{\'e}nault-Brunet}, {Howarth}, {Richter}, {Sana}, {Sim{\'o}n-D{\'\i}az}, {Taylor}, \& {Walborn}}]{2013A&A...550A.108V}
{van Loon}, J.~T., {Bailey}, M., {Tatton}, B.~L., {et~al.} 2013, \aap, 550, A108, \dodoi{10.1051/0004-6361/201220210}

\bibitem[{{Venzmer} {et~al.}(2012){Venzmer}, {Kerp}, \& {Kalberla}}]{2012A&A...547A..12V}
{Venzmer}, M.~S., {Kerp}, J., \& {Kalberla}, P.~M.~W. 2012, \aap, 547, A12, \dodoi{10.1051/0004-6361/201118677}

\bibitem[{{Walborn}(1977)}]{1977ApJ...215...53W}
{Walborn}, N.~R. 1977, \apj, 215, 53, \dodoi{10.1086/155334}

\bibitem[{{Walborn}(1991)}]{1991IAUS..148..145W}
{Walborn}, N.~R. 1991, in IAU Symposium, Vol. 148, The Magellanic Clouds, ed. R.~{Haynes} \& D.~{Milne}, 145

\bibitem[{{Walborn} \& {Blades}(1997)}]{1997ApJS..112..457W}
{Walborn}, N.~R., \& {Blades}, J.~C. 1997, \apjs, 112, 457, \dodoi{10.1086/313043}

\bibitem[{{Walker}(1999)}]{1999ASSL..237..125W}
{Walker}, A. 1999, in Astrophysics and Space Science Library, Vol. 237, Post-Hipparcos Cosmic Candles, ed. A.~{Heck} \& F.~{Caputo}, 125, \dodoi{10.1007/978-94-011-4734-7_8}

\bibitem[{{Wang} {et~al.}(2022){Wang}, {Hammer}, \& {Yang}}]{2022MNRAS.515..940W}
{Wang}, J., {Hammer}, F., \& {Yang}, Y. 2022, \mnras, 515, 940, \dodoi{10.1093/mnras/stac1640}

\bibitem[{{Zech} {et~al.}(2008){Zech}, {Lehner}, {Howk}, {Dixon}, \& {Brown}}]{2008ApJ...679..460Z}
{Zech}, W.~F., {Lehner}, N., {Howk}, J.~C., {Dixon}, W. V.~D., \& {Brown}, T.~M. 2008, \apj, 679, 460, \dodoi{10.1086/587135}

\bibitem[{{Zheng} {et~al.}(2024){Zheng}, {Tchernyshyov}, {Olsen}, {Choi}, {Bustard}, {Roman-Duval}, {Zhu}, {Di Teodoro}, {Werk}, {Putman}, {McLeod}, {Faerman}, {Simons}, \& {Peek}}]{2024ApJ...974...22Z}
{Zheng}, Y., {Tchernyshyov}, K., {Olsen}, K., {et~al.} 2024, \apj, 974, 22, \dodoi{10.3847/1538-4357/ad64d2}

\end{thebibliography}
\clearpage
\appendix

\section{Additional Results from Gaussian and Voigt Line Fitting}
In Table~\ref{tab:Voigt_results_app}, we present the results of GAKSAP Gaussian decomposition and the Voigt fitting analysis for sightlines~1--3 and~5--8. The results for sightline~4 are included earlier in Table~\ref{tab:Voigt_results_tab} as a representative example of our fits. The sightline IDs are given before the target name in the table header. Similar to Table~\ref{tab:Voigt_results_tab}, we report the ion transition in the first column, the center velocity in the second column, the logarithm of the column density in the third, the Doppler parameter in the fourth, and describe whether the absorbers are blended with the disk (labeled ``Disk/Offset Component"), or not (labeled ``Offset Component"). Any absorber that extends beyond the \hi~boundary for the offset gas, we label as ``Offset Component$^{*}$." We mark any saturated components with a superscript ``a." Saturated components have their column densities reported as lower limits as we utilized the AOD method to determine these values. The uncertainties for the velocity centroids, column densities, and Doppler parameters are given and features that have an uncertainty of $\pm~0.0~\kms$ were held constant in the Voigt fitting process.

\begin{longtable*}{ccccc}
\caption*{Same as Table~\ref{tab:Voigt_results_tab} but for Sightlines~1--2 and~5--8.} 
\label{tab:Voigt_results_app} \\
\hline
\hline
Ion & \multicolumn{1}{c}{v$_{\rm LSR}$}& \multicolumn{1}{c}{$\log{\left(N_x/\cm^{-2}\right)}$} & \multicolumn{1}{c}{$b$} & Notes \\
    & (\kms) & (dex) &  (\kms) &  \\
    \hline
\endfirsthead
\multicolumn{5}{l}{{\tablename\ \thetable{} -- continued from previous page}} \\
\hline\hline
Ion & \multicolumn{1}{c}{v$_{\rm LSR}$}  &  \multicolumn{1}{c}{$\log{\left(N_x/\cm^{-2}\right)}$} & \multicolumn{1}{c}{$b$} & Notes \\
    & (\kms) &  (dex) & (\kms)&  \\
\hline\hline
\endhead
\hline \multicolumn{5}{r}{{Continued}} \\ \hline
\endfoot
\hline
\endlastfoot

\multicolumn{5}{c}{ \textbf{(1) Sk$-$68$^\circ$140}} \\\hline
H\textsc{~i}&+253.8 $\pm$ 0.3 &20.98 $\pm$  0.06&10.1 $\pm$ 0.7& Disk Component\\
&+265.9 $\pm$ 1.0&21.43 $\pm$ 0.03&19.0 $\pm$ 0.5&Disk/Offset Component\\
S\textsc{~ii} &  $+229.5\pm 0.0$& $14.74 \pm 0.05$ & $14.9 \pm 3.3$&Disk Component \\
&$+267.5^{a}\pm 0.6$ & $>$15.52 &   $10.0 \pm 0.0$ &Disk/Offset Component \\
&  $+302.9 \pm 0.0$ & $14.72 \pm 0.06$ &  $17.5 \pm 3.8$ & Offset Component\\
C\textsc{~ii}$^{*}$         &  $+239.1^{a} \pm 0.0$& $>$14.84 &$ 29.0 \pm 3.6 $ &Disk/Offset Component\\

                       &  $+296.3^{a} \pm 0.0$ &  $>$ 13.99& $18.1 \pm 4.2$& Disk/Offset Component\\
 & $ +337.4 \pm 3.9 $&  $13.31 \pm 0.15$ & $8.0 \pm 0.0$ &Offset Component$^{*}$ \\
Si\textsc{~ii}&  $+246.2^{a} \pm 3.6 $& $>$15.53 & $12.1 \pm 2.8$&Disk Component \\
&  $+260.9 ^{a}\pm 0.0$ & $>$15.61 &  $ 12.9 \pm 0.0$ &Disk Component\\
&  $+291.2 \pm 0.0$ & $14.84 \pm 0.16$ & $11.2 \pm 5.8$ & Offset Component \\
Fe\textsc{~ii} &  $+256.4 \pm 0.3$ & $15.37 \pm 0.01 $& $15.4 \pm 0.4$ &Disk Component\\
                &  $+296.1 \pm 0.0$&  $14.21 \pm 0.10$ & $10.6 \pm 2.6$ &Offset Component$^{*}$\\
Al\textsc{~iii}& $+222.2^{a} \pm 2.1 $ & $>$13.61 & $22.3 \pm 3.5$ &Disk Component\\
& $+264.0^{a} \pm2.1 $ &$>$ 13.50 & $17.0 \pm 2.4$ &Disk/Offset Component\\
& $+293.5 \pm 0.0$ & $13.13 \pm 0.06$ & $26.1 \pm 4.7$ &Disk/Offset Component\\
                   & $+341.4 \pm 0.0$ & $12.44 \pm 0.16 $& $15.7 \pm 5.4$&Offset Component$^{*}$\\
\hline\multicolumn{5}{c}{ \textbf{(2) Sk$-$68$^\circ$129}} \\\hline
H\textsc{~i}&+261.7 $\pm$ 0.2 &21.50 $\pm$ 0.02&17.3 $\pm$ 0.3& Disk/Offset Component\\
&+263.9 $\pm$ 0.3 &20.50 $\pm$ 0.07&6.0 $\pm$ 0.7& Disk Component\\
S\textsc{~ii}
                    & $+265.6^{a} \pm 3.3$ & $>$ 15.17 &  $13.3 \pm 4.0$ & Disk/Offset Component\\
                    & $+291.7 \pm 0.0$ & $14.35 \pm 0.11$ &    $8.8 \pm 0.0$ & Offset Component$^{*}$\\
Ni\textsc{~ii}&  $+232.3 \pm 0.0$ &$ 13.03 \pm 0.26$ &   $10.3 \pm 0.0$&Disk Component \\
&  $+259.4 \pm 1.5 $ &  $13.90 \pm 0.04$ & $10.3 \pm 3.1$ &Disk/Offset Component\\
                       &  $+288.2\pm 0.0$ &  $13.01 \pm 0.25$ &   $10.0 \pm 0.0$& Offset Component \\
C\textsc{~ii}$^{*}$ 
                       &  $+256.0^{a} \pm 0.9$ &  $>$ 14.36 & $22.7 \pm 1.8$ &Disk/Offset Component\\
                       &  $+311.8  \pm 0.0$&  $13.84 \pm 0.07$ & $28.7 \pm 6.1$ & Offset Component$^{*}$\\

Si\textsc{~ii}    & $ +219.2\pm 0.0$  & $14.63 \pm 0.14$ &    $6.0 \pm 0.0$ &Disk Component\\
& $ +246.1^{a} \pm 1.8$ & $>$15.90 & $12.1 \pm 1.8$ &Disk Component\\                
& $+270.5 \pm 0.0$ & $15.51 \pm 0.07$ & $15.1 \pm 3.6$ & Disk/Offset Component\\
& $+310.4 \pm 0.0$ & $14.61 \pm 0.21$ &    $8.8 \pm 0.0$ & Offset Component$^{*}$ \\

Fe\textsc{~ii}
&$+256.3 \pm 0.3$ & $15.38 \pm 0.01$ &  $14.2 \pm 0.5$ & Disk/Offset Component\\
                  &  $+293.5 \pm 2.1$ &$14.25 \pm 0.10$ & $10.5 \pm 3.7$ & Offset Component$^{*}$ \\ 
Al\textsc{~ii}& $+221.7^{a} \pm 0.0 $& $>$13.20 &   $7.6 \pm 0.0 $&Disk Component\\
& $+264.2 \pm 0.0$& $>$12.98 &  $4.8 \pm 3.0$& Disk/Offset Component\\
                       & $+289.2 \pm 0.0$& $>$  12.58  & $6.8\pm 0.0$&Offset Component$^{*}$\\

Al\textsc{~iii}&  $+248.4 \pm 1.8$ & $13.37 \pm 0.12$ & $10.7 \pm 3.1$&Disk Component\\

                        &  $+274.3 \pm 3.9$ &$13.18 \pm 0.18$ &  $14.3 \pm 5.4$& Offset Component \\
                        &  $+314.3 \pm 0.0$ & $13.31 \pm 0.13$ &  $27.0 \pm 9.5$ &Offset Component$^{*}$\\

&  $+357.3 \pm 4.8$ & $12.79 \pm 0.19$ &   $15.0 \pm 0.0$ &Offset Component$^{*}$\\
&  $+386.3 \pm 0.0 $&$12.33 \pm 0.23$ &  $12.5 \pm 0.0$ &Offset Component$^{*}$\\
                
\hline\multicolumn{5}{c}{ \textbf{(3) Sk$-$68$^\circ$155}} \\\hline
 H\textsc{~i}&+246.8 $\pm$ 0.5&21.21 $\pm$ 0.04 &14.3 $\pm$ 0.7& Disk Component\\
 &+272.9 $\pm$ 1.4&21.51 $\pm$ 0.02 & 24.8 $\pm$ 1.2& Disk/Offset Component\\
 P\textsc{~ii}&  $+232.3\pm1.2 $& $13.21 \pm 0.17$ &   $3.9 \pm 0.0 $&Disk Component\\
                      &  $+260.4^{a} \pm 1.5 $ &$>$13.67 & $15.4 \pm 2.3 $&Disk/Offset Component\\
 
& $+290.4 \pm 2.4$  &  $12.80 \pm 0.23$ & $3.9 \pm 0.0$& Offset Component \\
S\textsc{~ii}&  $+253.0^{a} \pm 0.0$ & $>$15.57 & $23.8 \pm 0.8$&Disk/Offset Component \\
& $+278.8^{a} \pm 0.9$  &$>$ 15.45 &  $13.2 \pm 1.4$& Disk/Offset Component \\
                 & $+308.1^{a} \pm 1.8$  & $>$ 14.86 &  $17.6 \pm 1.7$& Offset Component \\
C\textsc{~ii}$^{*}$&  $+237.07^{a} \pm 0.0$ &$>$14.55&   $25.3 \pm 0.0$ &Disk/Offset Component\\
& $ +250.8^{a} \pm 3.3$ & $>$14.32 &  $13.6 \pm 0.0$&Disk/Offset Component\\
&  $+275.4^{a}\pm 3.6$ & $>$14.19&$ 18.7 \pm 4.6$ &Disk/Offset Component\\
& $+314.6 \pm 3.3$  & $13.47 \pm 0.13$ & $13.8 \pm 3.8$ & Offset Component\\
Si\textsc{~ii}&  $+248.5^{a} \pm 2.7$& $>$15.55& $14.0 \pm 2.5$ &Disk Component\\
& $+276.1^{a} \pm 1.5$ & $>$15.74 & $13.1 \pm 2.1$& Disk/Offset Component \\
                       & $+303.2 \pm 2.1$  & $15.03 \pm 0.08$ &   $8.1 \pm 0.0$ & Offset Component\\
                       & $+318.1 \pm 1.5$ & $14.87 \pm 0.14$ &  $6.4 \pm 2.1$ & Offset Component\\
 Fe\textsc{~ii}&  $+247.0 \pm0.9$ &   $14.60 \pm 0.04$ &   $7.6 \pm 1.3$&Disk Component\\
& $+277.5 \pm 0.3$ & $15.29 \pm 0.01$ & $14.2 \pm 0.6$&Disk/Offset Component \\
                       & $+313.3 \pm 0.0$ & $14.89 \pm 0.02$ & $12.0 \pm 0.9$ & Offset Component\\
                
Al\textsc{~ii} & $+249.9^{a} \pm 11.7$ & $>$13.32&  $24.2 \pm 0.0$ &Disk/Offset Component\\
 & $+294.8^{a} \pm 0.0$ & $>$13.12 &  $14.5 \pm 0.0$ & Disk/Offset Component\\
 & $+325.1^{a} \pm 0.0$ & $>$12.42 &  $4.8 \pm 2.7$& Offset Component\\
& $+342.2  \pm 0.0$&   $11.78 \pm 0.34$ &   $9.2 \pm 0.0$ & Offset Component\\

Al\textsc{~iii} & $+219.8 \pm 0.0$ & $12.62 \pm 0.15$ &   $ 7.7 \pm 0.0$ &Disk Component \\
                   & $+249.5\pm 0.0$ & $13.17 \pm 0.07$ &$ 16.0 \pm 4.1$ &Disk/Offset Component \\

& $+276.1 \pm 3.9$ & $12.42 \pm 0.24$ &    $6.0 \pm 0.0$ &Offset Component \\
                   & $+295.5 \pm 0.0$& $13.01 \pm 0.06$ &   $21.0 \pm 0.0$ &Offset Component\\

\hline\multicolumn{5}{c}{ \textbf{(5) BI184} }\\\hline
H\textsc{~i}&+223.6 $\pm$ 0.2&20.83 $\pm$ 0.03&9.3 $\pm$ 0.4& Disk Component\\
&+252.8 $\pm$ 1.5&21.01 $\pm$ 0.03 & 28.9 $\pm$ 1.9& Disk/Offset Component\\
P\textsc{~ii}
                      &  $+241.8 \pm 1.2$ &   $13.39 \pm 0.03$ & $18.9 \pm 2.1$ &Disk/Offset Component \\
                      &  $+274.0 \pm 0.0$ &   $13.03 \pm 0.07$ &  $20.1 \pm 4.2$ &Disk/Offset Component\\
S\textsc{~ii}&  $+232.7^{a} \pm 1.5$ & $>$ 15.26 & $ 9.0 \pm 1.2$ & Disk/Offset Component\\
                &  $+256.0^{a} \pm 1.5$&  $>$ 14.98 &  $9.6 \pm 3.2$ & Disk/Offset Component\\
                 &  $+279.4 \pm 2.1$&  $14.91 \pm 0.08$ &  $8.9 \pm 2.1$ & Offset Component\\

Ni\textsc{~ii}
                  &  $+251.9  \pm 3.0$ & $13.77 \pm 0.04$ & $18.6 \pm 2.2$ & Disk/Offset Component\\
                  &  $+274.8 \pm 0.0$ & $13.44 \pm 0.08$ &  $14.7 \pm 4.4$ &Disk/Offset Component \\
C\textsc{~ii}$^{*}$                
        &  $+236.3^{a} \pm 0.6$ & $>$ 14.24 & $20.2 \pm 1.0$ & Disk/Offset Component\\
        &  $+279.4 \pm 1.8$& $13.67 \pm 0.06$ & $12.3 \pm 2.3$ &Disk/Offset Component \\

Si\textsc{~ii}
                       &  $+230.5 \pm 2.1$&  $15.63 \pm 0.06$ & $16.8 \pm 3.0$ &Disk/Offset Component \\
                       &  $+262.4  \pm 5.1$&  $15.25 \pm 0.13$ &   $19.0 \pm 0.0$ &Disk/Offset Component \\
Fe\textsc{~ii}&  $+221.5 \pm 0.0$ &  $14.50 \pm 0.08$ &  $9.1 \pm 2.5$ &Disk Component\\
                  &  $+247.0 \pm 0.6$& $15.14 \pm 0.03$ & $12.3 \pm 1.3$ & Disk/Offset Component\\
                  &  $+273.4 \pm 0.9$ & $14.69 \pm 0.06$ &  $7.2 \pm 1.7$ & Offset Component \\
Al\textsc{~iii}&  $+212.8 \pm 1.8$ & $ 12.82 \pm 0.08$ &  $10.2 \pm 2.8 $&Disk Component\\
                        &  $+247.1  \pm 2.7$&   $12.80 \pm 0.10$ & $16.1 \pm 5.3$ & Offset Component  \\
                        &  $+283.7 \pm 3.3$ &  $12.37 \pm 0.14$ &  $9.9 \pm 5.1$ & Offset Component\\
\hline\multicolumn{5}{c}{ \textbf{(6) Sk$-$71$^\circ$45}} \\\hline
H\textsc{~i}&+224.5 $\pm$ 0.3&21.09 $\pm$ 0.02&13.2 $\pm$ 0.4& Disk Component\\
&+261.2 $\pm$ 1.6 &20.75 $\pm$ 0.05&23.2 $\pm$ 2.4&Offset Component\\
P\textsc{~ii} &  $+209.8 \pm 0.9$ & $ 13.56 \pm 0.02$ &   $14.0 \pm 0.0$ &Disk Component\\
&  $+232.9 \pm 1.5$ & $12.92 \pm 0.11$ &  $9.6 \pm 2.9$& Disk/Offset Component \\
 &  $+251.9 \pm 1.2$&  $12.72 \pm 0.07$ &    $6.0 \pm 0.0$ &Offset Component\\
                      &  $+267.2 \pm 0.0$ &  $12.42 \pm 0.1$ &    $3.9 \pm 0.0$ &Offset Component\\
S\textsc{~ii}&  $+223.2^{a} \pm0.0$ &$>$ 15.90& $20.9 \pm 0.7 $&Disk/Offset Component\\
& $+252.5^{a} \pm 1.5$  & $>$ 15.52 & $ 4.1 \pm 2.6$& Offset Component \\

                 & $+273.0^{a} \pm 0.0$  & $>$ 14.84  & $15.2 \pm 1.2$ & Offset Component\\
                 
Ni\textsc{~ii} &  $+236.5 \pm4.5 $ & $ 13.85 \pm 0.1$ & $22.1 \pm 5.7$ &Disk Component \\
& $+257.5 \pm 0.0$  & $12.91 \pm 0.28$ & $8.0 \pm 0.0$& Offset Component\\
& $+276.1 \pm 0.0$ & $13.21 \pm 0.11$ & $5.6 \pm 2.6$& Offset Component \\
C\textsc{~ii}$^{*}$ & $ +193.2^{a} \pm 0.0$& $>$14.52 &   $10.1 \pm 0.0$ &Disk Component\\
 &  $+234.9^{a}  \pm 0.9$& $>$14.41& $10.1 \pm 0.0 $&Disk/Offset Component\\
&$+275.3 \pm 0.0$ & $13.53 \pm 0.04$ &  $13.5 \pm 1.8$& Offset Component \\
Si\textsc{~ii}&  $+201.5^{a} \pm1.2$  & $>$15.42 &   $6.4 \pm 1.8$&Disk Component\\
&  $+225.3^{a} \pm 0.9$ & $>$15.75& $28.9 \pm 1.1$&Disk/Offset Component \\

& $+272.5 \pm 1.2$  &  $14.77 \pm 0.10$ &  $8.4 \pm 2.0$& Offset Component \\
                       & $+289.3 \pm 0.0$ & $14.63 \pm 0.15$ & $17.3 \pm 5.9$& Offset Component\\
Fe\textsc{~ii}&  $+199.9 \pm2.4 $&$13.94 \pm 0.15$ &    $6.0 \pm 0.0$ &Disk Component\\
& $+234.8 \pm 0.6$ & $15.17 \pm 0.02$ & $22.5 \pm 1.6$& Disk/Offset Component \\
                         & $+273.4 \pm 0.9$  & $14.57 \pm 0.05$ &  $8.8 \pm 1.5$ & Offset Component\\                
Al\textsc{~ii}&  $+215.5^{a} \pm2.1 $&$>$13.33 &   $16.3 \pm 0.0$ &Disk Component\\
& $+269.2^{a} \pm 0.0$&  $>$ 13.10 &  $ 18.8 \pm 3.8$& Offset Component \\

&$+310.6 \pm 0.0$&	$12.41 \pm	0.11$ &$6.0 \pm 1.3$& Offset Component\\
Al\textsc{~iii} & $ +214.7^{a}\pm 0.3$ & $>$13.48 & $19.1 \pm 0.0$ &Disk Component\\
& $+269.2 \pm 0.0$ & $12.76 \pm 0.05$ & $27.4 \pm 4.0$& Offset Component \\
\hline\multicolumn{5}{c}{ \textbf{(7) Sk$-$71$^\circ$41}} \\\hline
H\textsc{~i}&+227.1 $\pm$ 0.2&21.22 $\pm$ 0.01&13.6 $\pm$ 0.3& Disk Component\\
&+264.4 $\pm$ 1.3&20.71 $\pm$ 0.05 &19.3 $\pm$ 1.9& Offset Component\\
C\textsc{~i} & $ +220.4 \pm 0.9$ & $13.71 \pm 0.04$ & $12.2 \pm 1.5$&Disk Component \\
&  $+256.7 \pm 4.2$ & $12.89 \pm 0.22$ &   $ 8.2 \pm 0.0$ &Offset Component\\
&  $+283.0 \pm 1.8$& $13.57 \pm 0.07$ & $14.1 \pm 2.9$ &Offset Component\\
P\textsc{~ii}&  $+218.9^{a}\pm 0.9 $& $>$13.57 &  $7.6 \pm 1.5$ & Disk Component\\
&  $+236.7^{a} \pm 1.5$ & $>$ 13.26&    $6.8 \pm 0.0$& Disk Component \\
 &  $+250.7  \pm 2.1$& $12.77 \pm 0.18$ &    $3.9 \pm 0.0$ &Offset Component \\
 &  $+275.8 \pm 0.0$ & $13.17 \pm 0.12$ & $22.2 \pm 7.8$ &Offset Component \\

S\textsc{~ii}&$ +224.6^{a} \pm 0.0$ &$>$15.57 &  $14.3 \pm 0.6$ &Disk Component\\
                 & $+244.0^{a} \pm 0.6$ & $>$ 14.57 &   $4.5 \pm 1.4$&Disk/Offset Component\\
                 & $+262.8^{a} \pm 1.8$ & $>$ 15.16  &  $16.4 \pm 3.1$&Offset Component \\
                 & $+287.9 \pm 0.0$ & $14.52 \pm 0.09$ &  $17.5 \pm 3.6$&Offset Component \\
Ni\textsc{~ii}&  $+223.9\pm 0.0$ & $ 13.20 \pm 0.14$ & $10.7 \pm 4.1$ &Disk Component\\

                       &  $+245.1 \pm 1.8$ &$13.71 \pm 0.12$ &  $8.5 \pm 2.5$& Disk/Offset Component \\
                       &  $+268.6 \pm 1.5$ &$13.24 \pm 0.12$ &  $6.0 \pm 2.7$& Offset Component \\

C\textsc{~ii}$^{*}$ &  $+201.7^{a}\pm 0.0$ & $>$14.48& $15.3 \pm 1.0$ &Disk Component\\
                       &  $+235.0^{a} \pm 0.0$ &  $>$ 14.19 & $14.7 \pm 2.3$ & Disk/Offset Component\\
                       &  $+269.7 \pm 4.5$ &  $13.74 \pm 0.12$ & $18.2 \pm 4.5$ & Offset Component\\
                       &  $+306.9 \pm 1.5$& $12.95 \pm 0.12$ &    $5.6 \pm 0.0$ & Offset Component\\
Fe\textsc{~ii}& $ +212.7\pm 0.0 $ &$ 14.18 \pm 0.22 $&   $12.0 \pm 0.0$ &Disk Component \\
                       &  $+246.8^{a} \pm 2.1$ & $>$ 15.05 & $20.1 \pm 4.3$ &Disk/Offset Component \\
                       &  $+286.1 \pm 4.5$& $14.51 \pm 0.16$ & $11.0 \pm 6.0$ & Offset Component\\                   
\hline\multicolumn{5}{c}{ \textbf{(8) Sk$-$71$^\circ$50}} \\\hline
H\textsc{~i}&218.1 $\pm$ 0.3&21.20 $\pm$ 0.02 &14.9 $\pm$ 0.4&Disk Component\\
&254.1 $\pm$ 0.9 &21.20 $\pm$ 0.02 &25.5 $\pm$ 1.3& Disk/Offset Component\\
S\textsc{~i}&  $+253.7 \pm 1.2$&  $13.90 \pm 0.07$ & $7.8 \pm 1.9$& Offset Component \\
& $+270.0 \pm 0.0$& $13.76 \pm 0.12$ & $9.2\pm 0.0$& Offset Component\\
                     & $+281.8 \pm 2.7$  & $13.34 \pm 0.24$ &   $3.9 \pm 0.0$& Offset Component \\
C\textsc{~i}& $ +211.4 \pm 0.0$ & $13.27 \pm 0.12$ & $11.7 \pm 4.2$ &Disk Component\\
& $+259.6 \pm 0.9$ & $13.65 \pm 0.08$ &  $5.8 \pm 1.4$ & Offset Component\\
                & $+279.2 \pm 3.9$ &  $13.33 \pm 0.17$ & $11.3 \pm 5.7$ & Offset Component\\
                & $+316.0 \pm 0.0$&  $13.22 \pm 0.10$ &  $4.7 \pm 2.2$ & Offset Component\\
P\textsc{~ii}& $ +225.3^{a}\pm 0.0$ &$>$13.73 &   $25.0 \pm 0.0$&Disk/Offset Component\\
&  $+265.7 \pm 0.0$ & $13.28 \pm 0.12$ & $20.7 \pm 7.8$& Disk/Offset Component \\
                      & $ +306.8\pm 2.4$ & $12.91 \pm 0.16$ &   $8.0 \pm 4.3$& Offset Component \\
S\textsc{~ii}&  $+200.9 \pm 0.0$ & $14.21 \pm 0.06$ &    $3.9 \pm 0.0$ &Disk Component\\
& $ +219.2^{a} \pm 0.9$ & $>$15.21& $  10.0 \pm 0.0$ & Disk Component\\
& $+250.0 ^{a}\pm 0.6$  & $>$ 15.53 & $18.5 \pm 1.4$& Disk/Offset Component \\
&$ +283.8^{a} \pm 0.6$ &  $>$ 15.07  &  $7.6 \pm 0.7$ &Offset Component\\
Ni\textsc{~ii} &  $+189.6 \pm3.6 $ &$ 12.71 \pm 0.23$ &  $6.9 \pm 5.8$&Disk Component\\
& $ +218.3 \pm 3.0$ & $13.28 \pm 0.19$ & $ 9.0 \pm 3.9$& Disk Component \\
&  $+241.5 \pm 1.8$ & $13.65 \pm 0.16$ & $ 11.3 \pm 5.2$ &Disk/Offset Component\\
&  $+260.0 \pm 0.9$& $13.62 \pm 0.12$ & $ 5.9 \pm 1.9$& Offset Component \\
                  & $+277.2 \pm 1.2$ & $13.35 \pm 0.11$ &  $6.4 \pm 2.9$& Offset Component \\
                  & $+291.4 \pm 3.0$ & $12.67 \pm 0.26$ & $3.9 \pm 0.0$ & Offset Component\\
C\textsc{~ii}$^{*}$&  $+190.1 \pm 4.8$ & $  12.7 \pm 0.3$ &    $7.0 \pm 0.0$ &Disk Component\\
&  $+219.8^{a}  \pm1.5$& $>$14.11 &   $17.3 \pm 0.0$ &Disk Component\\
&  $+253.6^{a} \pm 0.9$ & $>$ 13.62 &   $13.1 \pm 1.5$ & Offset Component \\
                       &  $+281.7 \pm 0.0$  & $13.23 \pm 0.11$ &  $5.3 \pm 2.4$& Offset Component\\

Si\textsc{~ii} &  $+222.5 \pm2.1$ &$ 15.32 \pm 0.07 $& $14.7 \pm 2.7$&Disk Component \\
& $+254.7^{a} \pm 0.0$   & $>$ 15.26 & $15.9 \pm 1.9$& Disk/Offset Component \\
                       & $+285.0 \pm 0.0$ &   $14.90 \pm 0.1$ &    $7.0 \pm 0.0$& Offset Component \\
Fe\textsc{~ii}&  $+220.0 \pm 2.4 $  & $15.08 \pm 0.06$ & $19.8 \pm 2.7$ &Disk Component\\
&  $+240.2 \pm 2.1$ & $14.63 \pm 0.12$ &    $6.0 \pm 0.0 $ &Disk/Offset Component\\
&  $+256.4 \pm 1.2 $  & $15.06 \pm 0.13$ &  $8.6 \pm 2.8 $&Disk/Offset Component\\
&  $+276.6 \pm 3.9$  & $14.83 \pm 0.16$ & $12.2 \pm 3.8$ &Disk/Offset Component\\

Al\textsc{~ii} &  $+204.2^{a}\pm 0.0$&  $>$12.93 &  $9.8 \pm 3.2$ &Disk Component\\
&  $+240.9^{a} \pm 2.7$&  $>$ 13.19 &   $ 9.1 \pm 0.0$& Disk/Offset Component \\
                       &  $+279.0^{a} \pm 1.2$  & $>$ 12.79 &    $6.9 \pm 0.0$& Offset Component \\
                       &  $+304.7 \pm 5.1$ & $12.00 \pm 0.24$ &  $10.4 \pm 0.0$&Offset Component$^{*}$\\

Al\textsc{~iii}& $ +219.3 \pm0.0 $ &$ 12.92 \pm 0.04$ & $19.0 \pm 2.4 $&Disk Component\\
&  $+260.4 \pm 0.0$ &   $12.78 \pm 0.08$ & $17.5 \pm 3.9$ & Offset Component\\
&  $+284.1 \pm 0.0$ &  $12.01 \pm 0.22$ & $6.0 \pm 0.0$ & Offset Component\\
\end{longtable*} 

\section{Additional Position-Velocity Maps}\label{appendix_pv_section}
Below in Figure~\ref{fig:appendix_pv_plots}, we include the position-velocity maps for sightlines~1-3 and sightlines~5-8. The left and middle panels are slices through the \hi~emission from the GASKAP survey along a $0\fdg267$ region along the Galactic longitude and latitude of each sightline, respectively. However, for  sightlines~1 and~2, we extended the slice through 0\fdg5 of the Galactic longitude and latitude to observe gaseous material exterior to LMC's disk. We include the radial velocities of each of the background targets (except for sightline~2) on the plots to demonstrate their relation to the localized systemic velocity of the LMC's disk. We note that sightlines~3, 6, and~7 have radial velocities that do not align with the \hi~emission along the disk in the GASKAP observations. We emphasize that the sensitivity of GASKAP is 
$\log{\left(N_{\rm H\textsc{~i},\,3\sigma}/\mathrm{cm}^{-2}\right)} = 20.0$ for a $30~\kms$ wide line \citep{2022PASA...39....5P} and therefore, it cannot detect emission that is fainter but could be at the lower radial velocities of sightlines~3 and~6 and higher radial velocity of sightline~7. In order to check the placement of the background star relative to the disk, we created position-velocity maps using the GASS observations which have a sensitivity of $\log{\left(N_{\rm H\textsc{~i},\,3\sigma}/\mathrm{cm}^{-2}\right)} = 18.2$ for a $30~\kms$ wide line \citep{2009ApJS..181..398M}. We find in the GASS observations that the radial velocities of the background stars do kinematically overlap with fainter \hi~emission belonging to the disk. We emphasize that while GASS has detectable emission at lower and higher velocities, it has an angular resolution that is 32 times greater than GASKAP. Meaning, while GASS can investigate the global properties and behaviors of the gas along each sightline, GASKAP is needed to explore the small-scale structure and to distinguish individual gaseous features.

\begin{figure*}[ht]
    \centering
    \subfloat{\includegraphics[height=0.27\textwidth]{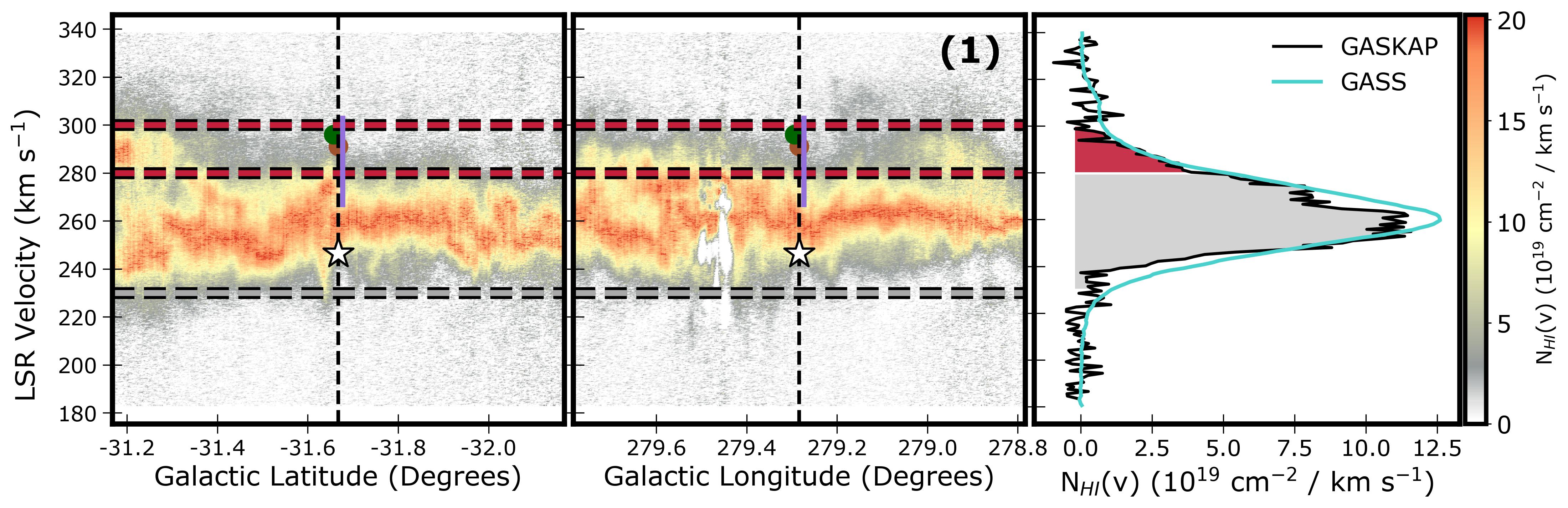}} \\
    \subfloat{\includegraphics[height=0.27\textwidth]{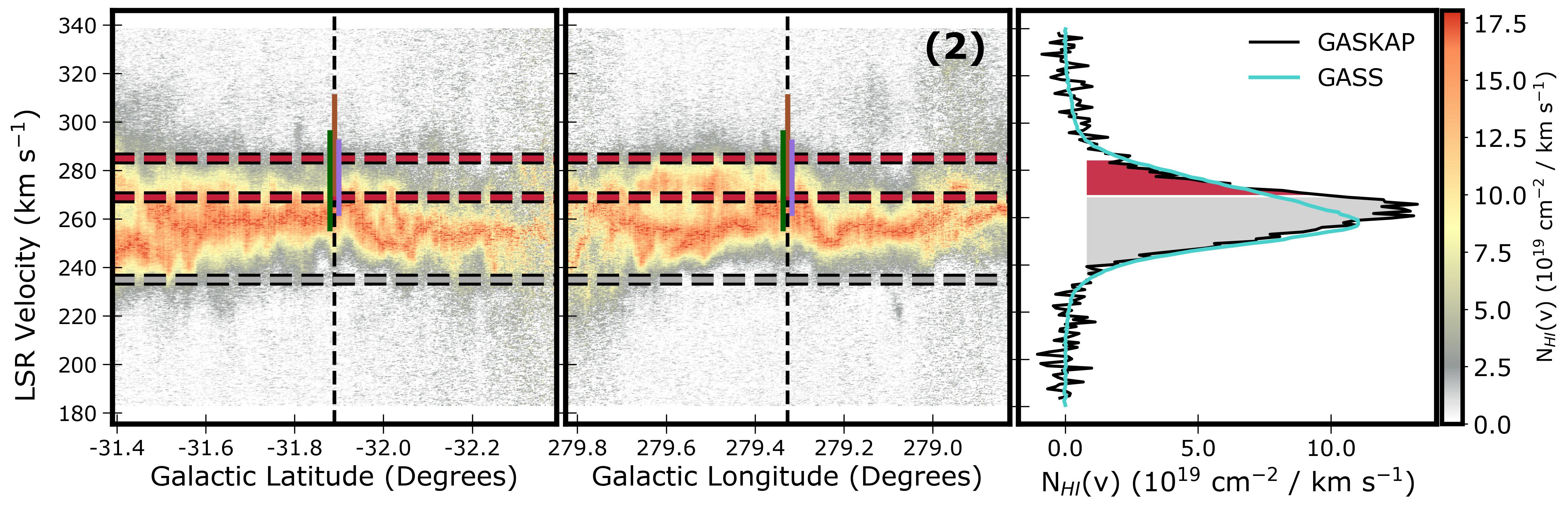}} \\
    \subfloat{\includegraphics[height=0.27\textwidth]{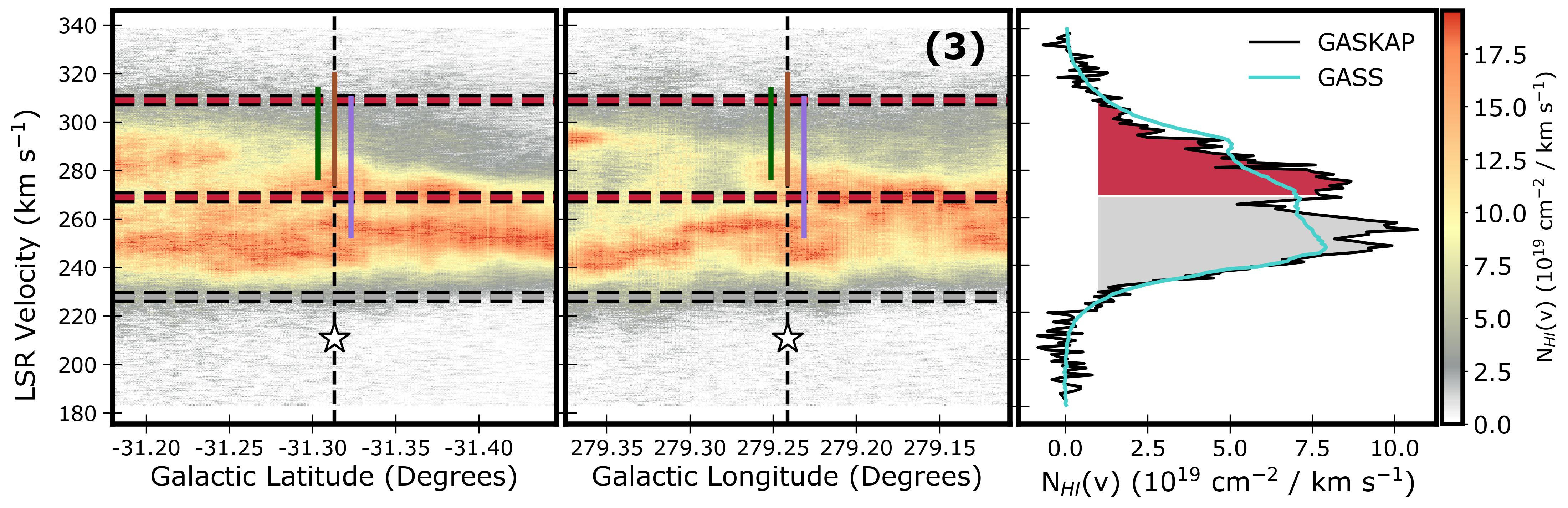}} \\
    \subfloat{\includegraphics[height=0.27\textwidth]{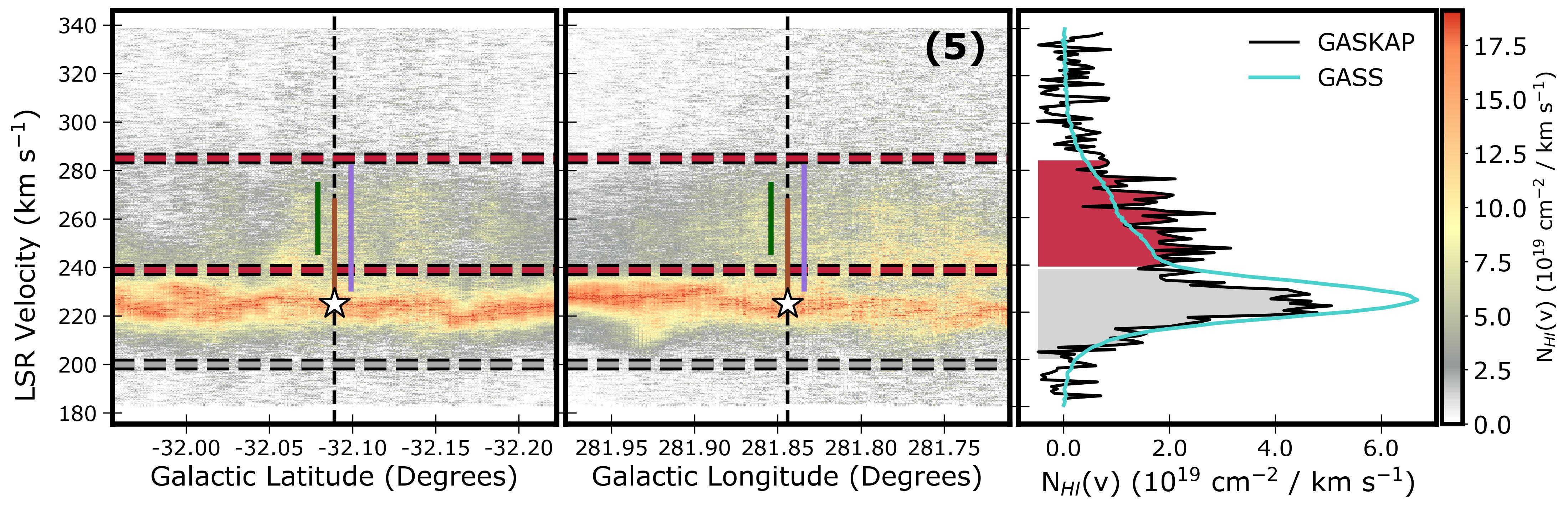}} 
    \caption{Position-velocity maps for sightlines~1-3 and~5-8, where the numbers in the upper right corner of the middle panel correspond to the sightline IDs in Table~\ref{tab:targets}. Sightline~4 is highlighted in Figure~\ref{fig:pv}. The vertical dashed lines in the left and middle panels are the locations of the sightline's Galactic latitude and longitude, respectively. We mark the radial velocity of the stellar sightline with a star symbol along the vertical dashed line. We note that sightline~2 does not have a recorded radial velocity and we do not include a star symbol on its maps. We highlight that the stars along sightlines~3,6,~and~7 do not have radial velocities that fall within the bulk motion of the \hi~gas in the disk. We explain this is due to the sensitivity limit of GASKAP (see Section~\ref{appendix_pv_section}). We find, using the more sensitive GASS observations, that sightlines~3, 6,~and~7 have fainter emission at the radial velocity of the star. We choose to show the GASKAP position-velocity maps because of its smaller angular resolution and therefore, more detailed small-scale structure of the gas. The right panel contains the \hi~emission from the GASS (cyan) and GASKAP (black) observations. We indicate the boundaries of the LMC's disk (gray) and the offset gas (red), as determined from the \hi~emission as discussed in Section~\ref{kinematic_bounds} and recorded in Table~\ref{tab:boundaries_hi}, by filling in the area under the emission profiles. We overlay the velocity extent of the Fe\textsc{~ii}, Si\textsc{~ii}, and S\textsc{~ii} components highlighted by the vertical green, brown, and purple lines, respectively, along each sightline. For sightline~1, we fixed the velocity centroids for the Fe\textsc{~ii} (green) and Si\textsc{~ii} (brown) components and symbolize this with the circles on the plot.  }\label{fig:appendix_pv_plots}
\end{figure*}

\addtocounter{figure}{-1}
\begin{figure*}[ht]
    \centering
    \subfloat{\includegraphics[height=0.27\textwidth]{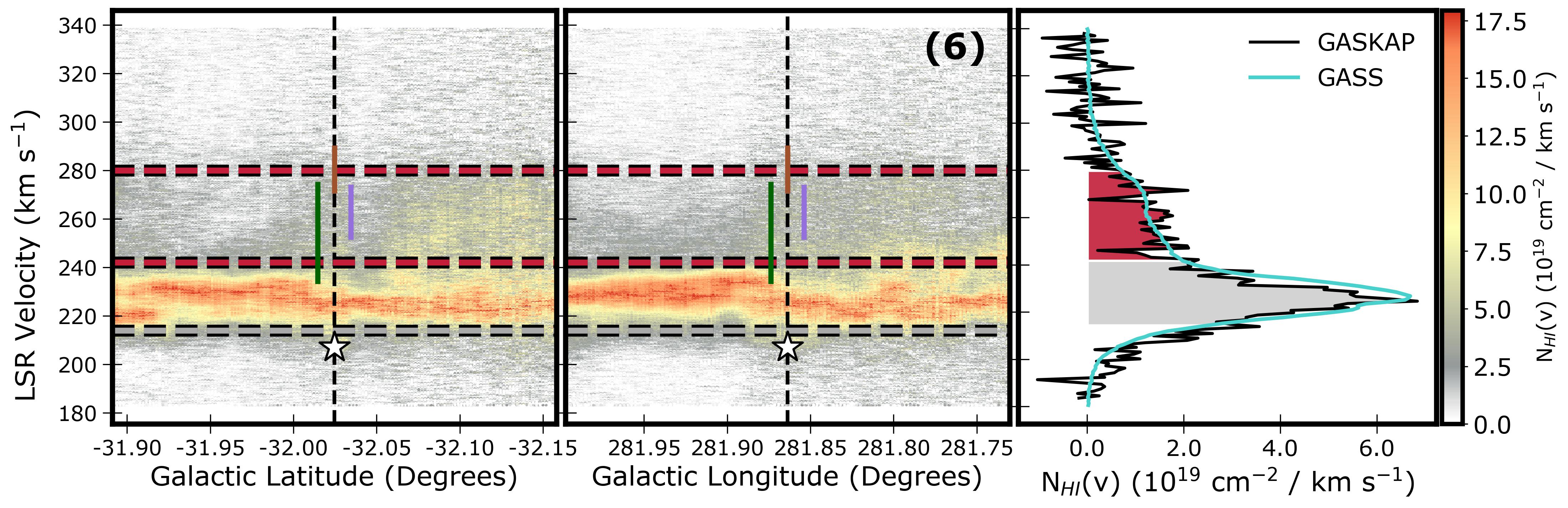}} \\
    \subfloat{\includegraphics[height=0.27\textwidth]{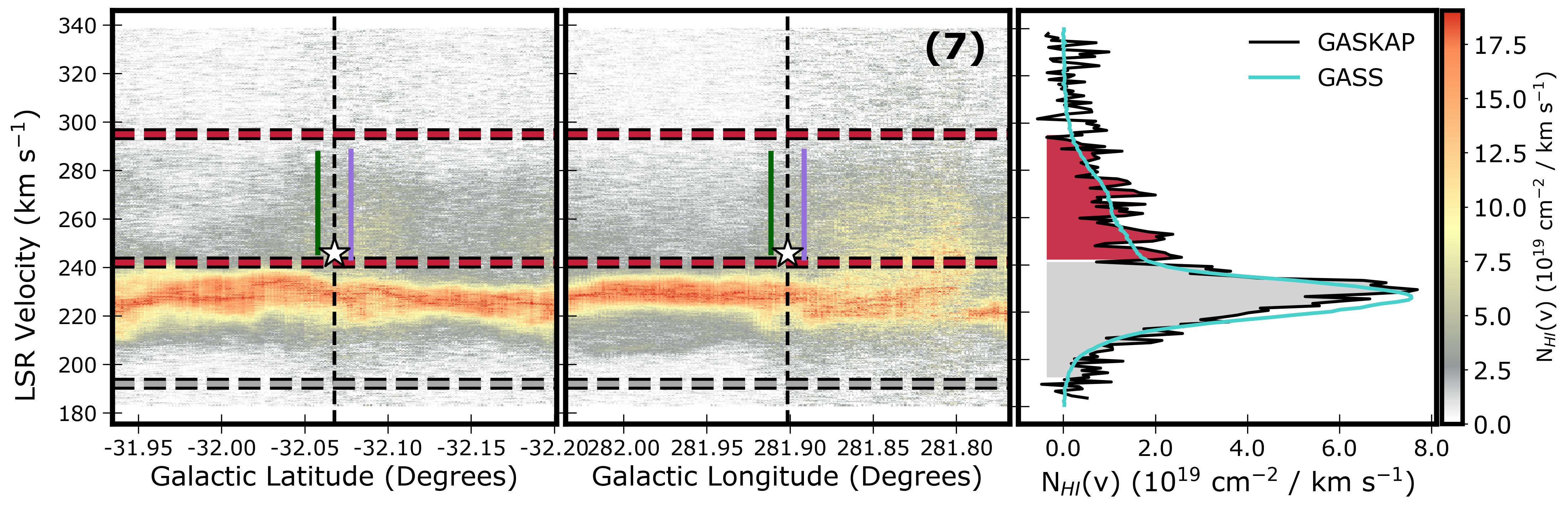}} \\
    \subfloat{\includegraphics[height=0.27\textwidth]{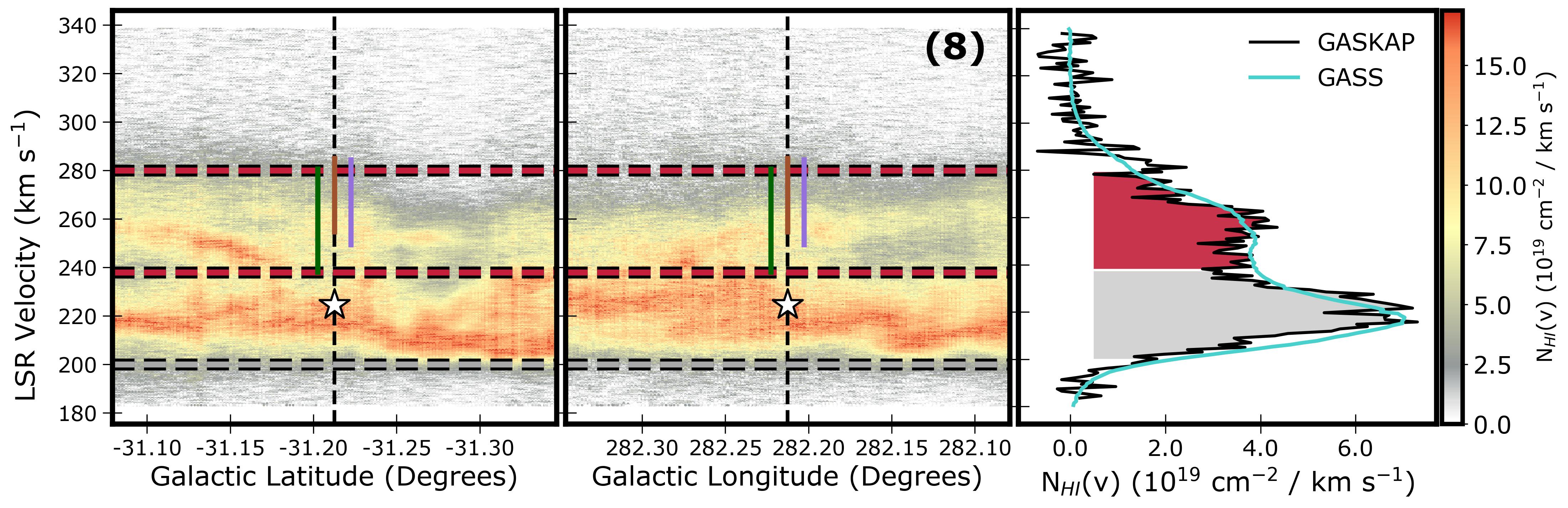}} 
    \caption{(Continued.)}
\end{figure*}

\section{Plotstacks with Voigt profiles}

Finally, we present the plotstacks for sightlines~1-3 and sightlines~5-8 in Figure~\ref{fig:appendex_plotstacks}. The top panel displays the GASS \hi~emission along each sightline. In the following panels, the absorption features of different low and intermediate ionization species are shown. Each panel is labeled with its transition and its oscillator strength. Again, we highlight the \hi~bounds of the LMC's disk and the offset gas in gray and red, respectively. The total Voigt profile fit is given by the light tan line and the individual components are marked with dashed black lines. The flux is normalized and the gray envelope is the 1$\sigma$ error. We label the predicted kinematic location of arm~B (purple) and arm~E (orange), given by the horizontal rectangles above the normalized flux and above the \hi~emission, from the velocity channel maps of \citet{2003ApJS..148..473K} and \citet{2003MNRAS.339...87S}. For sightlines~5--7, we only include the velocity extent of arm~B as these sightlines are mostly likely probing its material as they are spatially close by. Likewise, we only included arm~E on sightline~8 as arm~E is nearby. For the remaining sightlines, we are unsure which arm they are likely probing and place both arms as a reference on their plotstacks. 

\begin{figure*}[htp] 
    \centering
    \subfloat{\includegraphics[width=0.35\textwidth]{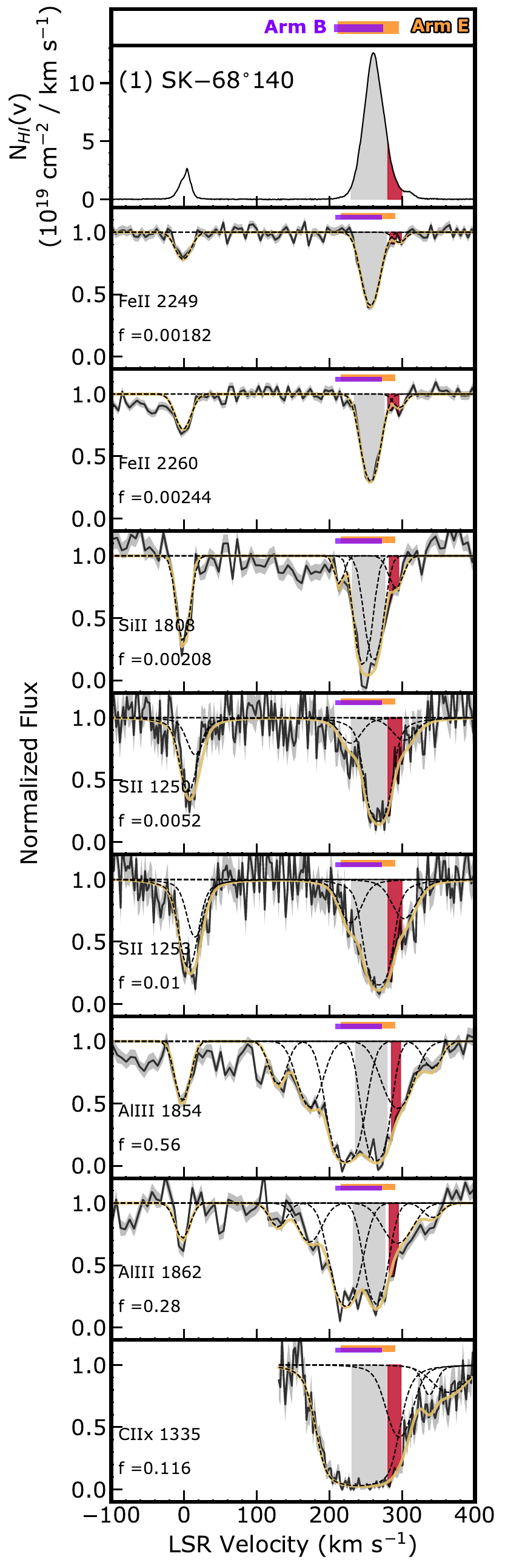}}
    \subfloat{\includegraphics[width=0.35\textwidth]{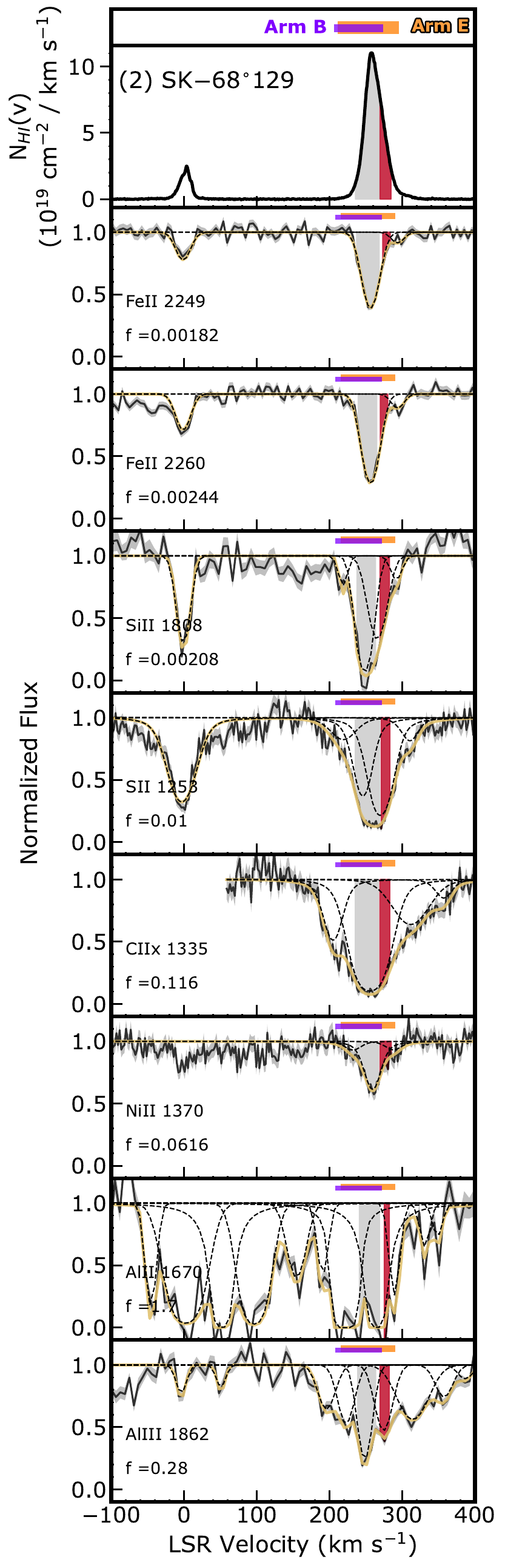}}
    \caption{Plotstacks for sightlines~1-3 and~5-8. We include 4~additional ionization species  for sightline~4 not highlighted in Figure~\ref{fig:plostacks}. The \hi~boundaries of the LMC's disk and the offset gas from Table~\ref{tab:boundaries_hi} are shaded in with gray and red, respectively. The individual Voigt fits are outlined by the black dashed lines. The total Voigt fit profile is highlighted in tan. The predicted velocities of arm~B and arm~E are labeled above each spectra. Each ionization species is labeled in the panel with their corresponding oscillator strength. The target name and ID are given in the \hi~emission panel.  }
    \label{fig:appendex_plotstacks}
\end{figure*}
\addtocounter{figure}{-1}
\begin{figure*}[htp] 
    \centering
    \subfloat{\includegraphics[width=0.35\textwidth]{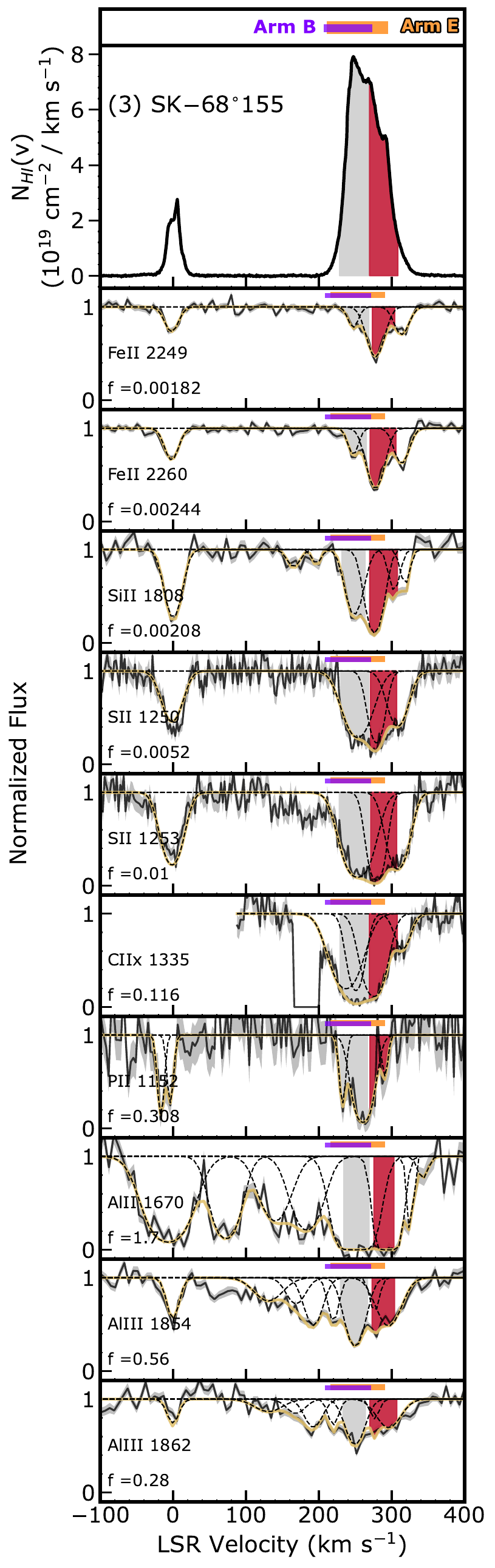}}
\raisebox{4.68cm}{\subfloat{\includegraphics[width=0.35\textwidth]{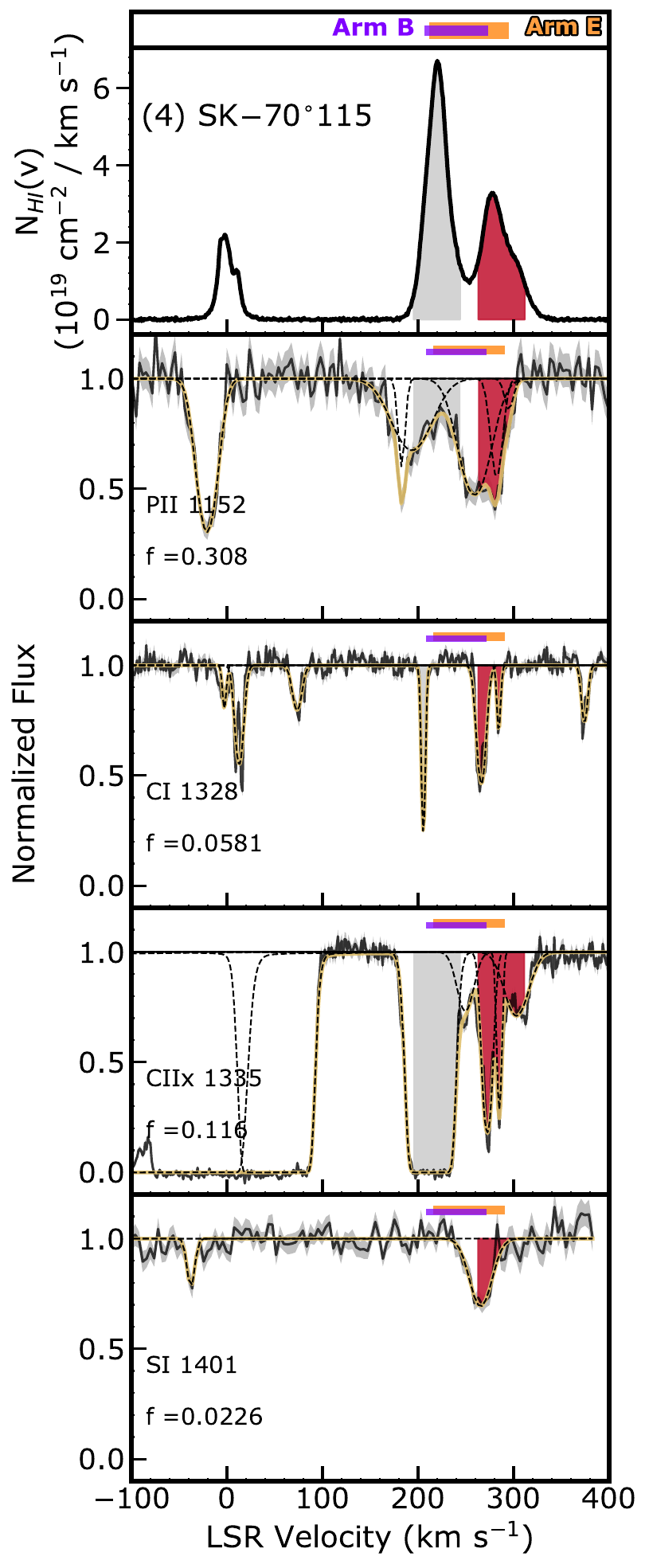}}}
    \caption{(Continued.)}
\end{figure*}
\addtocounter{figure}{-1}
\begin{figure*}[htp] 
    \centering
    \raisebox{4.68cm}{\subfloat{\includegraphics[width=0.35\textwidth]{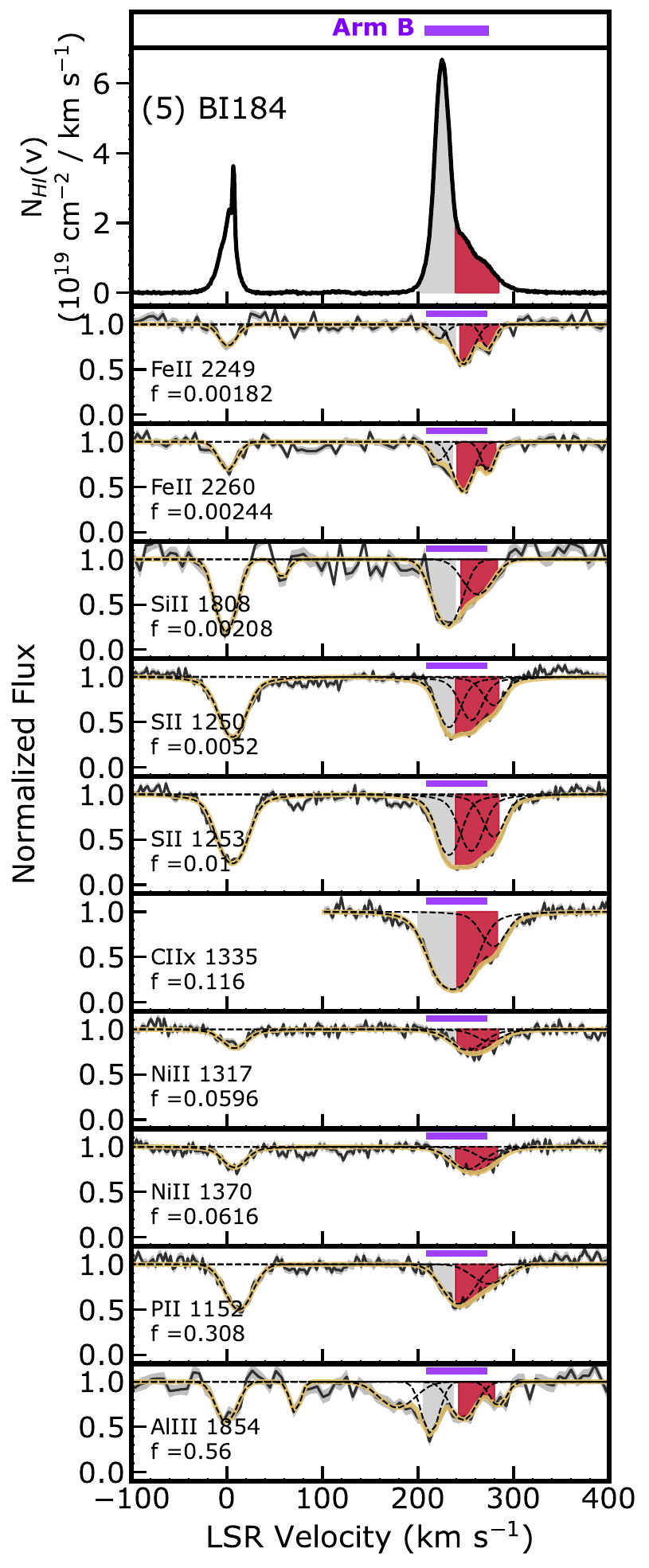}}} 
    \subfloat{\includegraphics[width=0.35\textwidth]{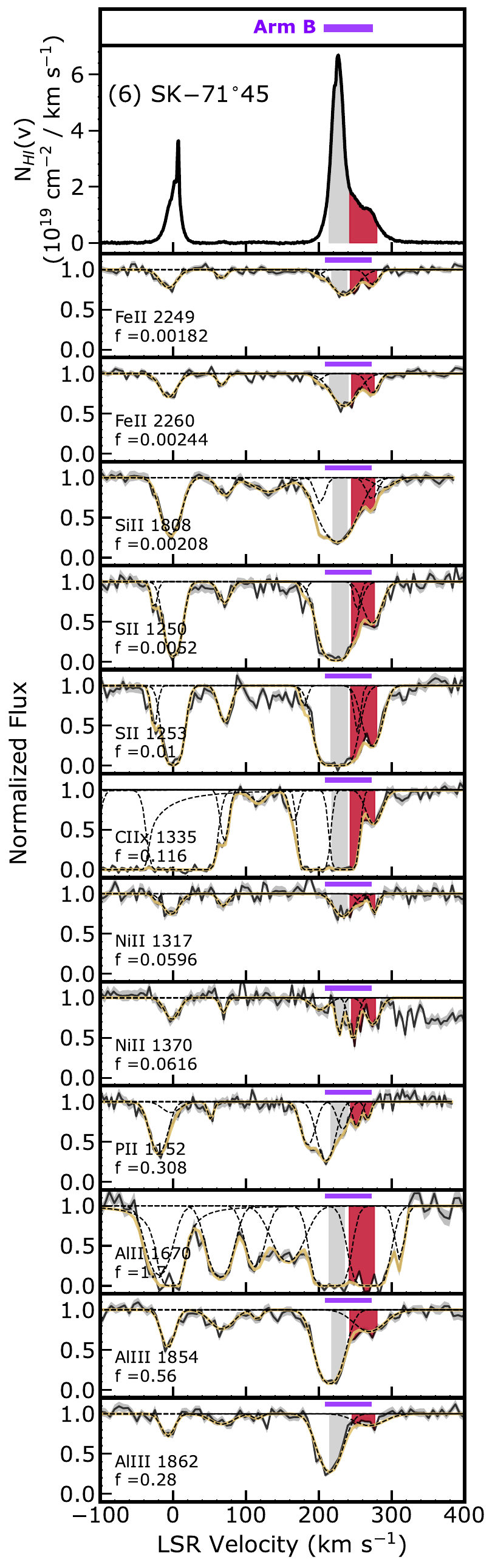}} 
    \caption{(Continued.)}
\end{figure*}
\addtocounter{figure}{-1}
\begin{figure*}[htp] 
    \centering
    \raisebox{4.68cm}{\subfloat{\includegraphics[width=0.35\textwidth]{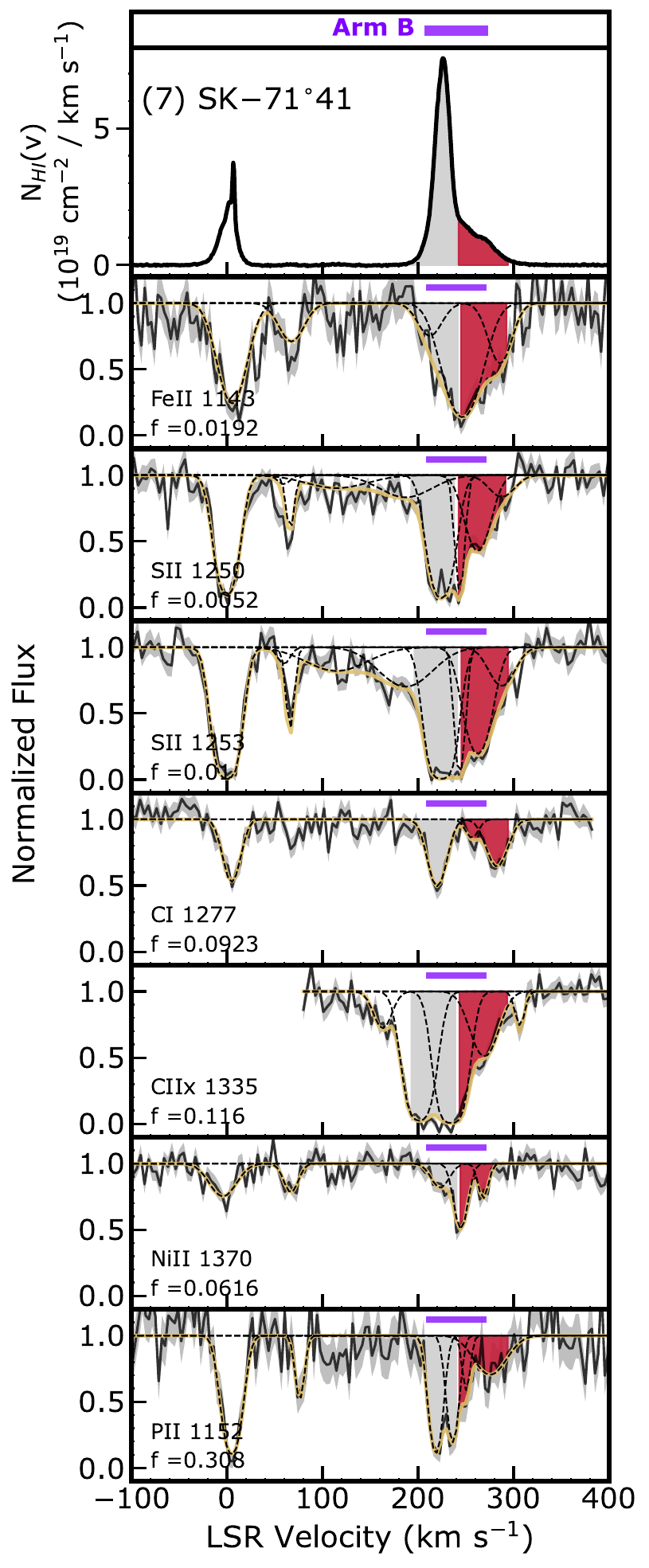}}} 
    \subfloat{ \includegraphics[width=0.35\textwidth]{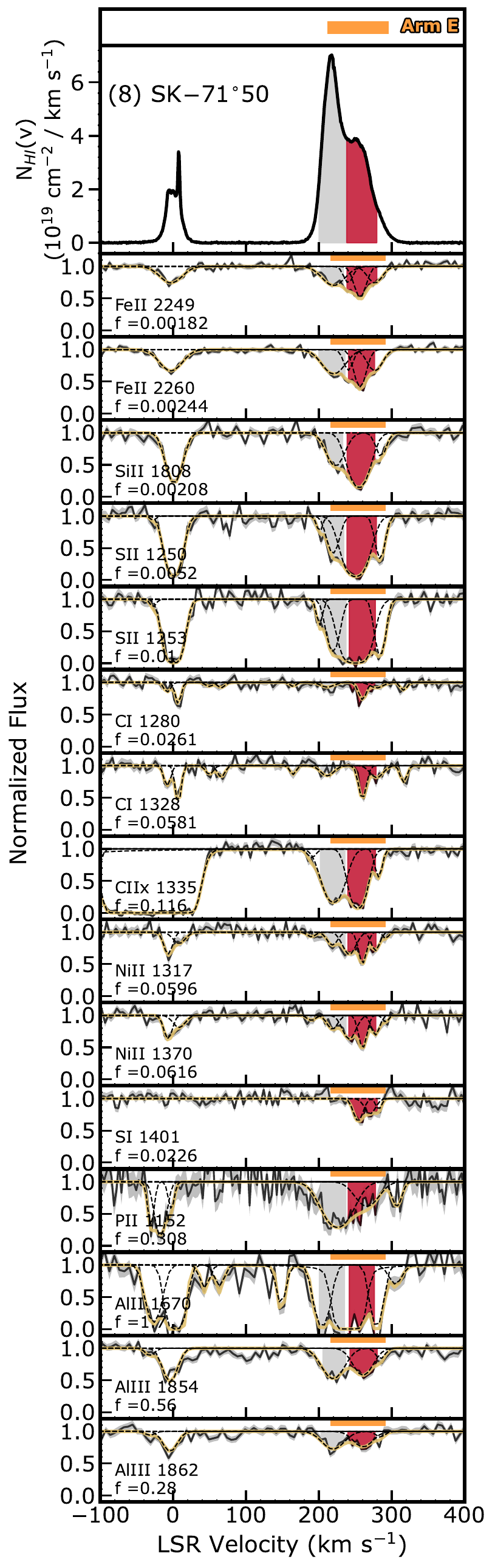}} 
    \caption{(Continued.)}
\end{figure*}

\section{Error Envelope}
For the Al\textsc{~iii} total integrated column density values, we performed a linear regression fit to the data points. We also incorporated an error envelope surrounding the best-fit line which follows the prediction interval as given by 

\begin{equation}\label{prediction_interval}
\hat{y}_{h} \pm t \times \sqrt{MSE \times \left( 1+ \frac{1}{n} + \frac{(x_{h}-\bar{x})^2}{\sum (x_{i}-\bar{x})^2}\right)}
\end{equation}
where $\hat{y}$ is the fitted line, $t$ is the t-multiplier from the Student's t-test, $n$ is the number of data points, $x$ is the angular offsets, $\bar {x}$ is the mean of the offsets, and $MSE$ is the Mean Square Error: 
\begin{equation}\label{MSE}
    MSE = \frac{1}{n} \sum_{i=1}^{n} (Y_{i} - \hat{Y}_{i})^2
\end{equation}
In Equation~\ref{MSE}, $Y_{i}$ is the total integrated column density of Al\textsc{~iii} and $\hat{Y_{i}}$ are the y-values from the linear regression fit. We can simplify Equation~\ref{prediction_interval} by identifying that 
\begin{equation}
    \sigma_{residuals} = \sqrt{\frac{\sum_{i=1}^{n} (Y_{i} - \hat{Y}_{i})^2}{n}}
\end{equation}
and therefore, 
\begin{equation}
    \hat{y}_{h} \pm t \times \sigma_{residuals}\times \sqrt{\left( 1+ \frac{1}{n} + \frac{(x_{h}-\bar{x})^2}{\sum (x_{i}-\bar{x})^2}\right)}
\end{equation}
The gray error envelope in  the bottom panel of Figure~\ref{tot_int_CD}, shows the 3$\sigma$ predication interval for the liner regression fit.  

\end{document}